\documentclass[printer]{tADP2e}

\pdfoutput=1
\newcommand{\suffix}[1]{#1.pdf}

\usepackage{amssymb}
\usepackage{graphicx}
\usepackage{amsmath} 
\usepackage[varg]{txfonts}
\usepackage{color}
\usepackage{bm}

\newcommand{\note}[1]{}
\newcommand{\NOTES}[1]{}
\def\NOTES{
\setlength{\marginparwidth}{6cm}
\setlength{\oddsidemargin}{1.5cm}
\setlength{\evensidemargin}{6.5cm}
\renewcommand{\note}[1]{\marginpar{\sf\small \color{blue} ##1}}}
\newcommand{\APSCR}[1]{\copyright\;#1 by The American Physical Society}

 \newcommand{\eref}[1]{(\ref{#1})} 
\newcommand{\eeref}[1]{equation (\ref{#1})} 
\newcommand{\fref}[1]{figure \ref{#1}}
\newcommand{\frefs}[1]{figures \ref{#1}}
\newcommand{\sref}[1]{section \ref{#1}}
\newcommand{\aref}[1]{Appendix \ref{#1}}
\newcommand{\tref}[1]{table \ref{#1}}
\newcommand{\etal}{\emph{et al.}}
\newcommand{\av}[1]{\langle#1\rangle}
\newcommand{\rC}{\text{\bf{C}}} 
\newcommand{\rI}{\text{\bf{I}}} 
\newcommand{\rL}{\text{\bf{L}}}    
\newcommand{\ncut}{n_{\rm cut}} 
\newcommand{\ecut}{\epsilon_{\rm cut}} 
\newcommand{\tecut}{\tilde{\epsilon}_{\rm cut}}

\newcommand{\myint}[1]{\int d^3 #1\;}
\newcommand{\EQ}[1]{\begin{eqnarray}#1\end{eqnarray}}

\newcommand{\PC}{\mathcal{P}_{\rC}}

\newcommand{\PI}{\mathcal{P}_{\rI}}
\newcommand{\DP}[2]{\frac{\bar{\delta}#1}{\bar{\delta}#2}}
\newcommand{\DDP}[1]{\frac{\bar{\delta}}{\bar{\delta}#1}}
\newcommand{\DPDP}[2]{\frac{\bar{\delta}^2}{\bar{\delta}#1\bar{\delta}#2}}
\newcommand{\intV}[1]{\int d^3 #1\;} 
\newcommand{\mbf}[1]{\mathbf{#1}}

\newcommand{\bff}{\hat{\Psi}}
\newcommand{\bef}{\hat{\psi}}
\newcommand{\bfC}{\bef_{\rC}}
\newcommand{\bfI}{\bef_{\rI}}
\newcommand{\cf}{\psi_{\rC}} 
\def\x{\mathbf{x}}

\newcommand{\CF}{c-field}
\newcommand{\xd}{\x '}

\newcommand{\xa}{(\x)}
\newcommand{\xad}{(\x ')}

\newcommand{\eb}{\epsilon^{B}}

\begin{document}

\title{ Dynamics and statistical mechanics of ultra-cold Bose gases\\ using c-field techniques}
\author{P. B. Blakie$^{\rm a}$$^\ast$\thanks{$^\ast$Corresponding author. Email: bblakie@physics.otago.ac.nz
\vspace{6pt}}, A. S. Bradley$^{\rm a,b}$, M. J. Davis$^{\rm b}$, R. J. Ballagh$^{\rm a}$, and C. W. Gardiner$^{\rm a}$
\\\vspace{6pt}  $^{\rm a}${\em{Jack Dodd Centre for Quantum Technology, Department of Physics, University of Otago, Dunedin,
New Zealand}}; \\$^{\rm b}${\em{The University of Queensland, School of Physical Sciences, ARC Centre of Excellence for Quantum-Atom Optics, Qld 4072, Australia}
}\\\vspace{6pt} 
\thanks{The first two authors acknowledge equal contributions to this review.}}

\maketitle 
\begin{abstract} 

We review phase space techniques based on the Wigner representation that provide an approximate description of dilute ultra-cold Bose gases. In this approach the quantum field evolution can  be represented using equations of motion of a similar form to the Gross-Pitaevskii equation but with stochastic modifications that include quantum effects in a controlled degree of approximation. 
These techniques provide a practical quantitative description of both equilibrium and dynamical properties of Bose gas systems. We develop versions of the formalism appropriate at zero temperature, where quantum fluctuations can be important, and at finite temperature where thermal fluctuations dominate. The numerical techniques necessary for implementing the formalism are discussed in detail, together with methods for extracting observables of interest. Numerous applications to a wide range of phenomena are presented. 
\end{abstract}
\bigskip

\bigskip
\newpage
\centerline{\bfseries \large{Contents} }\medskip

\noindent  
{1.}~Introduction \hfill \pageref{sintro}\\ 
{2.}~Background formalism \hfill \pageref{SEC:Formalism} \\
\hspace*{10pt}{2.1.}~Effective field theory for the dilute Bose
gas \hfill \pageref{SecEffFieldThry} \\
\hspace*{10pt}{2.2}~Projection into the \CF\ region\hfill \pageref{SEC:ProjOps} \\
\hspace*{24pt}{2.2.1.}~Projection operators\hfill \pageref{sssprojops} \\
\hspace*{24pt}{2.2.2.}~The Hamiltonian and equation of motion\hfill \pageref{sec:HEOM}\\
\hspace*{10pt}{2.3.}~Wigner formalism and the truncated Wigner approximation \hfill \pageref{WignerSec} \\
\hspace*{24pt}{2.3.1.}~Wigner representation of a single quantum mode\hfill \pageref{sec:Wigneruse} \\
\hspace*{24pt}{2.3.2.}~Operator correspondences and equations of motion\hfill \pageref{sec:OPcorresEOMs} \\
\hspace*{24pt}{2.3.3.}~Adaption to quantum field theory in the \CF\ region \hfill \pageref{WigQFT}\\
\hspace*{24pt}{2.3.4.}~Functional derivative notation \hfill \pageref{sssfuncderiv}\\
\hspace*{24pt}{2.3.5.}~Operator correspondences\hfill \pageref{sssopcorres} \\
\hspace*{24pt}{2.3.6.}~Truncated Wigner approximation\hfill \pageref{truncatedWig} \\
\hspace*{24pt}{2.3.7.}~Sampling the Wigner distribution\hfill \pageref {subs:initStates}\\
\hspace*{24pt}{2.3.8.}~Alternative methods for sampling the Wigner distribution\hfill \pageref{altsampling} \\
\hspace*{24pt}{2.3.9.}~Validity criteria for the truncated Wigner method \hfill \pageref{sec:validity} \\
\hspace*{24pt}{2.3.10.}~Features and interpretation of the truncated Wigner method \hfill \pageref{TWAfeatures} \\
{3.}~{The projected Gross-Pitaevskii equation\hfill \pageref{SEC:PGPE} \\
\hspace*{10pt}{3.1.}~Classical field description of thermal Bose
fields\hfill \pageref{CFthermBG} \\
\hspace*{24pt}{3.1.1.}~Importance of the projector and numerical methods\hfill \pageref{sec:Pimport} \\
\hspace*{24pt}{3.1.2.}~\CF\  region for the PGPE: the
``classical region"\hfill \pageref{secCR} \\ 
\hspace*{24pt}{3.1.3.}~PGPE formalism\hfill \pageref{PGPEformalismsec} \\
\hspace*{10pt}{3.2.}~Hands-on introduction to the PGPE formalism\hfill \pageref{SechandsOnPGPE} \\
\hspace*{24pt}{3.2.1.}~Simulation parameters\hfill \pageref{SEC:PGPEparams} \\
\hspace*{24pt}{3.2.2.}~Initial state preparation\hfill \pageref{SEC:InitStatePGPE} \\
\hspace*{24pt}{3.2.3.}~PGPE thermalization\hfill \pageref{sPGPEtherm} \\
\hspace*{24pt}{3.2.4.}~Equilibrium: Ergodicity, correlation functions and condensate fraction\hfill \pageref{SEC:PGPEdensitycondfrac} \\
\hspace*{24pt}{3.2.5.}~Thermodynamic quantities: temperature and chemical potential\hfill \pageref{Sec:PGPEtemperature} \\
\hspace*{24pt}{3.2.6.}~Including the incoherent region atoms\hfill \pageref{SEC:S3incohregion} \\
\hspace*{24pt}{3.2.7.}~Validity conditions\hfill \pageref{s3validity} \\
\hspace*{10pt}{3.3.}~Applications to the uniform Bose gas\hfill \pageref{sAppsuniformBG} \\
\hspace*{24pt}{3.3.1.}~Temperature and quasi-particle modes of the uniform system\hfill \pageref{Sec:TandNmodes} \\
\hspace*{24pt}{3.3.2.}~Shift of $T_c$ for the uniform Bose gas\hfill \pageref{shiftTcuniform} \\ 
\hspace*{10pt}{3.4.}~Applications to the trapped Bose gas\hfill \pageref{sAppstrappedBG} \\
\hspace*{24pt}{3.4.1.}~Shift in $T_c$ for a trapped Bose gas: Comparison with experiment\hfill \pageref{Tctrapped}\\ 
\hspace*{24pt}{3.4.2.}~Quasi-2D Bose gas\hfill \pageref{sec:q2D}\\ 
\hspace*{24pt}{3.4.3.}~Two point correlation functions\hfill \pageref{SEC:PGPEApp2ptcorrelns}\\  
\hspace*{10pt}{3.5.}~Applications of non-projected classical fields at finite temperature\hfill \pageref{sec:other_cfields} \\
\hspace*{24pt}{3.5.1.}~Homogenous gas\hfill \pageref{otherhomoBG}\\ 
\hspace*{24pt}{3.5.2.}~Trapped gas\hfill \pageref{othertrappedBG}\\ 
\hspace*{24pt}{3.5.3.}~Superfluid turbulence\hfill \pageref{otherSFturbulence}\\ 
 {4.}~{Applications of the TWPGPE to quantum matter wave dynamics\hfill \pageref{sec:TWPGPEapps} \\
\hspace*{10pt}{4.1.}~Condensate collisions in free space\hfill \pageref{sec:BECcollisions} \\
\hspace*{10pt}{4.2.}~Truncated Wigner treatment of three-body loss\hfill \pageref{s3Bloss} \\ 
\hspace*{10pt}{4.3.}~Quantum reflection of a Bose-Einstein condensate\hfill \pageref{sec:reflection} \\ 
\hspace*{10pt}{4.4.}~Applications to optical lattices\hfill \pageref{soptlatt} \\ 
\hspace*{10pt}{4.5.}~Dynamical instabilities and quasiparticle dynamics\hfill \pageref{sec:deLaval} \\ 
\hspace*{10pt}{4.6.}~Vortex formation in a stirred BEC\hfill \pageref{stirredBECs} \\ 
\hspace*{10pt}{4.7.}~Quantum statistical effects in Superchemistry\hfill \pageref{sec:superChem} \\ 
  \hspace*{10pt}{4.8.}~The quantum linewidth of an atom laser\hfill \pageref{satomlaser} \\  
 {5.}~{The stochastic projected Gross-Pitaevskii equation\hfill \pageref{sec:sgpe} \\
\hspace*{10pt}{5.1.}~Formalism\hfill \pageref{s5formalism} \\
\hspace*{24pt}{5.1.1.}~Background\hfill \pageref{s5syssep}\\ 
\hspace*{24pt}{5.1.2.}~The system and its separation\hfill \pageref{otherhomoBG}\\ 
\hspace*{24pt}{5.1.3.}~Treatment of the incoherent region\hfill \pageref{s5incohreg}\\ 
\hspace*{24pt}{5.1.4.}~Treatment of the \CF\ region: deriving the equation of motion\hfill \pageref{s5cregion}\\ 
\hspace*{24pt}{5.1.5.}~Stochastic projected Gross-Pitaevskii equation\hfill \pageref{s5SPGPE}\\ 
\hspace*{10pt}{5.2.}~Growth and scattering in the SPGPE\hfill \pageref{s5growscatt} \\
\hspace*{24pt}{5.2.1.}~Growth terms\hfill \pageref{sec:growth}\\ 
\hspace*{24pt}{5.2.2.}~Scattering terms\hfill \pageref{sec:scattering}\\ 
\hspace*{10pt}{5.3.}~Simple growth SPGPE\hfill \pageref{s5simplegrow} \\
\hspace*{10pt}{5.4.}~Applications to the dynamics of partially condensed Bose gases\hfill \pageref{SPGPEapps} \\
\hspace*{24pt}{5.4.1.}~Spontaneous vortex formation during Bose-Einstein condensation\hfill \pageref{s5SpontVortex}\\ 
\hspace*{24pt}{5.4.2.}~Rotating Bose-Einstein condensation\hfill \pageref{RBEC}\\ 
{6.}~{Conclusion\hfill \pageref{sconclus} \\
Acknowledgments\hfill \pageref{sAcknowledge} \\ 
 {A.}~{Numerical technique for the harmonically trapped system\hfill \pageref{SEC:Numerics} \\
\hspace*{10pt}{A.1.}~Numerical requirements\hfill \pageref{snumreq} \\
\hspace*{10pt}{A.2.}~Spectral representation of the PGPE\hfill \pageref{sspecrep}\\ 
\hspace*{10pt}{A.3.}~Mode evolution\hfill \pageref{smodeevol} \\
\hspace*{10pt}{A.4.}~Separability\hfill \pageref{sAsep}\\ 
\hspace*{10pt}{A.5.}~Evaluating the matrix elements\hfill \pageref{sec:matrixEval} \\
\hspace*{10pt}{A.6.}~Overview of numerical procedure\hfill \pageref{SEC:HARMnummeth}\\ 
{B.}~{Numerical technique for the uniform system\hfill \pageref{FTnumerics} \\
\hspace*{10pt}{B.1.}~Spectral representation\hfill \pageref{FTspecrep} \\
\hspace*{10pt}{B.2.}~Evaluating the matrix elements\hfill \pageref{FTmatelems}\\ 
\hspace*{24pt}{B.2.1}~Fourier interpretation\hfill \pageref{FTinterp}\\ 
\hspace*{10pt}{B.3.}~Overview of numerical procedure\hfill \pageref{SEC:PWnummeth} \\
{C.}~{Mapping to stochastic equations\hfill \pageref{sec:stochMapping} \\

\pagebreak

\section{Introduction}\label{sintro}
 
The dilute ultra-cold Bose gas presents a rare opportunity for 
theoretical physics: it has well-characterized interactions, and
it is feasible to begin with the full quantum theory and subsequently use
well-controlled approximations to develop formalisms suitable for
calculations.
These systems can be precisely manipulated and observed
in experiments and offer a unique chance to compare computational
quantum field theories directly with experiment. 

Several aspects of experiments present challenges for theory. First,
the experiments are usually non-equilibrium with long relaxation
times and are well-beyond any sort of linearized treatment. Second,
the harmonic trapping potentials used in experiments complicate the
traditional many-body methods which are more suitable for uniform
systems. The low energy collective dynamics and numerous finite-sized
aspects of this system critically rely on the external potential
being treated as a primary consideration of the theory.

At zero temperature an almost pure Bose-Einstein condensate (BEC)
forms, and for a wide range of situations its dynamics are well
described by the time-dependent Gross-Pitaevskii equation (GPE), e.g., see
\cite{Denschlag2000a,Simula2005a}. This approach assumes that all the
atoms are well-represented by a single condensate wavefunction, and the GPE
 describes the
coherent evolution of this wavefunction neglecting all spontaneous and incoherent
processes. However, experiments routinely operate in regimes where such processes 
are important and the GPE provides an inadequate physical description, for example: 
\begin{itemize} 
\item At higher temperatures, approaching the condensation
temperature, $T_c$, a sizable thermal cloud will be present.
Experiments examining collective oscillation frequencies of BECs
found that for temperatures higher than about $0.6T_c$ the GPE, due
to its neglect of the interplay between the condensate and thermal cloud, 
incorrectly predicts the collective mode frequencies and damping 
\cite{Jin1997a,Hutchinson1997a,Jackson2002a}. 
\item Two nearly-pure BECs colliding produce a halo of atoms
scattered onto a spherical shell in momentum space
\note{Cite NIST}
\cite{Chikkatur2000a}. Provided the phase-space density of the scattered atoms 
is low, this can be viewed as incoherent scattering of the individual atoms in 
the condensates, and the GPE can be augmented \cite{Band2000a}
 to account for these. However, at 
higher scattered densities Bose stimulation becomes important, and a theory 
which includes both Bose stimulation as well as incoherent scattering is 
required.
\end{itemize}
In order to treat these examples and many others it is necessary 
to formulate a description of
Bose gases that combines coherent and incoherent physics in a
general, yet tractable manner.  The key to a successful theoretical approach is 
the recognition that in all of these examples, even when there is no BEC, one 
or many modes of the system have an occupation which is much larger than 
one quantum. The systems are then highly Bose-degenerate, and the matter-wave 
field behaves much like a classical field. A set of theoretical approaches relying on the existence of significant Bose-degeneracy, known generically as
\emph{\CF} methods, provide a comprehensive solution to this
problem. 

An example simulation demonstrating such a scenario is shown in \fref{fig:cond-pic}. The averaged momentum density of a \CF\ simulation (see \sref{SEC:PGPE}) which describes many degenerate modes of a trapped Bose gas is shown for a range of temperatures spanning the critical temperature. The condensate is seen to emerge from the broad  thermal cloud  as the temperature decreases below $T_c$. 

\begin{figure}[t]  
 \centering{\includegraphics[width=5.25in]{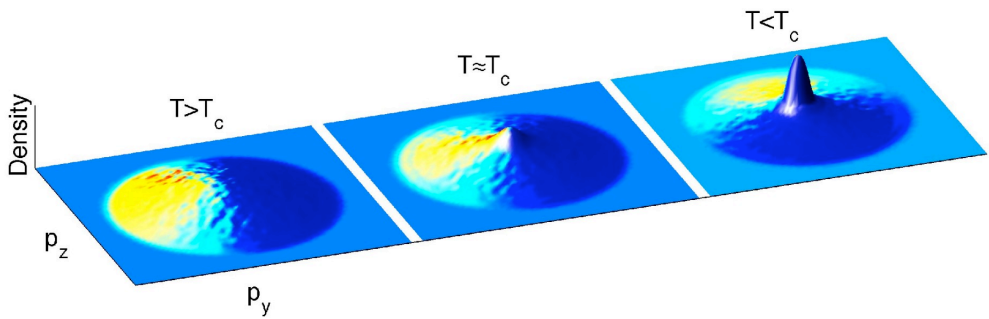} }
   \caption{(color online) Momentum space density (logarithmic) for classical field simulations at various temperatures. Emergence of the condensate is visible as a prominent spike at temperatures below $T_c$.  
    }
   \label{fig:cond-pic}
\end{figure} 

There are two main unifying features of the \CF\ techniques we present in this review.
The first is that the modes of the field theoretic description are divided into two regions. The precise way this is done depends on the approach, but generically we have:
\begin{enumerate}
\item[$\rC$:] {\bf  \CF\  region}.  This region is of primary importance in the description of the system and is so-named because it is simulated using classical
stochastic  
field equations. In the theories we develop here this region is chosen to contain not only the condensate, but all other highly Bose-degenerate modes. It may also contain modes of low occupation in which important dynamics occurs.
\item[$\rI$:] {\bf   Incoherent  region}. This region consists of the remaining modes which will individually be sparsely occupied (high energy) thermal  or vacuum modes.  Depending on the temperature and the density of states this region may contain a significant or even dominant fraction of the total number of particles in the system.  However, they have only a weak influence on the dynamics of the \CF\ region. In the techniques we develop in this review the static and dynamical properties of this region will be approximated as being incoherent.

\end{enumerate}
The second common feature to the \CF\ techniques is that their evolution equations are of similar
form to the GPE, but with 
 important  modifications.  
This is  the primary advantage of the
formalism --- its computational tractability, and capability to simulate
experimentally realistic parameter regimes.

This review is organised as follows. We begin in \sref{SEC:Formalism} where we outline the background theory relevant to the application of \CF\ techniques.  We then identify three separate implementations of the \CF\ techniques for different physical regimes, which are subsequently described in their own sections. 

 These techniques are\footnote{The term ``\CF" techniques has been coined to unify the methods discussed in this review. This terminology derives from the  name, ``\emph{classical field} method",  often  given to the pure PGPE formalism, 
 but which we have avoided because it can give the misleading impression that is not a quantum mechanical treatment.}:
 \begin{enumerate}
\item {\bf Projected Gross-Pitaevskii equation (PGPE) :} 
In all \CF\  approaches the GPE-like evolution is strictly limited to the $\rC$ region. 
This is implemented using a projection operator, and when this is the sole modification 
of the GPE we refer to the evolution equation  as the  {\em projected} Gross-Pitaevskii
equation. 

The PGPE is used to simulate the \CF\ region as a micro-canonical system, i.e.~as an isolated system of
fixed energy and number, with all couplings to the incoherent region
neglected. This approach is valid for high temperatures ($T\sim T_c$)
where the energy cutoff is chosen so that all \CF\ 
region modes are highly occupied, and quantum fluctuations can be
neglected. This theory is discussed in \sref{SEC:PGPE}, along
with applications of this formalism  to finite temperature
  phenomena.

\item {\bf Truncated Wigner PGPE (TWPGPE):} 
When there are modes with low occupation in the \CF\ region, additional noise terms must be included in the initial 
conditions to model the quantum-mechanical vacuum fluctuations.   Inclusion of 
quantum fluctuations cannot be done exactly, but can be well approximated by 
stochastic sampling of a Wigner
distribution for the initial state of the system.
The method
introduces spontaneous processes which are absent in the pure GPE
theory for which all scattering is stimulated. This formalism underlies all the \CF\ techniques  and is  
presented in \sref{WignerSec}, with applications of the theory to the non-equilibrium dynamics of systems at $T\ll T_c$  considered in \sref{sec:TWPGPEapps}.

\item {\bf Stochastic PGPE (SPGPE):} 
When exchange of energy and matter between the {\CF\ region} and the   {incoherent region} is important,  additional 
noise terms appear in the theory as well as in the initial conditions, via the 
truncated Wigner function method as above.

This approach is applicable in the same temperature regime as the PGPE
however it differs from that formalism in that scattering processes,
which couple to the incoherent region, are included. The theory is
implemented by solving the PGPE with additional dissipative and
stochastic terms. This transforms the description of the \CF\
region to a grand canonical form which includes the
exchange of particles and energy between the regions. This method,
discussed in \sref{sec:sgpe}, is well suited for modeling the
dynamics of evaporative cooling and, for example, vortex formation
during Bose-Einstein condensation. 

\end{enumerate}

In all the above \CF\ techniques it is important to ensure the numerical solutions of the 
equations inside the \CF\ region do not develop components 
outside that region.
Significant research has gone into developing numerical methods for
efficiently evolving projected equations, particularly the challenge of
implementing a projection operator efficiently and without
compromising the tractability of the equation compared to the usual
GPE. This is discussed further in \aref{SEC:Numerics} and \aref{FTnumerics}.
In this review we will show that a wide range of problems can be solved using 
these methods, and that accurate and reliable quantitative results can be 
computed.

\section{Background formalism}\label{SEC:Formalism}
\subsection{Effective field theory for the dilute Bose
gas}\label{SecEffFieldThry}
In this section we develop the basic formalism for the review.
We begin by restricting the full Hamiltonian to a low energy subspace, $\rL$, for which an \emph{effective} field theory provides an accurate description of the gas with a contact interaction. We then further divide this subspace into the $\rC$ and $\rI$ regions central to our development of the \CF\ techniques. Our basic approach here follows the derivation given in \cite{Gardiner2003a}.

Our starting point for describing a system of bosonic atoms interacting via an interatomic potential 
$ U\xa$ is the second quantized Hamiltonian
\begin{equation}
\hat{H} = \int d^3\x\,\bff^\dag\xa
H_{\rm{sp}}\bff\xa+\frac{1}{2}\int\!\!\!\int
d^3\x\,d^3\xd\,\bff^\dag\xa\bff^\dag\xad
U(\x-\xd)\bff\xad\bff\xa,\label{EQ:coldatomH} 
\end{equation}
where $\bff\xa$ is the bosonic field operator, and
\begin{eqnarray}
H_{\rm{sp}}&=&H_0+\delta V(\x,t),\label{EQ:Hsp}\\
H_0 &=&-\frac{\hbar^2\nabla^2}{2m}+V_0\xa,\label{EQ:H0}
\end{eqnarray}
are the \emph{single particle}  and \emph{basis}  Hamiltonians
respectively, with  $V_0(\x)$  the external  
potential. These Hamiltonians differ by the inclusion of a
``perturbation potential'' $\delta V(\x,t)$, which we include for
generality.
The basis Hamiltonian, $H_0$, is so named because we will use its
eigenstates as a basis for the low-energy description of the system, in particular to define the \CF\ region in \sref{SEC:ProjOps}. 
The inter-atomic potential, $U\xa$, has a size characterized by the
effective range parameter $r_0$, and only depends on the relative
separation of the atoms.

In typical ultra-cold atom experiments  the length scales of interest are much greater than $r_0$, and the full details
of the inter-atomic potential are unnecessary. It is desirable, therefore, to develop a theory that eliminates the need
to consider such small length scales, and hence the microscopic details of
the collisional interaction can be parameterized  in terms of the
S-wave scattering length alone -- such an approach is known as an
\emph{effective field theory}.

Formally this procedure can be implemented by restricting our
attention to a \emph{low-energy subspace}, $\rL$, that is spanned by
single particle states of energy less than an appropriately chosen
energy cutoff $E_{\max}$. This eliminates all momentum states with
momentum exceeding $\hbar\Lambda\xa\simeq\sqrt{2m(E_{\max}-V_0\xa)}$
 at $\x$, and in doing so effectively ``coarse grains'' our
description to a length scale of $1/\Lambda(\mathbf x)$. While our
choice of $E_{\max}$ is in principle arbitrary, the following
criteria ensure a simple and accurate effective field theory emerges: 
\begin{itemize}
\item[(i)]  $E_{\max}\ll {h^2}/{2m r_0^2}$, so that we eliminate the
need to include  short wavelength components of the wavefunction that
occur in the interaction region. Integrating out  these high energy
states allows the inter-atomic interaction to be replaced by the two body
$T$-matrix, which in the zero energy limit  becomes \cite{Gardiner2003a,Pethick2002a}
\begin{equation}T(0)\to g= \frac{4\pi a_s\hbar^2}{m}\end{equation}
where $a_s$ is the S-wave scattering length.
\item[(ii)] $E_{\max}\gg k_BT,\mu$, where $\mu$ is the chemical potential, so that the eliminated states will
not be occupied by thermal or interaction effects. This requirement ensures that the $T$-matrix does not depend on the population of states that are
eliminated in the theory, i.e., avoiding the need to consider a many-body $T$-matrix.  

\end{itemize}
As long as these conditions are satisfied, the effective field theory
derived should be insensitive to the precise value of $E_{\max}$
used. 

We can introduce a coarse-grained field operator, $\bef\xa$, which
only contains modes in $\rL$, and is described by the \emph{effective}
Hamiltonian
\begin{equation}
\hat{H}_{\rm{eff}} = \int d^3\x\,\bef^\dag\xa
H_{\rm{sp}}\bef\xa+\frac{u}{2}\int
d^3\x\;\bef^\dag\xa\bef^\dag\xa\bef\xa\bef\xa.\label{Eq:Heff}
\end{equation}
It must be emphasized that this resulting field theory has a cutoff,
so that the 
commutation relations of these new field operators are not precise
delta functions: 
\begin{eqnarray}
\left[\bef\xa,\bef^\dag\xad\right]&=&\delta_{\rL}(\x-\xd).
\label{EQ:psicommreln}
\end{eqnarray}
In \eeref{Eq:Heff} we have introduced the coupling constant
\begin{equation} 
u=\frac{g}{1-g\int_{\rL}d^3k\,\frac{\hbar^2k^2}{(2\pi)^3m}},\label{ueff}\end{equation}where  the integral  is taken over the momentum space of the $\rL$-region and accounts for the cutoff dependence of the coupling constant (e.g. see appendix A of \cite{Sinatra2002a}).

For the special case where the potential is slowly varying compared
to the local cutoff wavevector $\Lambda\xa$, we have
$\delta_{\rL}(\x-\xd)\simeq {\sin\left(\Lambda\xa|\xd-\x|\right)}/{2\pi^2|
\xd-\x|^3}$.
The function $\delta_{\rL}$ plays the role of a kind of
coarse-grained delta function which in general has a spatially
dependent width; however, it is also 
a projector into the subspace of non-eliminated modes. Using the
commutation relation \eref{EQ:psicommreln}
the Heisenberg equation of motion for the corresponding field operator takes the
form 
\begin{eqnarray}\label{psiEOM}
i\hbar\frac{\partial\bef\xa}{\partial t}&=&
\int d^3\xd\,{\delta}_{\rL}(\x-\xd)
\left\{H_{\rm{sp}}\bef\xad  
 + u\bef^\dag\xad\bef\xad\bef\xad\right\}.\label{EQ:psiEOM}
\end{eqnarray}
The main purpose of the methods discussed in this review is to
simulate this equation in various regimes. 

\begin{figure}[t]
\begin{center}
\includegraphics[height=6.75cm]{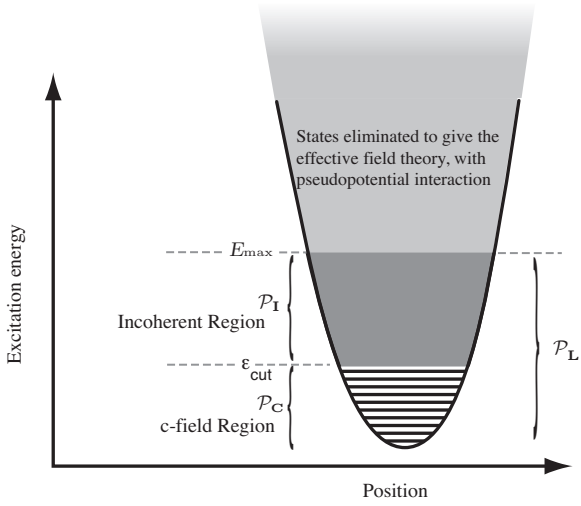}
\end{center}
\caption{\label{regions}%
Schematic view of the \CF\ region, the incoherent region, and
eliminated 
states for a harmonic trap. The \CF\ atoms require a quantum
description, and incoherent atoms may be treated using quantum
kinetic theory}
\end{figure}

\subsection{Projection into the \CF\ region}
\label{SEC:ProjOps}
\subsubsection{Projection operators}\label{sssprojops}

In \sref{SecEffFieldThry} we developed an effective field
theory description of the cold-atom Hamiltonian derived by
eliminating states outside of the  $\rL$ region. The resulting
effective Hamiltonian (\ref{Eq:Heff}) and equation of motion
(\ref{EQ:psiEOM}) are restricted to this space.

We now turn to a quantitative definition of the $\rL$ region. This is accomplished
by expanding the coarse-grained field operator as
\begin{eqnarray}
\bef\xa=\sum_{n\in{\rL}}\hat{a}_n\phi_n\xa,\label{EQ:psimodes}
\end{eqnarray} 
where  $\phi_{n}\xa$ are single particle eigenstates of the basis Hamiltonian with energy $\epsilon_n$, i.e.
\begin{equation}
\epsilon_n\phi_n\xa=H_0\phi_n\xa.
\end{equation} 
The operators $\hat{a}_n$ satisfy the usual Bose
commutation relations, $[\hat{a}_i,\hat{a}_j]=0$, and
$[\hat{a}_i,\hat{a}_j^\dag]=\delta_{ij}$. The restriction of the summation in \eeref{EQ:psimodes} to modes in $\rL$ is defined by
${\rL}=\{n:\epsilon_n\le E_{\max}\}$.


In general the requirements put on $E_{\max}$ for a useful effective
field theory to emerge lead to an $\rL$-space that is far too large to
simulate. Furthermore, the validity conditions for the \CF\  methods typically restrict their application to describing a sub-system of $\rL$.
Thus it is necessary to  further subdivide
 $\rL$  into two regions:
 \begin{enumerate}
\item The \emph{\CF\ region} ($\rC$), which
will normally consist of the lowest energy modes in $\rL$ and will be
numerically simulated using classical fields.\footnote{It is worth noting that the $\rC$ region has also been referred to as the \emph{coherent region}, the \emph{condensate band}, or the \emph{classical region} in the literature, although some of these names also imply additional restrictions on $\rC$. } 
\item The \emph{incoherent region} ($\rI$), consisting of
all the modes of $\rL$ not in $\rC$. The choice of $\rI$ will be such that any atoms occupying this region will be best described by a particle-like description.\end{enumerate}

For all cases considered in this review, these regions are defined
by a single particle energy $\ecut$, such that $\rC$ is spanned by
the single particle modes with energy $\epsilon\le\ecut$ and $\rI$ is
spanned by the single particle modes with energy
$\ecut<\epsilon<E_{\max}$.
We define projectors for these regions as
\begin{eqnarray} 
\PC\{ F\xa\}&\equiv&\sum_{n\in\rC}\phi_{n}\xa\int
d^{3}\xd\phi_{n}^{*}\xad F\xad,\label{eq:projectorC}\\
\PI\{ F\xa\}&\equiv&\sum_{n\in\rI}\phi_{n}\xa\int
d^{3}\xd\phi_{n}^{*}\xad F\xad,\label{eq:projectorI} 
\end{eqnarray}
where $\rC=\{n:\epsilon_n\le \epsilon_{\rm{cut}}\}$ and
$\rI=\{n:\epsilon_{\rm{cut}}<\epsilon_n\le E_{\max}\}$,
such that $\rL=\rC+\rI$ and $\PI \PC\equiv 0$.


We define quantum field operators for the \CF\ and 
incoherent regions as
\begin{eqnarray}
\bfC\xa&\equiv&\PC\{\bef\xa\}=\sum_{n\in\rC}\hat{a}_n\phi_n\xa,\label{psiCx}
\\
\bfI\xa&\equiv&\PI\{\bef\xa\}=\sum_{n\in\rI}\hat{a}_n\phi_n\xa,
\\ \label{sums}
\hat\psi\xa&=&\bfC\xa+\bfI\xa.
\end{eqnarray}
Most of the theoretical development in this review will be made in terms of the $\bfC\xa$ and $\bfI\xa$ operators.

\paragraph{Properties of projectors}
It is important to note that the kernel of the $\PC$-projector, namely
\begin{equation}\label{deltaC}
\delta_\rC(\x,\x^\prime)\equiv\sum_{n\in\rC}\phi_{n}\xa\phi^*_{n}\xad,
\end{equation}
is the commutator of the \CF\ operator, i.e. 
\begin{equation}
[\bfC(\x),\bfC^\dag(\xd)]=\delta_\rC(\x,\xd).\label{deltarC}
\end{equation}
The  function $\delta_{\rC}$  plays the role of a  Dirac-delta function for any function in $\rC$, e.g.,
\begin{equation}
\myint{\x'}\delta_\rC(\x,\xd)\psi_\rC(\xd)=\psi_\rC(\x),
\end{equation} 
which also follows from the idempotence property, $\PC\PC\equiv \PC$. The imposition of an energy cutoff thus has major consequences for the field theory. Imposing a cutoff in momentum leads to a discretized form of the continuous field theory with $\delta_\rC(\x_j,\x_k)=\delta_{jk}/\Delta V$ for a lattice representation with volume $\Delta V$ per lattice point. Imposing it in another basis leads to a field theory with a position dependent commutator \eref{deltaC}. 

\subsubsection{The Hamiltonian and equation of motion}\label{sec:HEOM}

The effective Hamitonian can now be rewritten in terms of $\bfC$ and
$\bfI$, and because the projection is in terms of eigenfunctions of
the single-particle Hamiltonian, cross terms in the result will appear
only in the quartic interaction term.   The different ways of approaching the
implementation of the \CF\ theory depend on the way in
which these cross-terms, which connect \rC\ and \rI\, are dealt with.
The three cases are:
\begin{enumerate}
    \item \textbf{Projected Gross-Pitaevskii equation:}  The cross terms are dropped, but
    it is assumed that all mode occupations in \rC\ are significantly
    greater than one throughout the dynamical evolution. Thus, there is significant occupation of \rI, but
    this is taken as fully thermalized, and conditions are chosen  so 
    that when the motion in \rC\ reaches equilibrium it matches
    smoothly to \rI.
    \item \textbf{Truncated Wigner PGPE:} Here one deals with
    processes in which many modes in \rC\ are unoccupied, including
    all the higher modes.  In this case \rI\ is unoccupied, and the
    effect of the cross terms is negligible, and they are dropped.
    However, the quantum fluctuations in  \rC\ have a significant
    effect, and this is taken into account by including a random element
    corresponding to half a quantum occupation in each mode. 
    
    
    \item \textbf{Stochastic PGPE :} Here one accounts for interactions
    between \rC\ and \rI\ by assuming \rI\ is thermally occupied, and
    by using quantum stochastic techniques, terms which involve
    dynamic noise and damping are introduced.  When the truncated Wigner PGPE is also
    used, the resulting equation of motion is a modified PGPE with
    noise and damping.
\end{enumerate}

These methods differ in both their conditions for validity and in the details of their numerical implementation. We stress that in general a system may evolve from a regime where (ii) is the best description (zero temperature BEC), through to a regime where (i) is applicable (a thermalised classical field), and finally through sufficient heating, to the realm where (iii) may be appropriate. The parameters determining this crossover, and therefore the appropriate definition of \rC\ and \rI, are the mode occupancies, strength of interactions, and temperature of the system which are in general time dependent. Thus, some care must be taken when applying the methods and we shall return to this point when discussing validity criteria in \sref{sec:validity}. 

\subsection{Wigner formalism and the truncated Wigner approximation}\label{WignerSec}
The justification for all three \CF\ techniques can be made using a Wigner distribution 
methodology.
Discussion of the properties of the Wigner
distribution can be found in many places (see for
example~\cite{Gardiner1998a,Walls1994a}). Here we provide
an introduction to the theory using the single mode case in sections \ref{sec:Wigneruse} and \ref{sec:OPcorresEOMs}, before discussing the quantum field case in section \ref{WigQFT}.

\subsubsection{Wigner representation of a single quantum mode}\label{sec:Wigneruse}
The Fourier transform of a classical probability distribution is known as its characteristic function, and moments of the distribution are proportional to derivatives of the characteristic function. This same procedure can be generalized to a system described by a quantum density operator: For a single bosonic mode with density operator
 $\hat{\rho}$, the \emph{symmetrically ordered}\footnote{There are several different operator orderings that are commonly used to define the quantum characteristic function, with the symmetric case being the standard choice for defining the Wigner function.}  characteristic function is
defined as
\begin{equation}
\chi_W(\lambda,\lambda^*)={\rm tr}\left\{\hat{\rho}e^{\lambda
\hat{a}^\dag-\lambda^*\hat{a}}\right\},\label{chiW1mode}
\end{equation}
where $\lambda$ is a complex variable. The {symmetrically ordered} moments are given by the derivative of $\chi_W$ at $\lambda=0$, i.e.
\begin{equation}
\left\langle\left\{\hat{a}^s\left(\hat{a}^\dagger\right)^r\right\}_{\rm{sym}}\right\rangle= \left(\frac{\partial}{\partial\lambda}\right)^r\left(-\frac{\partial}{\partial\lambda^*}\right)^s\chi_W(\lambda,\lambda^*)\big|_{\lambda=0},\label{Xwderivs}
\end{equation}
where $\left\{\right\}_{\rm{sym}}$ means a symmetrical product of the operators, which is a average of all ways of ordering the operators, e.g.
\begin{eqnarray}
\left\{\hat{a}\hat{a}^\dagger\right\}_{\rm{sym}}&=& \frac{1}{2}\left\{\hat{a}\hat{a}^\dagger+\hat{a}^\dagger\hat{a}\right\}, \\
\left\{\hat{a}^2\left(\hat{a}^\dagger\right)^2\right\}_{\rm{sym}}&=& 
\frac{1}{6}\left\{\left(\hat{a}^\dagger\right)^2\hat{a}^2 + \hat{a}^\dagger\hat{a}\hat{a}^\dagger\hat{a} +\hat{a}^\dagger\hat{a}^2\hat{a}^\dagger+\hat{a}\left(\hat{a}^\dagger\right)^2\hat{a}+\hat{a}\hat{a}^\dagger\hat{a}\hat{a}^\dagger\right.\\
&&\left.+\hat{a}^2\left(\hat{a}^\dagger\right)^2
 \right\}.\nonumber \end{eqnarray}

The Wigner function was introduced by Wigner in 1932 \cite{Wigner1932a} and is defined 
as a Fourier transform of the symmetrically ordered quantum characteristic function
\begin{equation}
W(\alpha,\alpha^*)=\frac{1}{\pi^2}\int d^2\lambda\;
e^{\lambda^*\alpha-\lambda\alpha^*}\chi_W(\lambda,\lambda^*).\label{WigDef}
\end{equation}
The Wigner function exists for any density matrix (see \cite{Gardiner1998a} for a proof), and in \sref{subs:initStates} we give the Wigner functions for several standard quantum states.

Integrating \eeref{WigDef} by parts we see that
\begin{equation}
\int d^2\alpha\;
\alpha^s(\alpha^*)^rW(\alpha,\alpha^*) =\left(\frac{\partial}{\partial\lambda}\right)^r\left(-\frac{\partial}{\partial\lambda^*}\right)^s\chi_W(\lambda,\lambda^*)\big|_{\lambda=0}.
\end{equation}
Thus (from \eeref{Xwderivs}) the moments of the Wigner function give symmetrically ordered operator
averages
\begin{equation}
\langle \{\hat{a}^s(\hat{a}^\dag)^r\}_{\rm
sym}\rangle=\overline{\alpha^s(\alpha^*)^r}\equiv\int d^2\alpha\;
\alpha^s(\alpha^*)^rW(\alpha,\alpha^*),\label{Wigmoms}
\end{equation}
where we have introduced the notation $\overline{F(\alpha,\alpha^*)}$ for averaging a function of phase space variables $F(\alpha,\alpha^*)$ over the Wigner distribution. 
This suggests that the  Wigner function acts like a probability distribution, indeed $W(\alpha,\alpha^*)$ is commonly refereed to as a \emph{quasi-probability}  since it need not be positive. 
However, for many important classes of quantum states the Wigner function is either positive (or is well-approximated by a positive function) and can be interpreted as a probability distribution. In these cases the average $\overline{F(\alpha,\alpha^*)}$ is equivalently calculated by statistically sampling $\alpha$ as a random variable from this distribution  and calculating the average of $F(\alpha,\alpha^*)$ over many such samples.

Correlation functions of experimental interest are often normally
ordered, requiring some tedious reordering in order to calculate them
from symmetrically ordered Wigner averages. However, for a normally ordered operator in the
form
\begin{eqnarray}
O(\hat{a},\hat{a}^\dag)\equiv\sum_{n,m}c_{nm}\hat{a}^{\dag\;n}\hat{a}^m,
\end{eqnarray}
it can be shown that the kernel, $O_W(\alpha,\alpha^*)$, for the equivalent stochastic average over the
Wigner function, $\overline{O_W(\alpha,\alpha^*)}=\langle
O(\hat{a},\hat{a}^\dag)\rangle$ is given by~\cite{Scully1997a}
\begin{eqnarray}
O_W(\alpha,\alpha^*)=\sum_{n,m}
c_{nm}(-1)^m\left(\frac{\partial}{\partial
z}+\frac{z^*}{2}\right)^n\left(\frac{\partial}{\partial
z^*}+\frac{z}{2}\right)^m\;e^{z\alpha^*-z^*\alpha}\Big|_{z=z^*=0},
\end{eqnarray}
giving, for example 
\begin{eqnarray}
\langle\hat{a}^{\dag}\hat{a}\rangle&=&\overline{|\alpha|^2}-\frac{1}{2},\\
\langle
\hat{a}^{\dag\;2}\hat{a}^2\rangle&=&\overline{|\alpha|^4}-2\overline{|
\alpha|^2}+\frac{1}{2}.\end{eqnarray} 
An explicit form for $O_W(\alpha,\alpha^*)$ can also be found by evaluating the expression
\begin{eqnarray}
O_W(\alpha,\alpha^*)=\frac{1}{2\pi}\int d^2\eta\; e^{-|\eta |^2/2}O(\alpha-\eta/2,\alpha^*+\eta^*/2),
\end{eqnarray}
which can be useful in certain circumstances, e.g. for reordering exponential operators~\cite{Polkovnikov2004b}. For example, choosing $O(\hat{a},\hat{a}^\dag)=\hat{a}^\dag\hat{a}$, we obtain 
\begin{equation}
O_W(\alpha,\alpha^*)=\frac{1}{2\pi}\int d^2\eta\; e^{-|\eta |^2/2}(\alpha^*+\eta^*/2)(\alpha-\eta/2)=|\alpha|^2-\frac{1}{2}.
\end{equation}

\subsubsection{Operator correspondences and equations of motion}\label{sec:OPcorresEOMs}
The equation of motion for the density operator under Hamiltonian evolution is von Neumann's equation
\begin{equation}
\label{vonNeumann1mode}
i\hbar\frac{\partial\hat\rho}{\partial t} = \left[\hat{H},\hat{\rho}\right],
\end{equation}
where $\hat{H}$ is the Hamiltonian. For typical Hamiltonians the right hand side of \eeref{vonNeumann1mode} will involve products of operators and the density operator, and here we discuss how this equivalently maps onto a differential operator acting on the Wigner function. Consider for example the operator product $\hat{a}\hat{\rho}$. Using \begin{equation}
e^{\lambda\hat{a}^\dagger-\lambda^*\hat{a}}=e^{\lambda\hat{a}^\dagger}e^{-\lambda^*\hat{a}}e^{-|\lambda|^2/2},
\end{equation}
(e.g. see  Baker-Hausdorff formula discussion in \cite{Gardiner1998a}) 
and the invariance of the trace under cyclic permutation we have (c.f. \eeref{chiW1mode})
 \begin{eqnarray}
 {\rm tr} \left\{\hat{a}\hat{\rho}e^{\lambda\hat{a}^\dagger-\lambda^*\hat{a}}\right\} &=&
  {\rm tr} \left\{\hat{\rho}e^{\lambda\hat{a}^\dagger}e^{-\lambda^*\hat{a}}e^{-|\lambda|^2/2}\hat{a}\right\} \\
  &=&\left(\frac{1}{2}\lambda-\frac{\partial}{\partial \lambda^*} \right)\chi_W(\lambda,\lambda^*).\label{arho}
 \end{eqnarray}
 Fourier transforming \eeref{arho} and integrating by parts we obtain the correspondence
\begin{equation}
\hat{a}\hat{\rho}\leftrightarrow \left(\alpha+\frac{1}{2}\frac{\partial}{\partial\alpha^*}\right)W(\alpha,\alpha^*).\label{opcor1}
\end{equation}
Similarly, the other correspondences are
\begin{eqnarray} 
\hat{a}^\dagger\hat{\rho}&\leftrightarrow& \left(\alpha^*-\frac{1}{2}\frac{\partial}{\partial\alpha}\right)W(\alpha,\alpha^*),\\
\hat{\rho}\hat{a}&\leftrightarrow& \left(\alpha-\frac{1}{2}\frac{\partial}{\partial\alpha^*}\right)W(\alpha,\alpha^*),\\
\hat{\rho}\hat{a}^\dagger&\leftrightarrow& \left(\alpha^*+\frac{1}{2}\frac{\partial}{\partial\alpha}\right)W(\alpha,\alpha^*).\label{opcor4}
\end{eqnarray}
We now have a set of mappings from operator equations involving the density matrix to a partial differential equation for the Wigner function.
\paragraph{Application to the damped and driven harmonic oscillator}
As a demonstration of the Wigner formalism we consider the driven harmonic oscillator with Hamiltonian
\begin{equation}
\hat{H} = \hbar\omega\hat{a}^\dagger\hat{a}+\hbar (g\hat{a}^\dagger+g^*\hat{a}),
\end{equation}
where $\omega$ is the oscillator frequency, and $g$ is the driving strength. Such an Hamiltonian arises when considering a single mode of an optical resonator driven by a coherent laser field. Damping to a vacuum field outside the cavity adds additional non-Hamiltonian terms, leading to the master equation
\begin{equation}
\frac{\partial \hat{\rho}}{\partial t}=-\frac{i}{\hbar}[\hat{H},\hat{\rho}]+\frac{\gamma}{2}\left(2\hat{a}\hat{\rho}\hat{a}^\dag-\hat{a}^\dag\hat{a}\hat{\rho} -\hat{\rho}\hat{a}^\dag \hat{a}\right).
\end{equation}
Using the operator correspondences \eref{opcor1}-\eref{opcor4} we obtain the equivalent equation of motion for the Wigner function
\begin{eqnarray}
\frac{\partial W}{\partial t}&=&\left[\frac{\partial}{\partial \alpha}\left(i\omega\alpha+\frac{\gamma}{2}\alpha+ig\right)+\frac{\partial}{\partial \alpha^*}\left(-i\omega\alpha^*+\frac{\gamma}{2}\alpha^*-ig^* \right) \right]W(\alpha,\alpha^*,t)\nonumber\\
&+&\frac{\gamma}{2}\frac{\partial^2}{\partial \alpha\partial\alpha^*}W(\alpha,\alpha^*,t)\label{WigEOM}
\end{eqnarray}
This evolution equation is of the form of a Fokker-Planck equation (FPE) with a drift (first derivative) term and a diffusion (second derivative) term. Here an important tool emerges which is central to the techniques in this review: if the initial Wigner distribution $W(\alpha,\alpha^*,0)$ is positive and our interest is in moments of the Wigner distribution at some later time (e.g. \eeref{Wigmoms}) then we can avoid solving the partial differential equation \eref{WigEOM} and instead simulate a large number of \emph{trajectories} governed by the (It\^{o}) stochastic differential equation (SDE)
\begin{equation}
d\alpha=\left(-i\omega-\gamma/2\right)\alpha dt-i g dt+\sqrt{\gamma/2}dw(t),\label{dhosde}
\end{equation}
with the  initial conditions $\alpha(0)$ sampled from $W(\alpha,\alpha^*,0)$. Here the dissipative process has generated a diffusive term in the SDE which is a complex Gaussian noise satisfying $\overline{dw(t)}=0$ and $\overline{dw^*(t)dw(t)}=dt$.
The justification for this procedure is the standard mapping of a Fokker-Planck equation onto a stochastic differential equation \cite{Gardiner1985a}. In the case where dissipation is absent, the SDE simplifies to an ordinary differential equation. This is the situation for much of the work described in this review, but the reasons for the simplification to an ordinary differential equation of motion are more subtle and are discussed in detail in \sref{truncatedWig}. The SDE \eref{dhosde} also contains the essential technical elements of the phase space mapping required for deriving the SPGPE described in \sref{sec:sgpe}. A more complete discussion of the correspondence between SDEs and FPEs is presented in \aref{sec:stochMapping}. Moving to spatially continuous, modally finite dimensional field theories adds some further complexities which are addressed in the following sections.

\subsubsection{Adaption to quantum field theory in the \CF\ region}\label{WigQFT}
The extension from the single mode case to quantum field theory is accomplished using the projected
functional generalization of the single mode formalism. Because there are only 
a finite number of modes,   the theory can be generalized  in a form which
involves minimal additional calculus. For a system with $M$ modes in the \CF\ region, we define the vector of mode amplitudes $\bm{\alpha}=[\alpha_0,\alpha_1, \dots,\alpha_{M-1}]^T$ and the notation
\begin{equation}
\int d^2\bm{\alpha}\equiv\prod_{n\in \rC}
\int d^2\alpha_n.
\end{equation}
The multimode Wigner function is then given by
\begin{eqnarray}
W_\rC(\bm{\alpha},\bm{\alpha}^*)&=&\int\frac{d^2\bm{\lambda}}{\pi^{2M}}
\exp{\left(\bm{\lambda}^\dag
\bm{\alpha}-\bm{\alpha}^\dag\bm{\lambda}\right)}\chi_{\scriptscriptstyle
W}(\bm{\lambda},\bm{\lambda}^*),
\end{eqnarray}
where $\bm{\alpha}^\dag=(\bm{\alpha}^*)^T$, and $\chi_{\scriptscriptstyle
W}$ is the characteristic function for the \CF\ region density operator, $\hat{\rho}_{\rC}$.
Moments of the Wigner distribution give symmetrically ordered
operator averages, for example
\begin{eqnarray}\label{wigav1}
\int d^2\bm{\alpha}\;
|\alpha_q|^2W_\rC(\bm{\alpha},\bm{\alpha}^*)=\left\langle\frac{\hat{a}_q^
\dag\hat{a}_q+\hat{a}_q\hat{a}_q^\dag}{2}\right\rangle.
\end{eqnarray}
Introducing the c-number \CF\ (c.f. \eeref{psiCx})
\begin{eqnarray}\label{wigav101}
\psi_\rC(\x) &=&\sum_{n\in \rC}\alpha_n\phi_n(\x),
\end{eqnarray}
the field density average corresponding to \eeref{wigav1} is 
\begin{eqnarray} 
\int d^2\bm{\alpha}\;|\psi_\rC(\x)|^2W_\rC(\bm{\alpha},\bm{\alpha}^*)&=&
\left
\langle\frac{\hat{\psi}_\rC^\dag(\mathbf{x})\hat{\psi}_\rC(\mathbf{x})+\hat{
\psi}_\rC(\mathbf{x})\hat{\psi}_\rC^\dag(\mathbf{x})
} {2}\right\rangle,\label{Wigden1}\\
&=&\left\langle\hat{\psi}_\rC^\dag(\mathbf{x})\hat{\psi}_\rC(\mathbf{x})\right
\rangle+\frac{\delta_\rC(\mathbf{x},\mathbf{x})}{2}.\label{Wigden2}
\end{eqnarray}
The contribution from the projector  in \eeref{Wigden2} represents a
central result of the Wigner representation of quantum field theory. Physics beyond mean-field theory arises in the truncated Wigner method because of vacuum noise evident in the commutator term
$\delta_\rC(\mathbf{x},\mathbf{x})$  \eref{deltarC}, which accounts for half a quantum per mode of noise present in
the theory. This noise mimics the role of vacuum fluctuations, but would render
 the theory ultraviolet divergent if all physically allowed modes were 
included.  However, the projection into the \CF\ region  
involves a finite number of basis functions, so that in this situation the term 
is a well defined finite contribution to the stochastic average. 
 
\subsubsection{Functional derivative notation}\label{sssfuncderiv}
There is a useful connection between projectors and functional
calculus that greatly simplifies multimode calculations while still
including all the necessary projectors into low energy modes. We
 define 
the projected derivative operators as
\begin{eqnarray}\label{alphadef}
\DDP{\psi_{\rC}(\mathbf{x})}&\equiv&\sum_{n\in\rC}
\phi_n^*(\mathbf{x})\frac{\partial}{\partial \alpha_n},\\
\DDP{\psi_{\rC}^*(\mathbf{x})}&\equiv&\sum_{n\in\rC}
\phi_n(\mathbf{x})\frac{\partial}{\partial \alpha_n^*}.
\end{eqnarray}

\subsubsection{Operator correspondences}\label{sssopcorres}
Using the projected functional derivatives, one then finds functional operator
correspondences between the density operator, $\hat{\rho}_{\rC}$, and the Wigner function~\cite{Gardiner1998a}
\begin{eqnarray}\label{opcorr1}
\hat{\psi}_\rC(\mathbf{x})\hat{\rho}_\rC&\longleftrightarrow&\left(\psi_\rC(
\mathbf{x})+\frac{1}{2}\frac{\bar{\delta}}{\bar{\delta}\psi_\rC^*(\mathbf{x})}
\right)W_\rC,\\
\label{opcorr2}\hat{\psi}_\rC^\dag(\mathbf{x})\hat{\rho}_\rC&\longleftrightarrow&
\left(\psi_\rC^*(\mathbf{x})-\frac{1}{2}\frac{\bar{\delta}}{\bar{\delta}\psi_
\rC(\mathbf{x})}\right)W_\rC,\\
\label{opcorr3}\hat{\rho}_\rC
\hat{\psi}_\rC(\mathbf{x})&\longleftrightarrow&\left(\psi_\rC(\mathbf{x})-
\frac{1}{2}\frac{\bar{\delta}}{\bar{\delta}\psi_\rC^*(\mathbf{x})}\right)W_\rC,\\
\label{opcorr4}\hat{\rho}_\rC
\hat{\psi}_\rC^\dag(\mathbf{x})&\longleftrightarrow&\left(\psi_\rC^*(
\mathbf{x})+\frac{1}{2}\frac{\bar{\delta}}{\bar{\delta}\psi_\rC(\mathbf{x})}
\right)W_\rC,
\end{eqnarray}
which are used to map the equation of motion for the density operator
to an equation of motion for $W_{\rC}$. 
\subsubsection{Truncated Wigner approximation}\label{truncatedWig}
From \eeref{Eq:Heff} we see that the time development of $\hat{\psi}_\rC$ in isolation\footnote{We consider the description of coupled $\rC$ and $\rI$ regions in \sref{sec:sgpe}.} is governed by the Hamiltonian
\begin{equation}
\hat{H}_{\rC} = \int d^3\x\;{\hat\psi^\dag_\rC(\bf x)}
H_{\rm{sp}}{\hat\psi_\rC(\bf x)}
+\frac{u}{2}\int d^3\x\;{\hat\psi^\dag_\rC(\bf x)}{\hat\psi^\dag_\rC(\bf x)}
{\hat\psi_\rC(\bf x)}{\hat\psi_\rC(\bf x)}.
\label{Eq:HRC}
\end{equation} 
The equation of motion for the density operator in \rC\ is then
von Neumann's equation
\begin{eqnarray}\label{vonN}
i\hbar{\partial \hat \rho_\rC(t)\over \partial t}
=[\hat{H}_\rC,\hat \rho_\rC(t)].
\end{eqnarray}
Using   the operator correspondences (\ref{opcorr1}) --
(\ref{opcorr4}), the Hamiltonian (\ref{Eq:HRC}) generates the time
evolution equation
\begin{equation}\label{HCthirdorder}
\begin{split}
\frac{\partial W_\rC}{\partial t}\Bigg{|}_{\hat{H}_\rC}=&\intV{\mathbf{x}}\Bigg\{\frac{iu}{4\hbar}\DPDP{\psi_\rC(\mathbf{x})}{\psi_\rC^*(\mathbf{x})}
\psi_\rC^*(\mathbf{x})\DDP{\psi_\rC^*(\mathbf{x})}+{\rm h.c.}\\
&\frac{i}{\hbar}\DDP{\psi_\rC(\mathbf{x})}\Big(H_{\rm{sp}}+u[|\psi_\rC(\mathbf{x})|^2-\delta_\rC(\mathbf{x},
\mathbf{x})]
\Big)\psi_\rC(\mathbf{x})+{\rm h.c.}\Bigg\}W_\rC,
\end{split} 
\end{equation}
where ${\rm h.c.}$ represents the Hermitian conjugate. 
Equation \eref{HCthirdorder} as it stands is very difficult to solve. However, if we are able to neglect the first line of right hand side terms, i.e., those containing third order derivatives, then progress can be made.
This approximation, which is referred to as the \emph{truncated Wigner
approximation} (TWA), is valid over a wide regime for the quantum degenerate
gas. The resulting description is also obtained formally in the classical limit
which we describe below. (We discuss the basic validity conditions for this approximation further in \sref{sec:validity}).
As discussed in \aref{sec:stochMapping}, a mapping to ordinary stochastic differential equations is not possible for \eeref{HCthirdorder}.
However, 
making the TWA, the Wigner function evolution takes the form of a Fokker-Planck equation with drift but no diffusion terms, i.e., 
\begin{equation} 
\frac{\partial W_\rC}{\partial t}\Bigg{|}_{\hat{H}_\rC}\approx\intV{\mathbf{x}}\Bigg\{\frac{i}{\hbar}\DDP{\psi_\rC(\mathbf{x})}\Big(H_{\rm{sp}}+u[|\psi_\rC(\mathbf{x})|^2-\delta_\rC(\mathbf{x},
\mathbf{x})]
\Big)\psi_\rC(\mathbf{x})+{\rm h.c.}\Bigg\}W_\rC.
 \label{TWEOM}
\end{equation}
The Fokker-Planck  evolution can be equivalently mapped to a stochastic partial differential equation \cite{Gardiner1985a} that describes the trajectory of a single realisation of the field $\psi_\rC(\mathbf{x})$, which we refer to as the \emph{truncated Wigner projected Gross-Pitaevskii
equation} (TWPGPE)
\begin{equation}
i\hbar\frac{\partial \psi_{\rC}(\mathbf{x})}{\partial
t}=\PC\left\{ \Big(H_{\rm{sp}}+u[|\psi_\rC(
\mathbf{x})|^2-
\delta_\rC(\mathbf{x},\mathbf{x})]\Big)\psi_\rC(\mathbf{x})\right\}.\label{TWAPGPE}
\end{equation}
The lack of a diffusion term in \eref{TWEOM} means that no explicit noise term appears in the TWPGPE, however as we shall discuss further in \sref{subs:initStates} the initial conditions are stochastic and need to be appropriately sampled from the initial Wigner function. We remark that the equation of motion \eref{TWEOM} is also known as a Liouville equation. A formal property of the Liouville equation is that the distribution function is constant along any trajectory in phase space. This can be seen by applying the method of characteristics to \eref{TWEOM}, which shows that, within the TWA, the Wigner function is constant along classical trajectories given by \eref{TWAPGPE}.

\paragraph{Classical limit}
While we consider the validity conditions for the truncation in \sref{sec:validity}, here we show that the truncation is exact in the \emph{classical limit}, which we define as 
\begin{equation}
N_\rC\to \infty,\quad u\to 0,\quad uN_\rC=\rm{constant},\label{classicallimit}
\end{equation}
where $N_\rC$ is the number of \CF\ region particles
\begin{equation}\label{PN}
N_\rC=\intV{\x}|\psi_\rC(\mathbf{x})|^2, \qquad\mbox{(classical limit)}.
\end{equation}
This expression for $N_\rC$ is only valid in the classical limit, and the general case, obtained from \eeref{Wigden2}, is $N_\rC=\intV{\x}[\overline{|\psi_\rC(\mathbf{x})|^2}-\delta_\rC(\x,\x)/2]$. The integral   $\intV\x\delta_\rC(\x,\x)/2 = M/2$, representing the half quantum per mode vacuum noise included in the Wigner description.

Renormalising the \CF\ according to $\phi_\rC=\sqrt{N_\rC}\psi_\rC$, \eeref{HCthirdorder} becomes
\begin{equation}\label{HCthirdorderRenorm}
\begin{split}
\frac{\partial W_\rC}{\partial t}\Bigg{|}_{\hat{H}_\rC}=&\intV{\mathbf{x}}\Bigg\{\frac{iu}{4\hbar N_\rC}\DPDP{\phi_\rC(\mathbf{x})}{\phi_\rC^*(\mathbf{x})}
\phi_\rC^*(\mathbf{x})\DDP{\phi_\rC^*(\mathbf{x})}+{\rm h.c.}\\
&\frac{i}{\hbar}\DDP{\phi_\rC(\mathbf{x})}\Big(H_{\rm{sp}}+uN_\rC|\phi_\rC(\mathbf{x})|^2
\Big)\phi_\rC(\mathbf{x})+{\rm h.c.}\Bigg\}W_\rC,
\end{split}
\end{equation}
In the classical limit we have $u/N_\rC\to0$, so that the third order derivatives vanish in \eeref{HCthirdorderRenorm}, and the TWPGPE  \eref{TWAPGPE} is the asymptotically exact equation of motion for the system.
However, stochastic initial conditions are still present, reflecting quantum or thermal fluctuations, or uncertainties in the initial data of the problem. 

A purely \emph{deterministic} classical description is recovered when the initial state approaches a delta function in phase space which is precisely the limit obtained for a zero temperature BEC in a coherent state. For $N_\rC$ atoms in a single mode coherent state with mode function $\xi_0(\x)$, the renormalised field can be written as $\phi_\rC(\x)=\alpha\xi_0(\x)$, with phase space distribution
\begin{equation}W(\alpha,\alpha^*)=\frac{2N_\rC}{\pi}\exp{\left(-2N_\rC\left|\alpha-\frac{\alpha_0}{\sqrt{N_\rC}}\right|^2\right)}, 
\end{equation}
where $|\alpha_0|^2=N_\rC$. In the classical limit $W(\alpha,\alpha^*)\to \delta^{(2)}(\alpha-1)$, giving the TWPGPE for the system dynamics with non-stochastic initial conditions. In general thermal fluctuations will be present even in the classical limit, and the problem remains a stochastic one, subject to deterministic evolution. We note that in the classical limit vacuum noise is unimportant, but it can play an important role in BEC physics where $N_\rC\sim 10^3 - 10^9$. Indeed, the effect of zero point fluctuations can be rather striking and even the dominant effect in certain circumstances. An important example is given by the dynamics of condensate collisions, described in \sref{sec:BECcollisions}.
\paragraph{Classical mechanics treatment}\label{sec:Hcfield}

For future reference, we note that replacing the field operator by the classical field $\psi_\rC$ in Hamiltonian \eref{Eq:HRC} yields the   Hamiltonian 
\begin{eqnarray}\label{PH}
H_{\rC}&=&\intV{\mathbf{x}}
\psi_{\rC}^*(\mathbf{x})H_{\rm{sp}}\psi_{\rC}(\mathbf{x})+\frac{u}{2}\intV{\mathbf{x}}
|\psi_{\rC}^*(\mathbf{x})|^4,
\end{eqnarray}
which we shall also refer to as the energy functional for the field $\psi_\rC$.
The classical equation of motion can then be found by defining the \emph{Poisson bracket} $\{F,G\}$ for any two functionals $F$ and $G$ of the classical field $\psi_\rC(\x)$ as
\begin{eqnarray}
\left\{F,G\right\}=\intV{\x}\DP{F}{\psi_{\rC}(\mathbf{x})}\DP{G}{\psi_{\rC}^*(\mathbf{x})}-\DP{F}{\psi_{\rC}^*(\mathbf{x})}\DP{G}{\psi_{\rC}(\mathbf{x})}.
\end{eqnarray}
The equation of motion is then found as
\begin{eqnarray}\label{PHeom}
i\hbar\frac{\partial \psi_{\rC}(\mathbf{x})}{\partial
t}=\left\{\psi_{\rC}(\mathbf{x}),H_\rC\right\}=\DP{H_{\rC}}{\psi_{\rC}^*(\mathbf{x})},
\end{eqnarray}
which yields the projected Gross-Pitaevskii equation
\begin{equation}\label{Pgpe}
i\hbar\frac{\partial \psi_{\rC}(\mathbf{x})}{\partial
t}=\PC\left\{\left(H_{\rm{sp}}+u|\psi_{\rC} (\mathbf{x})|^2\right)\psi_{\rC}(\mathbf{x})\right\},
\end{equation} 
as the classical equation of motion of the system. 
This equation is equivalent to the TWPGPE \eref{TWAPGPE} in the classical limit where the $\delta_\rC$ term can be neglected. The theory can be cast in standard Hamiltonian form by introducing the canonical position and momentum variables 
\begin{eqnarray}
Q_n&=&1/\sqrt{2\epsilon_n}(\alpha_n+\alpha_n^*), \label{Qdef}\\
P_n&=&i\sqrt{\epsilon_n/2}(\alpha_n^*-\alpha_n), \label{Pdef}
\end{eqnarray}
where the $\alpha_n$ are the basis amplitudes of $\psi_\rC\xa$  (see \eref{wigav101}) and $\epsilon_n$ are the energies of the modes comprising the basis. The Poisson bracket then takes the Hamiltonian form and any function $F(P_n,Q_n,t)$ obeys the equation of motion
\begin{equation}
\frac{dF}{dt}=\frac{1}{\hbar}\sum_n\left(\frac{\partial F}{\partial Q_n}\frac{\partial H_\rC}{\partial P_n}-\frac{\partial F}{\partial P_n}\frac{\partial H_\rC}{\partial Q_n}\right)+\frac{\partial F}{\partial t}.
\end{equation}
We emphasize that the Hamiltonian classical mechanics formulation is not only recovered in the continuum limit, but holds generally as a consequence of including the projection operator formally in the definition of the classical field $\psi_\rC(\x)$. This Hamiltonian property is used in \sref{Sec:PGPEtemperature} to determine microcanonical thermodynamic quantities of the \CF.

\subsubsection{Sampling the Wigner distribution}\label{subs:initStates}
We have shown that by making the truncated Wigner approximation,
simulations of ultra-cold 
Bose gas dynamics under the Hamiltonian $\hat{H}_{\rC}$ \eref{Eq:HRC} are reduced to simulations of the PGPE (or more
accurately the TWPGPE (\ref{TWAPGPE})) for an ensemble of samples of
the initial state of the system. The equation of motion is quite easy to
solve, but sampling the Wigner distribution for a general  many-body system
is difficult. 
However, sometimes this sampling issue can be avoided,  e.g., in the PGPE method  a random initial field can be used and allowed to thermalize by evolution (see \sref{SEC:InitStatePGPE}).

\paragraph{Bogoliubov formalism}
Here our basic aim is to present a procedure for sampling the
Wigner distribution for a cold ($T\ll T_c$) Bose condensed cloud in
thermal equilibrium. In this regime the Bogoliubov method provides an
appropriate many-body description of the system, provided that the number
of non-condensate particles, $N_{\rm ex}\equiv N_{\rC}-N_0$, satisfies
$N_{\rm ex}\ll N_{\rC}$.  We briefly review the Bogoliubov formalism, and refer to references \cite{Gardiner1997a,Hutchinson1997a,Castin1998a,Hutchinson1998a,Hutchinson2000a,Morgan2003a,Proukakis2008a} for a more complete discussion. 
The basic Bogoliubov approach is to expand the
field operator in the form
\begin{equation}
\bfC\xa=\frac{\hat{a}_0}{\sqrt{N_0}}\xi_0\xa+\sum_{j>0}\left[u_j\xa\hat
b_j+v_j^*\xa\hat b_j^\dag\right],\label{psibogform}
\end{equation}
where $\xi_0$  is the condensate mode normalized to $N_0$ atoms,  
$\{u_j\xa,v_j\xa\}$ are the quasi-particle amplitudes, and
$\{\hat b_j$, $\hat{b}^\dag_j\}$ are quasiparticle operators that satisfy the
usual Bose commutation relations.
The standard procedure  is to take the condensate
mode as a solution to the time-independent Gross-Pitaevskii equation
\begin{equation}
\mu\xi_0\xa=H_{\rm
sp}\xi_0\xa+u\left|\xi_0\xa\right|^2\xi_0\xa,\label{EQ:tiGPE}
\end{equation}
where $\mu$ is the chemical potential, and then determine  $\{U_j,V_j\}$ 
and the quasi-particle eigenvalues $\eb_j$ from
the Bogoliubov-de Gennes equations
\begin{eqnarray}
\eb_jU_j\xa&=&\left[H_{\rm
sp}+2u\left|\xi_0\xa\right|^2-\mu\right]U_j\xa+u\xi_0\xa^2V_j\xa,\label{EQ:BdG1}\\
-\eb_jV_j\xa&=&\left[H_{\rm
sp}+2u\left|\xi_0\xa\right|^2-\mu\right]V_j\xa+u\xi_0^*\xa^2U_j\xa\label{EQ:BdG2}.
\end{eqnarray}
The expansion in \eeref{psibogform} diagonalizes the
many-body Hamiltonian \eref{Eq:HRC} to quadratic order in the quasi-particle
operators, which is adequate in the regime of  small condensate depletion,
so that the quasiparticle levels are thermally occupied according to 
\begin{eqnarray}
\langle
\hat{b}^\dagger_i\hat{b}_j\rangle&=&\delta_{ij}\,\frac{1}{e^{\eb_j/k_BT}-1},\\
&=&\delta_{ij}\,\bar{n}_j.
\end{eqnarray}
We note that the Bogoliubov modes, $\{U_j\xa,V_j\xa\}$, are in general not orthogonal to the condensate. Even though the correct eigenfrequencies are obtained, orthogonality is automatic only for the special case of a uniform system. The correct modes for the expansion of the field operator \eref{psibogform} can be recovered from \eref{EQ:BdG1} and \eref{EQ:BdG2} by taking the projection orthogonal to the condensate~\cite{Morgan2000a}. Defining the projector
\begin{equation}
{\cal P}_0\psi(\x)=\psi(\x)-N_0^{-1}\xi_0(\x)\intV{\xd}\xi_0^*(\xd)\psi(\xd)
\end{equation}
the orthogonal modes are given by $\{u_i\xa,v_i\xa\}=\{{\cal P}_0U_i\xa,({\cal P}_0^*V_i^*\xa)^*\}$.

 \paragraph{Wigner sampling of the Bogoliubov state}

By introducing the random variables $\alpha_0$, and
$\bm{\beta}$ (an $M-1$ element vector) in place of the operators $\hat{a}_0$ and $\{\hat{b}_j\}$, respectively, the Wigner distribution for the Bogoliubov state is appropriately sampled  as the stochastic \CF\
\begin{equation}
\cf\xa=\frac{
{\alpha}_0}{\sqrt{N_0}}\xi_0\xa+\sum_{j>0}\left[u_j\xa\beta_j+v_j^*\xa\beta_j^*\right].\label{psibogformsample}
\end{equation}
In the Bogoliubov theory outlined above, the condensate and quasi-particle occupations are uncorrelated, i.e. the Wigner distribution is of the separable form $W_{\rC} =W_{0}(\alpha_0,\alpha_0^*)W_{qp}(\bm{\beta},\bm{\beta}^*)$, and in the following paragraphs we discuss how these can be independently sampled. We note that within a number conserving Bogoliubov approach additional correlations between the condensate and quasi-particles arise \cite{ Sinatra2001a,Sinatra2002a}, providing a better description of the low temperature manybody state of the gas.
\begin{figure}[t]
\begin{center}
\includegraphics[width=12cm]{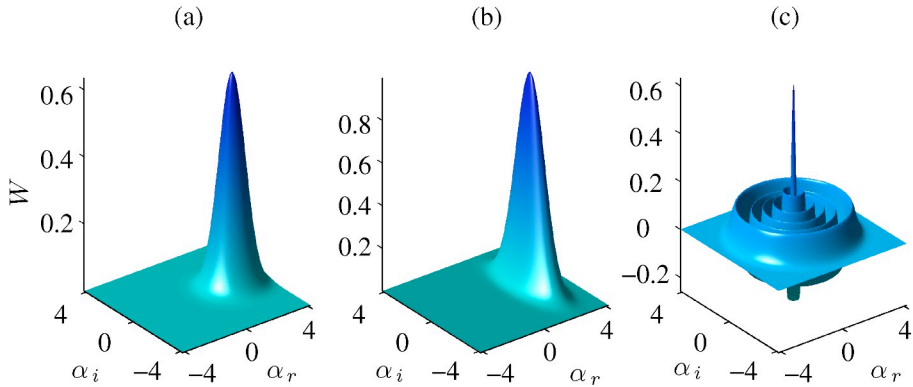}
\end{center}
\caption{ \label{Fig:wignerfun} Some possible Wigner functions for the condensate mode
of a small BEC. (a) Coherent state, (b)
squeezed state, and (c) number state. Wigner distribution for   $\langle N_0\rangle=10$ atoms, and $\alpha=\alpha_r+i\alpha_i$.}
\end{figure}
\paragraph{Condensate mode: coherent state}
To a good approximation the condensate can be regarded as being in a
coherent state, for which the Wigner function is
\EQ{\label{Wcoh}
W_0(\alpha,\alpha^*)=\frac{2}{\pi}\exp{\left(-2|\alpha-\alpha_0|^2\right)},
}
where $N_0=|\alpha_0|^2$.
For large condensate occupation the finite
width of $W_0$ can be neglected and for all samples of the condensate amplitude we can take
$\alpha_0\approx\sqrt{N_0}$.

\paragraph{Quasi-particle modes: thermalized states}
The quasi-particle levels are thermalized modes, with a Wigner distribution of the form of a product of uncorrelated Gaussian quasi-probability distributions, i.e.,
\begin{eqnarray}
W_{qp}(\bm{\beta},\bm{\beta}^*) &=& \prod_{j>0}W_j(\beta_j,\beta_j^*),\\
W_j(\beta_j,\beta_j^*)&=& \frac{2}{\pi}\tanh\left(\frac{\eb_j}{k_BT}\right)\exp\left[-2|\beta_j|^2\tanh\left(\frac{\eb_j}{k_BT}\right)\right].
\end{eqnarray}
This distribution  is sampled by the Gaussian
complex random variables, $\{\beta_j\}$, with properties
\begin{eqnarray}
\overline{\beta_j} &=& \overline{\beta_i\beta_j} = 0,\\
\overline{\beta_i^*\beta_j}&=& \delta_{ij}\,\left(\bar{n}_j+\frac{1}{2}\right).
\end{eqnarray}
In practice these variables can be generated as
\begin{equation}
\beta_j= \sqrt{ \bar{n}_j+{1}/{2}}\,\left(\frac{x_j+iy_j}{\sqrt{2}}\right),
\end{equation}
where $x_j$, $y_j$   are normally distributed Gaussian random variates
with mean zero and unit variance.
Sampling the field in this way, we recover the correct symmetrically ordered moments, e.g
\begin{eqnarray}
\overline{\cf\xa}&=& \langle \hat{\psi}_\rC(\mbf{x})\rangle,\\
&=&\xi_0(\mathbf{x}),\\
\overline{|\cf\xa|^2}&=&\langle
\{\hat{\psi}_\rC^\dag(\mbf{x})\hat{\psi}_\rC(\mbf{x})\}_{\rm
sym}\rangle,\\
&=&|\xi_0(\mathbf{x})|^2+\sum_{j>0}\frac{1}{2}(\langle
\hat{\beta}_j^\dag\hat{\beta}_j\rangle+\langle
\hat{\beta}_j\hat{\beta}^\dag_j\rangle) (|u_j(\mbf{x})|^2
+|v_j(\mbf{x})|^2).
\end{eqnarray}

\paragraph{Vacuum occupation}
We note that even in the zero temperature limit, where $\bar{n}_j\to0$,
 the random variables $\beta_j$  have the finite variance $\overline{|\beta_j|^2}=1/2$,
i.e., each mode of the system has on average half an atom of \emph{vacuum noise},
necessary to ensure the symmetrically ordered interpretation of Wigner moments. Thus an attribute of the Wigner method is that for a simulation with $M$ modes, $M/2$ virtual particles (i.e. vacuum noise) are included in the field  in addition to the $N_{\rC}$ real particles.

\subsubsection{Alternative methods for sampling the Wigner distribution}\label{altsampling}
\paragraph{Efficient sampling of a number-conserving Bogoliubov state}
Sinatra \etal\  have shown how a number conserving version of the
Bogoliubov formalism can be implemented via a Brownian motion simulation.
This approach, developed in references
\cite{Sinatra2000a,Sinatra2001a,Sinatra2002a}, has the advantage that it
does not require diagonalization of the Bogoliubov de-Gennes  equations.

\paragraph{Approximate ground state construction}
For  a nearly pure condensate  the appropriate ground state of the system
is sampled as the $T=0$ limit of expression \eref{psibogformsample}.
However, for many non-equilibrium scenarios, the quasi-particle properties
of the low energy modes are unimportant, and the vacuum noise can be added
in any basis orthonormal to the condensate. That is
\begin{equation}
\cf\xa=\frac{
{\alpha}_0}{\sqrt{N_0}}\xi_0\xa+\sum_{j}\bar\xi_j\xa\alpha_j,\label{psivacformsample}
\end{equation}
where $\{\bar\xi_j\xa\}$ is an orthonormal basis. The condensate amplitude, $\alpha_0$, is sampled as described below \eeref{Wcoh}, and the other mode amplitudes are sampled as Gaussian random variables with $\overline{\alpha_j^*\alpha_j}=\delta_{ij}/2$. 
This type of construction is  useful in collision experiments where the vacuum fluctuation of the high energy modes drive scattering events.

\paragraph{Ideal gas ground state}
In the absence of interactions, the ground state Wigner function can be sampled as
\begin{equation}
\cf\xa=\sum_{j}\phi_j\xa\alpha_j,\label{spwig}
\end{equation}
where $\phi_n\xa$ are the single particle basis states and the $\bm{\alpha}$ are sampled according to $\alpha_0= \sqrt{N_\rC}$ and
$\overline{\alpha_i^*\alpha_j}={\delta_{ij}}/{2}$, for $i,j>0$.

\paragraph{Ideal gas high temperature state}
For temperatures above $T_c$  expansion \eref{spwig} also suffices to describe the  thermalized state of the system but with all ${\alpha}_j$ sampled as Gaussian random variables with 
\begin{equation}
\overline{\alpha_i^*\alpha_j}= \delta_{ij}\,\left(\bar{n}_{\rm BE}(\epsilon_j)+\frac{1}{2}\right),
\end{equation}
where $\bar{n}_{\rm BE}(\epsilon_j)=\{\exp[(\epsilon_j-\mu)/k_BT]-1\}^{-1}$ is the Bose-Einstein distribution.

\paragraph{More general condensate states}
It is possible to consider more general states for the condensate, e.g. the
number state $|N_0\rangle$, which has the
Wigner function
\EQ{
W_{0}(\alpha,\alpha^*)=\frac{2(-1)^{N_0}}{\pi}\exp{\left(-2|\alpha|^2\right)}
\mathrm{L}_{N_0}(4|\alpha|^2),
}
where $\mathrm{L}_n(x)$ is the Laguerre polynomial. The number state
Wigner function for $N_0=10$ is shown in \fref{Fig:wignerfun}c.
It is non-positive-definite, and is highly oscillatory for
large numbers which makes exact stochastic sampling difficult.
However, for large $N_0$ the radial distribution is well approximated by a
delta function~\cite{Gardiner2002a}. A
Gaussian approximation is thus suitable in this regime and a method for
sampling the number state Wigner function has been developed and
shown to reproduce all moments with error $\sim O(1/N_0)$~\cite{Olsen2004}. 
However, more simply we can take
$\alpha_0\approx\sqrt{N_0}e^{i\theta}$, where $\theta$ is a uniformly
distributed random phase, $\theta\in[0,2\pi]$. Sampling $\theta$ this way preserves the $U(1)$ symmetry of the system. 

Other quantum states, such as squeezed states (see \fref{Fig:wignerfun}b), and crescent states can be sampled (e.g. see
reference \cite{Olsen2004c}) to investigate their influence on the many-body dynamics.

\paragraph{Adiabatic mapping}
In reference \cite{Polkovnikov2004a}  Polkovnikov \etal\ sampled the
Wigner function of an ideal Bose gas in an optical lattice at $T=0$. In
the ideal case the bare modes, $\{\phi_j\xa\}$, form the appropriate basis
and the Wigner function can be sampled as discussed below \eeref{spwig}. The interactions are slowly ramped up to the
desired value to generate samples of the interacting system (justified by the quantum adiabatic theorem).

\subsubsection{Validity criteria for the truncated Wigner method}\label{sec:validity}
The only approximation made in deriving the TWPGPE has been the neglect of third order derivatives in the evolution equation for the multimode Wigner function \eref{HCthirdorder}. 
The complete set of validity conditions for the truncation is still the subject of current research. Polkovnikov, who derived the TWA using a path integral method, has obtained expressions for the next order corrections in quantum scattering processes~\cite{Polkovnikov2004b,Polkovnikov2008a} that can be used to assess the validity of any simulation.  This approach represents a fundamental advance in the formulation and application of the truncated Wigner method: it promotes the truncated Wigner method to a \emph{controlled} approximation theory since corrections to the TWA can, at least in principle, be explicitly calculated. In practice evaluating the corrections is a challenging task for large multi-mode problems, and applications have thus far been restricted to discrete lattice systems where the relative strength of interactions to linear system evolution is straightforwardly defined. We also make note of comparisons that have been made between the TWA and exact results \cite{Kinsler1989,Kinsler1991,Plimak2001} to characterize the limitations of the approach for quantum optical systems.
 
Several practical conditions for ensuring the reliability of truncated Wigner simulations in a variety of regimes have emerged in the literature, and we summarize these here. Broadly these conditions fall into two categories: (i) those required to ensure consistency of short-time dynamics (relative to the thermalization timescale),  and (ii) those required for simulations over longer timescales where the system may thermalize.

\paragraph{Short time evolution: Quantum dynamics}
The strict condition of validity of the TWPGPE is that all modes of the  \CF\ region are highly occupied, so that the classical limit discussed in \sref{truncatedWig} is approached.
In general this condition is rather restrictive, especially for systems well-below the critical temperature.

Another criterion has been derived by Norrie \etal\cite{Norrie2006a}  for 
factorizable Gaussian states. 
In particular, those authors considered the class of states with a Wigner function of the form
\begin{equation}
W_{\rC}(\bm{\alpha},\bm{\alpha}^*)=\prod_j \frac{\Gamma_j}{\pi}\exp{\left(-\Gamma_j|\alpha_j-\alpha_{j_0}|^2\right)},
\end{equation} 
where  the random variable $\alpha_j$,  with corresponding orthonormal basis mode $\xi_j(\x)$, has mean value $\alpha_{j0}$, and variance $\Gamma_j^{-1}$.
A sufficient condition for the  validity of the TWPGPE for these states is
\begin{equation}
\left|\langle \hat{\psi}^\dag_\rC(\x)\hat{\psi}_\rC(\x)\rangle-\frac{1}{2}\delta_{\rC}(\x,\x)\right|\gg\sum_j\frac{\Gamma_j}{4}|\xi_j(\x)|^2,\label{NorrieCond}
\end{equation}
i.e., the system must have sufficiently high density in position space. Generally, this condition is much more readily satisfied than the condition of high mode occupancy.

Sinatra \etal \cite{Sinatra2002a} have made detailed comparisons of the truncated Wigner and time-dependent Bogoliubov approaches for a uniform Bose condensate in the regime $T\ll T_c$, where the non-condensate population is much smaller than the condensate. They find that the truncated Wigner predictions become inaccurate if the quantum noise sampled in the initial condition dominates the number of particles in the system, i.e., the condition for validity is
\begin{equation}
\frac{M}{2}\ll N_{\rC}.\label{denrequniform}
\end{equation} 
We note that this condition can be rewritten in terms of the spatial density as 
$n\xa\,\Delta V\gg1$, 
where $n\xa=N_\rC/V$ and $\Delta V=V/M$. 
Since  $\delta_{\rC}(\x,\x)=1/\Delta V$ and $|\xi_j|^2=1/V$ for the uniform gas,
we see that results \eref{denrequniform} and \eref{NorrieCond} are equivalent for this system.

\paragraph{Long time evolution: Thermalization} 
As discussed earlier, when modeling a system of $M$ modes an additional half quantum per mode of noise is added on average to the initial condition. This introduces $M/2$ virtual particles into the TWPGPE simulation which should be subtracted to recover the correct operator averages. However, under evolution these virtual particles may thermalize, and change the equilibrium properties of the system.

 Sinatra \etal \cite{Sinatra2002a} have proposed the condition 
 \begin{equation}
 |T-T_{\rm class}|\ll T, \label{Sinatra_Tcond}
 \end{equation}
 to ensure the long-time validity of a truncated Wigner calculation for a system, where $T$ is the initial temperature of the system, and $T_{\rm class}$ is the temperature once the noise has thermalized.
In practice they find that \eeref{Sinatra_Tcond} is best ensured by limiting the number of modes in the numerical calculation, and that an acceptable description is obtained if the largest single particle (quasi-particle) energy in the calculation is no more than a few $k_BT$.

\paragraph{Interactions and dimensionality}
It is clear that the conditions listed above cannot be considered complete as they do not  explicitly involve interactions or dimensionality.  
 In regimes where interactions dominate the system can evolve into a strongly correlated state. Important examples for ultra-cold atoms are the Mott-insulating and Bose-glass  states that emerge when a system of repulsive bosonic atoms is confined in a deep optical lattice \cite{Jaksch1998a,Buonsante2007a}. Generally speaking the truncated Wigner approximation does not provide a good representation for such strongly correlated systems.

\begin{figure}[t]
\begin{center}
\includegraphics[width=6.75cm]{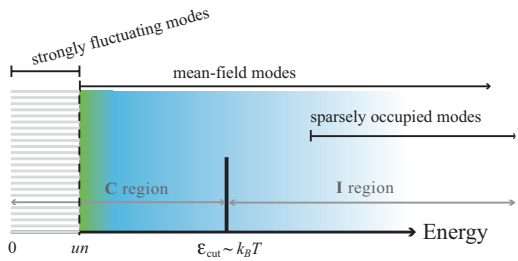}
\end{center}
\caption{\label{Figcut}%
Schematic view of the modes in a critical system showing the appropriate choice of $\ecut$.}
\end{figure}

Another regime in which interactions play an important role is in the critical region where the system undergoes strong fluctuations. These \emph{critical fluctuations} are classical in nature and thus amenable to the truncated Wigner treatment, however care is needed in choosing $\rC$ to describe this regime. Typically strong fluctuations occur in the infra-red modes up to the energy scale  $un$ where $n$ is the density, above which the modes are well-described by mean-field theory (e.g. see the discussion in \cite{Kashurnikov2001a,Prokofev2001a}). To provide an accurate description the \CF\ region must be chosen to include these modes, i.e. we must have $\ecut>un$, as shown schematically in \fref{Figcut}.

The occurrence of critical fluctuations can be identified by the Ginzburg criteria (e.g. see \cite{Fedichev1998b}), which predicts that fluctuations are important for the three-dimensional Bose gas  only in a narrow temperature range about  $T_c$. Outside of this range a pure mean-field description provides a good description of the equilibrium system.  In contrast, for low dimensional systems strong fluctuations prevail over a broad temperature range, and inhibit the formation of phase coherence according to the Mermin-Wagner-Hohenberg theorem \cite{Mermin1966a,Hohenberg1967a}. For these systems a pure mean-field analysis is of limited use, yet a \CF\ description of the low energy modes ($\rC$ region) with a mean-field description of the high energy states ($\rI$ region) can be used to provide a comprehensive description (see \cite{Prokofev2001a,Prokofev2002a,Proukakis2003a,Proukakis2006a} and \sref{sec:q2D}).

Low dimensional systems exhibit several additional properties that make them well-suited to simulation by \CF\  techniques. First, in lower dimensionality the rate of thermalization is significantly reduced with respect to the three dimensional case. This suggests that the thermalization of the vacuum noise that occurs in  truncated Wigner evolution will happen more slowly and simulations in the low temperature regime should be valid over longer time scales. Second, the density of states increases more slowly with energy as the dimensionality of the system decreases, and thus for 1D and 2D systems at finite temperature a larger portion of atoms reside in  low energy, highly occupied modes.

In general the study of 1D and 2D Bose gases is an active area of research, and many aspects of their behaviour, particularly dynamics, are not well-understood. While the \CF\ techniques are widely applicable to describing these systems, a significant challenge remains to develop techniques for sampling the Wigner distribution when the Bogoliubov description fails. A few procedures have been developed for quasi-1D lattice systems at $T\approx0$ in \cite{Polkovnikov2004a}, and for finite temperature quasi-2D trapped gases in \cite{Simula2006a}.

\paragraph{Role of projection in ensuring validity} 
Both the short-time and long-time validity conditions are sensitive to the number of modes in the \CF\ region and the maximum energy of modes represented.  For this reason it is essential to have a projector in the formal theory to exert as much control as possible over the modes retained in the \CF\ description. 

A natural question that arises is how dependent are the results of a simulation on the energy cutoff we use to define the $\rC$ and $\rI$ regions? For the case of a system in the critical regime it is clear that a lower bound for $\ecut$ is provided by the energy scale $un$ as discussed above (see \fref{Figcut}). More generally, this lower bound is also appropriate for inhomogeneous systems outside of the critical regime.  This is because we have chosen to implement the projector in the single particle basis (see \sref{SEC:ProjOps}), which provides a good representation of the low energy modes and a well-defined energy cutoff only when $\ecut>un$.

The criteria for setting an upper bound to $\ecut$ varies greatly according to application. A large value of $\ecut$ can be used for short time simulations, subject to the condition given in \eeref{denrequniform}. For equilibrium situations (or long time simulations where relaxation occurs) typically the condition $\ecut\sim k_BT$ \cite{Sinatra2002a} ensures that all the modes in the \CF\ region are appreciably occupied, and thus accurately described by the Wigner approach (we discuss this case in more detail in \sref{SEC:PGPE}). 

Within the aforementioned guides for choosing the cutoff, there is still an appreciable degree of freedom. For example, consider the situation shown in \fref{Figcut}, where a particular choice of $\ecut$ is indicated that splits our system of interest into the $\rC$ and $\rI$ regions, which we take to be described by \CF\ and mean-field descriptions respectively. In this case it is clear that an appreciable number of modes appropriately described by mean-field theory are included in the $\rC$ region.
Moderate shifts in the value of $\ecut$ result in the transfer of some modes, well-described by mean-field theory, between the regions. Clearly this scenario will have no effect on the physical predictions for the complete system as the modes shifted between the regions are equally well described by both formalisms. However, large changes in $\ecut$ will lead to problems: If $\ecut$ is set too low, then strongly fluctuating modes will inappropriately appear in the $\rI$ region. If $\ecut$ is too high then many sparsely occupied modes will be included in $\rC$.

While the theoretical motivation for choosing the cutoff is clear, there are only a few studies that have examined the dependence of simulation results on the cutoff. Of most note:
\begin{itemize}
\item Sinatra \etal \cite{Sinatra2002a} studied the damping rates for coherent excitations as the number of $\rC$ modes was varied by an order of magnitude. They found that the damping rate varied by a factor of two over this range, and the best results (i.e. those agreeing with the Beliaev-Landau damping result) were obtained for low cutoffs.
\item Bradley \etal \cite{Bradley2005b} derived Ehrenfest relations for a \CF\ system, which display an explicit dependence on the projector. They also presented results showing that the macroscopic properties of density distribution and the condensate fraction varied appreciably when the numerical method used was changed from a spectral approach using oscillator states (i.e. implementing the $\rC$ region using a well-defined energy cutoff in the single particle basis) to a uniform grid (i.e. implicit projection that only provides a momentum cutoff at the inverse of the grid point spacing).
 \end{itemize}

\subsubsection{Features and interpretation of the truncated Wigner method}\label{TWAfeatures}
\paragraph{Single trajectory interpretation}
Phase space methods provide a practical means for calculating correlation functions, which  can only be compared with the equivalent quantities calculated as an ensemble average of experimental measurements.
However, for highly occupied fields, the behaviour observed in each trajectory of the TWA seems to be typical of that seen in single realizations of experiments, (e.g. see the results in  \sref{sec:BECcollisions} and \sref{sec:reflection}). Thus it is plausible that single realizations of Wigner trajectories should approximately correspond to a possible outcome of a given experiment. This is no more surprising than the observation that the GPE is a remarkably good predictor of the dynamics of individual experimental runs, and follows from taking the classical limit of the full quantum evolution, with the important addition of initial fluctuations. 
\paragraph{Restriction to states with positive Wigner function} However, this is not the full story as the Wigner function can be negative for some states (see \fref{Fig:wignerfun}c). Negative Wigner functions are not amenable to exact treatment by diffusive processes and so there are in fact certain quantum states that are inaccessible to the stochastic sampling methods described in this review. There are actually two points here. Firstly, negative Wigner functions are difficult to sample as an initial condition. Secondly, a positive Wigner function will not become negative under diffusive TWA evolution. While it may predict the mean fields accurately, it may not (and cannot) give the correct correlation functions for some processes. These are however, fairly rare with BEC. This restriction limits the range of quantum phenomena to states with positive Wigner functions, ruling out superpositions of number states and demonstrations of the non-locality of quantum mechanics (violation of a Bell inequality).
 
\paragraph{Spontaneous scattering} The GPE is
fundamentally a theory of stimulated (Bose-enhanced) scattering, which does not include
spontaneous processes.
In particular, scattering into initially
unoccupied modes will not occur, although this may eventually occur in computational simulations due to the gradual accumulation of numerical errors.
The Wigner method, however, sets an irreducible level of initial
fluctuations in all modes of the \CF\, i.e. half an atom of vacuum fluctuations. 
In effect, spontaneous scattering
becomes modeled by weakly seeded stimulated scattering. 
\paragraph{Multimode averaging} 
The  \CF\ used in truncated Wigner simulations usually consists of some large number of stochastically sampled modes $M$. 
Many observables of interest (e.g. column density, cloud rms-width) depend on the values of a significant portion of these modes, so that the statistical fluctuations in the value of such observables can be quite small. Often the behaviour and evolution of these observables exhibit little difference between independent trajectories.

\paragraph{Long time dynamics} Sampling the initial state  introduces fictitious population into the system, i.e. the vacuum noise. In ensemble averaged calculations this
is subtracted when constructing operator averages from trajectory
averages. In single trajectory dynamics of the truncated Wigner
method, the extra population becomes dynamically thermalized and
indistinguishable from the rest of the field. There are two effects
at work here. First, the truncation of the equation of motion means
that  quantum mechanical corrections, which prevent this thermalization, have been neglected. 
Second, by considering single trajectories, the
formal correspondence to operator averages is lost and the results
must be interpreted within the context of classical field theory. For
long times, the advantage of the truncated Wigner method is that it
provides a more complete physical picture of the
system evolution than the GPE, but it must be interpreted with some care. The
primary gain is the inclusion of spontaneous effects in the dynamics
from the outset.

\section{The projected Gross-Pitaevskii equation}
\label{SEC:PGPE}

The projected Gross-Pitaevskii equation (PGPE) formalism is valid for degenerate Bose gases at finite temperature, a system for which many excited modes (in addition to
the condensate) of the atomic field may have a  high mean occupation,
 i.e.,~satisfying the criterion $\langle \hat{a}_j^\dagger\hat{a}_j\rangle \gg 1$.
In this section we describe the PGPE  
formalism, and
show how it can be formulated to make quantitative comparisons with
experiments in this regime.  Unlike the other \CF\ techniques described in this review, the PGPE formalism can be described as a ``classical field theory" in the sense of the classical limit taken in \sref{truncatedWig}.

\subsection{Classical field description of thermal Bose
fields}\label{CFthermBG}
The suggestion that the Gross-Pitaevskii equation could be used to describe the dynamical evolution of the Bose field in the limit of large mode occupation was first made by Svistunov in 1991 \cite{Svistunov1991}, and later by Kagan \etal\ in references \cite{Kagan1992,Kagan1994,Kagan1997}.  
Damle \etal~\cite{Damle1996a} first investigated this using numerical calculations in 1996.  They used the homogeneous GPE with a very weak
nonlinearity to study the phase-ordering kinetics of a Bose gas in two and three dimensions on small grids, and performed a scaling analysis of the growth of the condensate fraction in a temperature quench.

Subsequently Marshall \etal~\cite{Marshall1999a} studied  equilibration of a harmonically trapped Bose gas in 2D using the GPE.  They observed  changes in the population distribution of the bare harmonic oscillator states, and relaxation of the density profile from an initial asymmetric  form to a radially symmetric one, and interpreted these changes as thermalization.

The introduction of a projection operator to restrict the modes represented by the GPE  was first reported by Davis \etal~\cite{Davis2001b} for the case of a 3D homogenous gas. At a similar  time thermalization for a homogeneous system was  demonstrated by Goral \etal~\cite{Goral2001b}, who solved the equations of motion for the mode amplitudes explicitly in the classical approximation.

\subsubsection{Importance of the projector and numerical methods}\label{sec:Pimport}

It has been long known that applying classical field theory to the electromagnetic field results in the ultra-violet (UV) catastrophe in which an infinite number of modes each have the equipartition share of energy, $k_BT$.
Thus it would seem that the effects of a UV catastrophe would also impact a classical field description of the Bose gas.
However, the manifestation of the catastrophe is rather different. The GPE is the equation of motion for a classical microcanonical field in which the total energy and particle number (field normalization) is conserved.  
In thermal equilibrium this energy is shared equally (equipartitioned) between all system modes.  
Any numerical solution of the GPE is constructed from some finite basis, e.g. choosing an equally spaced grid (equivalent to choosing a basis of planewaves in the first Brillouin zone).  Increasing the number of grid points on which the thermally equilibrated
GPE solution is constructed means the fixed energy is now shared between a larger number of modes, so that the average energy per mode (i.e. temperature of the system) will decrease. In this sense, the results of calculations are dependent on the numerical basis, or more correctly the number and nature of the modes contained in the calculation. However, as we discuss in \sref{secCR}, if the modes of the calculation  correspond to only the highly occupied modes of the physical system, then a quantitative description of the system can be made. For this reason  the use of an explicit projector in the PGPE equation \eref{Pgpe} is of great importance, because it precisely defines the calculation without reference to the numerical implementation. In addition, great care must be taken in  implementing numerical methods for propagating the PGPE so that \emph{all} modes of the \CF\ region are evolved accurately in order to avoid spurious dynamics which can lead to an incorrect representation of the physical system of interest. We note that a classical field formalism  has been developed using unprojected grid methods, summarized in Brewczyk \etal \cite{Brewczyk2007a}. While such an approach seems suitable for investigating qualitative behaviour of Bose gases in various regimes, it has not been applied to quantitative comparison with experiments.

\paragraph{Use of grid methods}
Grid methods  are ubiquitous in the solution of the GPE, but care must be taken in using these methods as the basis of classical field simulations. For example, the cubic nonlinearity in the PGPE can generate momentum components up to three times larger than those present in the classical field.  In a grid representation of the field this leads to \emph{aliasing}, which corresponds to (unphysical) collisions between modes that do not conserve momentum.  
Grid methods can be used effectively for numerically solving the PGPE for a uniform gas, if several adjustments are made: 
(i) The projector needs to be explicitly implemented. E.g. the single particle energy cutoff, discussed in \sref{SEC:ProjOps}, can be implemented by setting to zero all modes outside a sphere of radius $\hbar k_{\rm cut}=\sqrt{2m\ecut}$ in momentum space.
(ii) A large number of momentum states outside the projected region need to be retained to avoid the aliasing problem. The equivalent position space requirement is that if $k_{\rm cut}$ is the largest wavevector retained by the projector, then a spatial grid of spacing $\Delta x=\pi/2 k_{\rm cut}$ needs to be used to evaluate the nonlinear term, which is twice as dense as the Nyquist sampling requirement, $\Delta x_N=\pi/k_{\rm cut}$. Additional discussion of these issues is given in \aref{FTnumerics}. 

The experimentally relevant harmonically trapped system poses a more formidable challenge since the natural modes of grid representation (i.e. planewave modes) bear little resemblance to the harmonic oscillator modes, making projection difficult. Also, we note that in typical experimental regimes there should be of order $10^2-10^4$ modes in the \CF\ region (see \sref{secCR}), whereas grid methods usually require  $\gtrsim10^5$ points to accurately simulate the GPE in 3D. Details of an efficient numerical algorithm for the PGPE in a harmonic trap is summarized in \aref{SEC:Numerics}.

\subsubsection{\CF\  region for the PGPE: the
``classical region"}\label{secCR}
For the PGPE formalism, the occupations of all the modes  of the \CF\ region satisfy the mean occupation
requirement
\begin{equation}
\langle \hat{a}_n^\dagger\hat{a}_n\rangle \approx
\overline{\alpha_n^*\alpha_n}\ge \ncut,\label{EQ:PGPECSR}
\end{equation} 
where $\ncut$ is a number of order one (typical choices range from
about 1 to 10). 
Loosely, $\ncut$ is the degree of coherence of the mode, and should
be compared to the  basic level of quantum fluctuations set by the Wigner 
function requirement $\overline{(\alpha_n^*\alpha_n)_{vac}}={1\over 2}$ for vacuum modes. We hence
refer to the modes satisfying this condition as constituting the
\emph{classical region}, in the sense that quantum corrections to
\CF\  equation of motion  \eref{Pgpe}   for these modes are small (see \sref{truncatedWig}). 
\begin{figure}[htbp]  
 \centering{\includegraphics[width=3in]{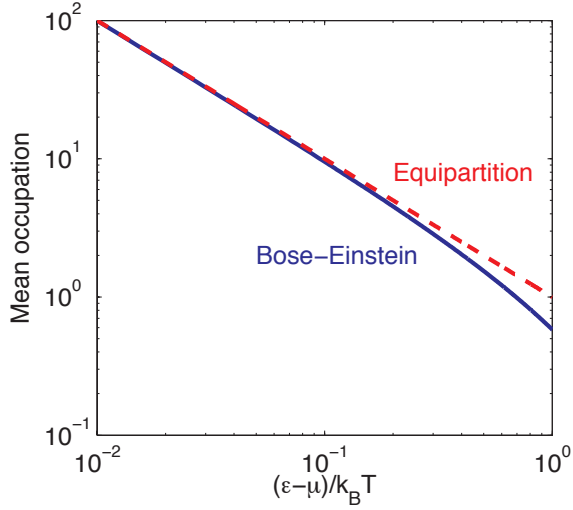} }
   \caption{(color online). Comparison of the quantum Bose-Einstein
and classical equipartition predictions for the mean occupation of a
single particle mode. 
    }
   \label{fig:classmodes}
\end{figure} 
A further justification  for setting $\ncut$ can be obtained from  
\fref{fig:classmodes}, where the mean mode occupation is examined as a function of the scaled single
particle energy.  When the parameter $(\epsilon-\mu)/k_BT \ll 1$, then the exponential in the quantum Bose-Einstein distribution
\begin{equation}
\bar{n}_{BE}(\epsilon)=\frac{1}{\exp[(\epsilon-\mu)/k_BT]-1},\label{nBE}
\end{equation}
can be expanded to first order to give the classical equipartition distribution 
\begin{equation}
\bar{n}_{EQ}(\epsilon)=\frac{k_BT}{\epsilon-\mu}.\label{nEQ}
\end{equation}
In \fref{fig:classmodes} it can be seen that these two distributions are in good agreement for highly occupied modes, i.e.  modes satisfying
$\epsilon-\mu\lesssim k_BT$ with mean occupation $\bar{n}\gtrsim1$.  

The properties and size of the classical region for typical
experimental parameters are not \emph{a priori} obvious, especially
for the case of an interacting gas.
In \fref{fig:PGPE_CR} we consider the case of a harmonically trapped system and estimate the number of classical region
modes and the number of particles occupying those  modes using a
Hartree-Fock mean-field calculation  \cite{Blakie2005b}. Those
results reveal that the number of classical modes is maximum at the condensation transition, with of order several thousand modes satisfying 
condition \eref{EQ:PGPECSR} for the parameters of this calculation.

Strictly, \eeref{EQ:PGPECSR} is only applicable to the
noninteracting modes of the  gas, but can be generalized to the interacting system, e.g. by analysing the occupation of the natural orbitals of the one-body density matrix.
However,  typically $\ecut$  is sufficiently large compared to the
interaction energy scale, that the highest energy noninteracting modes (i.e. $\phi_n\xa$ with $\epsilon_n\approx\ecut$) in $\rC$ are a good approximation to the modes of the interacting system. In this case \eeref{EQ:PGPECSR} can be directly applied to these high energy modes.

\begin{figure}[htbp]  
 \centering{\includegraphics[width=5in]{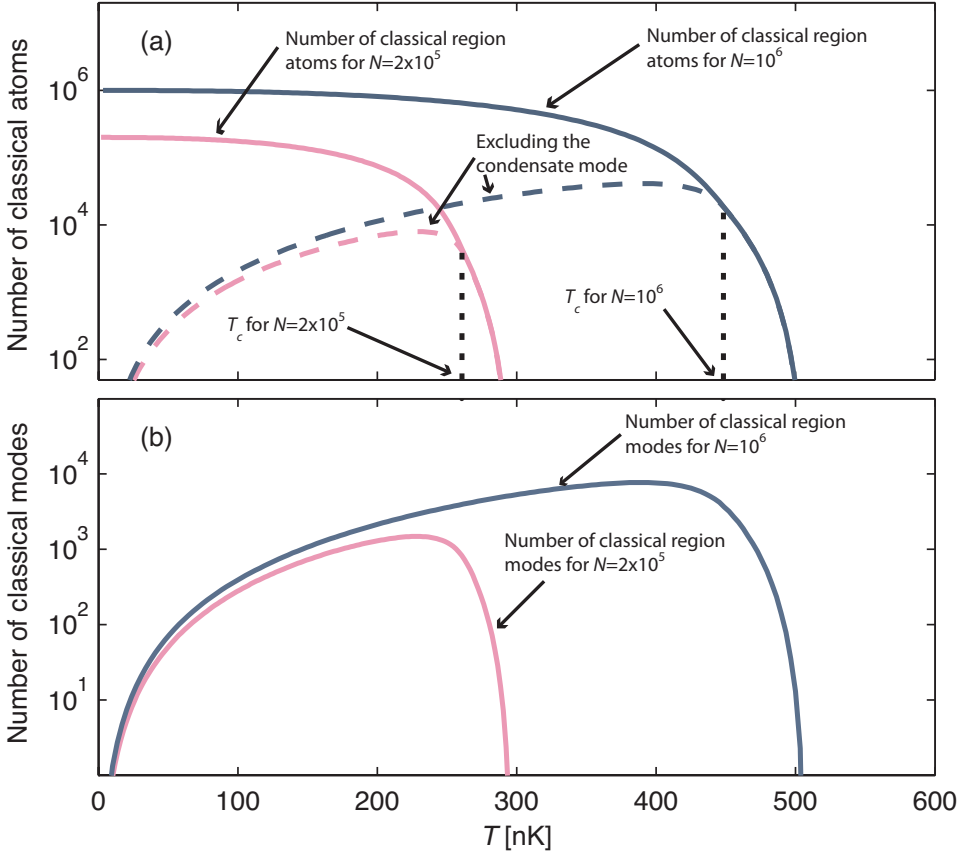} }
   \caption{(color online). Size and population of the classical
region for a harmonically trapped system. (a) Classical region population including (line)   and
excluding  (dashed) the condensate occupation. (b) Number of classical region
modes. Results: pink
(grey) $N=2\times10^5$ atoms, and blue (black) $N=1\times10^6$ atoms.
Results calculated using Hartree Fock theory (see 
\cite{Blakie2005b}) for rubidium-87 atoms in an isotropic harmonic trap of frequency 100 Hz with $\ncut=3$ used to
define the classical region}. 
   \label{fig:PGPE_CR}
\end{figure}

\subsubsection{PGPE formalism}\label{PGPEformalismsec}
The appropriate equation of motion for the \CF\  is the PGPE \eref{Pgpe}.
Since all the modes in $\rC$ are highly occupied, in a three-dimensional system we  find an appreciable number of atoms residing in the incoherent region.
Thus, the detailed non-equilibrium dynamics of the system
will in general depend on the exchange of energy and particles
between $\rC$ and $\rI$. A consistent 
formalism for including these processes is described in \sref{sec:sgpe}.
However, for the purposes of this section, we will assume that for near equilibrium scenarios the $\rC$ and $\rI$ regions are weakly coupled, and we can treat each region in isolation as long as we match their temperatures and chemical potentials.
Building on this assumption, in the remainder of \sref{SEC:PGPE} we mainly concern ourselves with the properties and
interpretation of the PGPE (\ref{Pgpe}) as a micro-canonical means for describing the classical region of a
finite temperature Bose gas. In \sref{SEC:S3incohregion} we discuss a 
mean-field treatment of the incoherent region, $\rI$, as a means to providing a quantitative description of the full system.

\subsection{Hands-on introduction to the PGPE formalism}\label{SechandsOnPGPE}
In this section we  introduce the basic ideas of the PGPE formalism by example. To do this we present a hands-on case study of how to prepare initial states and evolve them to thermal equilibrium. We introduce various tools for analyzing PGPE simulations as needed.

\subsubsection{Simulation parameters}\label{SEC:PGPEparams}
To guide our presentation we illustrate the PGPE method using simulations for an experimentally realistic system. We take this system to be a harmonically trapped gas of rubidium-87 atoms in a potential
\begin{equation}
V_0(\mathbf{x}) = \frac{1}{2}m(\omega_x^2x^2+\omega_y^2y^2+\omega_z^2z^2),
\end{equation}
where $\{\omega_x,\omega_y,\omega_z\}$ are the  oscillation frequencies  (and with ${\delta V}=0$) .
For definiteness we take   $\omega_x\!=\!2\pi\!\times\!120$ Hz, $\omega_{y,z}=\!2\pi\!\times\!30$ Hz, i.e. the trap has a \emph{fat pancake} geometry with the $x$-direction being tightly confined. 
For the calculations we fix the  cutoff defining the \CF\  region at ${\epsilon}_{\rm{cut}}\!=\!33\hbar\omega_z$, so that there are  $M\!=\!1560$ single particle modes in $\rC$.
We take the number of  atoms in this region to be fixed at $N_{\rC}= 10^4$ (see \eeref{PN}), and will   verify \emph{a posteriori} that all the modes are highly occupied as required for the validity of the PGPE formalism. 

\subsubsection{Initial state preparation}\label{SEC:InitStatePGPE}
The nonlinearity of the PGPE causes its evolution to be ergodic and many of the issues involved with appropriate sampling of initial conditions in the truncated Wigner approach can be avoided if we are only interested in the equilibrium properties of the system.  

Thus the generic method to study finite temperature regimes is to begin with a randomized initial state with some definite energy, as specified by the \CF\ Hamiltonian $E_{\rC}=H_{\rC}[\psi_{\rC}]$ \eref{PH}.
The \CF\ energy is a constant of motion for the PGPE (\ref{Pgpe}) and forms a convenient macroscopic constraint for specifying the thermal state of the system. The procedure for making such energy states is rather arbitrary. We choose to make use of the Thomas-Fermi approximation to the condensate mode 
\begin{equation}
{\xi}_{\rm TF}(\mathbf{x})=\sqrt{\frac{\mu_{\rm TF}-{V}_{0}(\mathbf{x})}{u}}\,\,\theta\left(\mu_{\rm TF}-{V}_{0}(\mathbf{x})\right),
\end{equation}
where $\theta(x)$ is the unit step function and 
\begin{equation}
\mu_{\rm TF}=\frac{\hbar\bar{\omega}}{2}\left(\frac{15 a N_0}{\bar{a}}\right)^{2/5},\label{muTF}
\end{equation}   is the Thomas-Fermi chemical potential \cite{Dalfovo1999a}, with with $\bar{\omega}^3=\omega_x\omega_y\omega_z$  and $\bar{a}=\sqrt{\hbar/m\bar\omega}$.  We can generate a state of desired energy by superimposing the Thomas-Fermi state with a (high energy) randomized state, $\xi_r\xa$, according to 
\begin{equation}
\xi_E\xa = p_0\xi_{\rm TF}\xa+p_1\xi_r\xa,\label{Estategen}
\end{equation}
where the variables $p_0$ and $p_1$ are adjusted to obtain the desired energy. In practice, $\xi_r$ is approximately orthogonal to $\xi_{\rm TF}$ and we can take $p_1=\sqrt{1-|p_0|^2}$. 

For reference, under the constraint of fixed \CF\ normalization, $N_{\rC}$, the minimum energy configuration corresponds to the zero temperature case with all atoms residing in the condensate mode $\xi_0\xa$ (i.e.~$N_{0}=N_{\rC}$), which can be obtained by solving the time-independent Gross-Pitaevskii equation \eref{EQ:tiGPE}.

\subsubsection{PGPE thermalization} \label{sPGPEtherm}
\begin{figure}[t]  
 \centering{\includegraphics[width=5in]{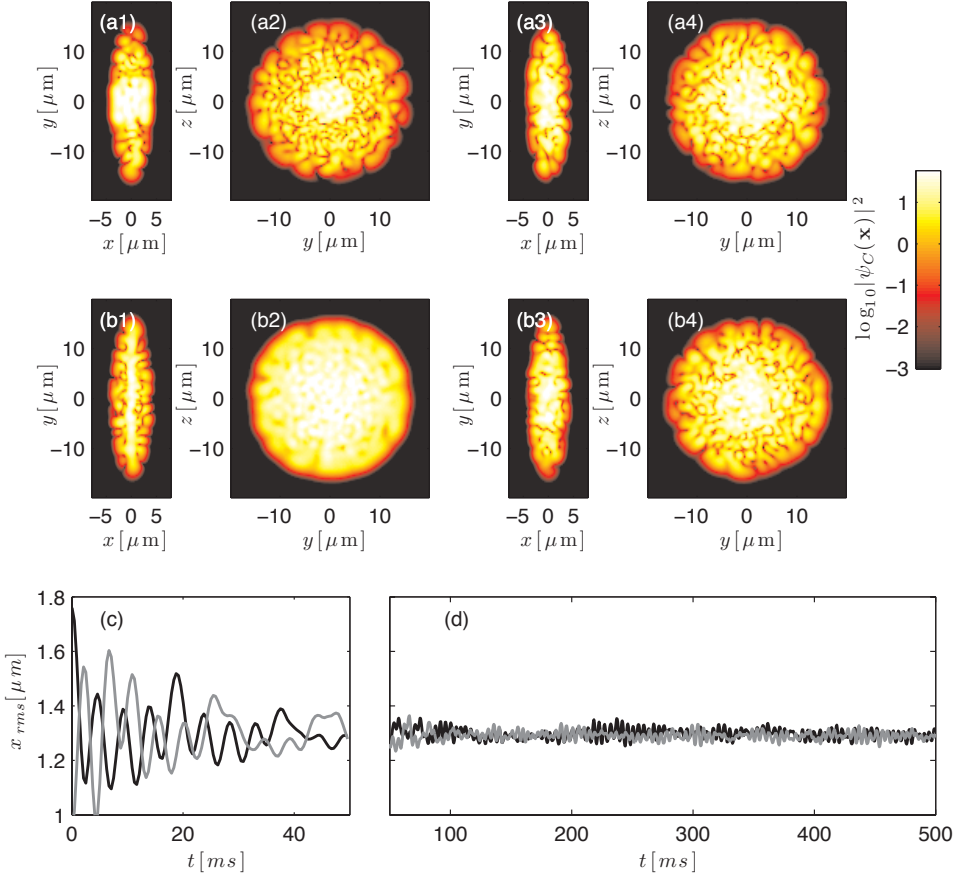} }
   \caption{(color online) Relaxation to equilibrium. Density slices of the two non-equilibrium initial states: $\xi_E^{(a)}\xa$  (a1)-(a2), and $\xi_E^{(b)}\xa$ (b1)-(b2), and the respective states they evolve to at $t=1000ms$ (a3)-(a4) and (b3)-(b4). Both states have $E_{\rC}=20.0N_{\rC}\hbar\omega_z$.  The rms-width of the \CF\  in the
    $x$-direction (c) during the first few trap periods and (d) after 15 trap periods. Width of simulation ``case(a) 
    "   (black line) and the ``case (b)"  (grey line).  Note: other parameters given in \sref{SEC:PGPEparams}, and colourmap in density plots corresponds to $\log_{10}$ of the density measured in units of  $(\mu m)^{-3}$.
    }
   \label{fig:retherm}
\end{figure}


Here we present evidence for thermalization of the PGPE in the trapped system.
To do this we consider two initial microstates of the \CF: $\xi_E^{(a)}\xa$ and   $\xi_E^{(b)}\xa$, which we will refer to as cases (a) and (b) respectively.   Both of these initial states have the same energy $E_{\rC}=20.0N_{\rC}\hbar\omega_z$, and are constructed according to the procedure outlined in \sref{SEC:InitStatePGPE} but with a modified  choice of the Thomas-Fermi state, $\xi_{\rm TF}$, as we discuss below.
Such initial states will not be equilibrium states, and during PGPE evolution will thermalize. To emphasize the initial non-equilibrium dynamics and the role of thermalization we choose to use distorted Thomas-Fermi states in \eref{Estategen}:   initial state $\xi_E^{(a)}\xa$ is produced using a Thomas-Fermi state that has been squeezed in the $x$ direction;  initial state $\xi_E^{(b)}\xa$ is produced using a Thomas-Fermi state that has been squeezed in the $y$ direction.  These two initial states, while having the same energy,  are clearly very distinct in spatial character as revealed by the density slices shown in \fref{fig:retherm}(a1)-(a2) and (b1)-(b2).  The final states after   PGPE evolution for 1000 ms are shown in  \fref{fig:retherm}(a3)-(a4) and (b3)-(b4). 

These results show that the system thermalizes, in the sense that the system evolves to more-likely microstates. Indeed, while the states in \frefs{fig:retherm}(a3)-(a4) and \frefs{fig:retherm}(b3)-(b4) are not identical (differ by fluctuations), they are much more similar than their initial states.

To examine the dynamics of thermalization more carefully we show the $x$-widths of the states, as characterized by the rms value $x_{rms}\equiv\langle x^2\rangle_t-\langle x\rangle^2_t$ (where $\langle\,\rangle_t$ is the expectation at time $t$), in \fref{fig:retherm}(c) and (d). Initially, systems (a) and (b) exhibit large oscillations and differ strongly in their width dynamics (see Fig~\ref{fig:retherm}(c)), reflecting the differences in the intial non-equilibrium states. After approximately  20 ms the large scale width oscillations have damped significantly, leaving much smaller fluctuations. In \fref{fig:retherm}(d) we show the width dynamics   from 50 ms to 500 ms. Here the width fluctuations are about an order of magnitude smaller than the initial  oscillations, and despite both systems beginning from very distinct initial states, these dynamics are consistent with both systems thermalizing to the same equilibrium, i.e.~the same mean width and fluctuation properties.

There are a large variety of observables that we could compute to examine the thermalization of the system. However in general we typically find that the system relaxes towards equilibrium appreciably within a few trap periods. For typical simulations, where we are interested in equilibrium properties and start from the (undistorted) initial state described in \sref{SEC:InitStatePGPE}, we evolve the \CF\ for several tens of trap periods to thermalize before sampling for system properties.

\subsubsection{Equilibrium: Ergodicity, correlation functions and condensate fraction}\label{SEC:PGPEdensitycondfrac}
\paragraph{Ergodic averaging correlation functions} The \CF\  energy, given by the functional \eref{PH}, is a constant of motion under PGPE evolution. Indeed, the energy and other such constants of motion, e.g. field normalization  and angular momenta (when the trap has rotational symmetry), take the form of the macroscopic constraints on the thermal state of the system. In principle, the equilibrium properties of the system could be determined by ensemble averaging over all fields consistent with these constraints. The nonlinearity of \eeref{PH} makes finding all functions $\cf$, for a given normalization and energy, impossible without approximation. If we were to move beyond the microcanonical ensemble, some form of Monte Carlo sampling could be used, although we do not pursue this possibility here.

In contrast, numerical methods for evolving the PGPE are well-developed and allow a different means to sample the ensemble:
we can make use of the ergodic hypothesis (that a system will in time
visit every accessible configuration in phase space without bias) to sample microstates of the system. 
 Thus, for an ergodic system, an ensemble average of an  observable, $O$, can be
calculated by a time average over a sufficiently long period of dynamical evolution, i.e.  
\begin{equation}
\langle O\rangle =\lim_{\theta\to\infty}\left\{ \frac{1}{\theta}\int_{\theta_i}^{\theta_i+\theta}dt\, O\right\}\approx\frac{1}{M_s}\sum_{s=1}^{M_s}\,O(t_s),\label{EQ:timeave}
\end{equation}  
where $\{ t_{s}\} \in[\theta_i,\theta+\theta_i\}$ is a set of $M_{s}$ time instances
at which the system  evolution has been sampled \cite{Davis2002a,Goral2002a}. For this choice
to be a accurate estimates of the ensemble average we require $M_{s}\gg1$,
and the time span over which averaging is done to be long compared
to the slowest time scale in the problem, e.g. the longest harmonic
oscillator period.  

In general the observables of interest are of the form of a
correlation function of the field, typically  a product of quantum
field operators such as 
\begin{equation}
\langle \hat{O}\rangle\equiv\langle\hat{\psi}^{\dagger}_{\rC}(\mathbf{x}_{1})\ldots\hat{\psi}^{\dagger}_{\rC}(\mathbf{x}_{j})\hat{\psi}_{\rC}(\mathbf{x}_{j+1})\ldots\hat{\psi}_{\rC}(\mathbf{x}_{n})\rangle.\label{OQ}
\end{equation}
This expression could also be  generalized to multi-time correlations, although we will not do so here.  Note that here the correlation functions
only involve the \CF\ operator: We discuss correlations involving incoherent region operators in \sref{SEC:S3incohregion} and in \sref{SEC:PGPEApp2ptcorrelns}.
 To evaluate \eref{OQ}
we make the substitution $\hat{\psi}_{\rC}(\mathbf{x})\to\cf(\mathbf{x})$, transforming the expression to the general classical field  form, and then replace the ensemble average with a time-average according to \eref{EQ:timeave}.  We note that this procedure is in accordance with that outlined for the truncated Wigner approach (see \sref{sec:Wigneruse}) as we can neglect the commutation relations of the operator fields  for the highly occupied modes described by the PGPE. 

\begin{figure}[htbp]  
 {  \centering
   \includegraphics[width=5in]{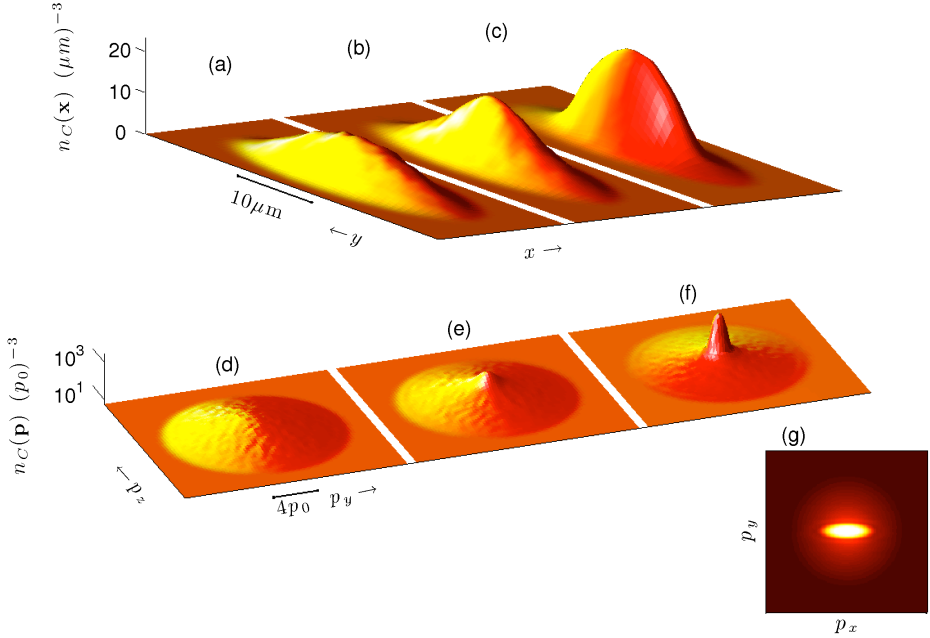} }
   \caption{(color online) \CF\  position (a)-(c) and momentum (d)-(f) density.
   Simulation parameters: (a), (d) $E_{\rC}=24N_{\rC}\hbar\omega_z$, (b), (e) $E_{\rC}=22N_{\rC}\hbar\omega_z$, (c),(f),(g) $E_{\rC}=15N_{\rC}\hbar\omega_z$. Other parameters: $N_{\rC}=10^4$ $^{87}$Rb atoms, simulations are evolved for $t=10^3\pi/\omega_z$, with $M_s=2500$ samples taken over the last half of the simulation used to time-average. Trap parameters and cutoff are given in \sref{SEC:PGPEparams}.
    }
   \label{fig:posmtmden}
\end{figure} 

\paragraph{Position space density} 
In \frefs{fig:posmtmden}(a)-(c)  we show the time-averaged density of the \CF\ in the $z=0$ plane, i.e.
\begin{equation}
n_{\rC}(\mathbf{x}) = \langle \hat{\psi}^\dagger_{\rC}(\mathbf{x})\hat{\psi}_{\rC}(\mathbf{x})\rangle\approx  \frac{1}{M_{s}}\sum_s|\cf({\mathbf{x}},{t}_s)|^2.
\end{equation}
While the instantaneous density, e.g. see \fref{fig:retherm}(a3)-(a4), exhibits spatial fluctuations and a random appearance of no particular symmetry, the averaged density is smooth and highly symmetric. 
The cases  in \fref{fig:posmtmden}(a)-(c) vary from a condensate fraction of $\lesssim0.5\%$ (\fref{fig:posmtmden}(a)) to $56\%$ (\fref{fig:posmtmden}(c)), yet the spatial density profiles change rather gradually and do not provide clear evidence for condensation. We discuss our procedure for quantifying the condensate  later in this section. We also note the work of Krauth who has developed a path integral quantum Monte Carlo scheme for the trapped Bose gas in reference \cite{Krauth1996a} and has computed density profiles for systems with up to $10^4$ atoms.
 
\paragraph{Momentum space density} 
It is also desirable to be able to calculate correlation functions of the momentum space field operator,
\begin{equation}
\hat{\phi}_{\rC}(\mathbf{k})=\frac{1}{(2\pi)^{3/2}}\int d^3\mathbf{x}\,\hat{\psi}_{\rC}(\mathbf{x})e^{-i\mathbf{k}\cdot\x}.
\end{equation}
A particularly useful example is the momentum density, 
\begin{equation}
n_{\rC}(\mathbf{k}) = \langle \hat{\phi}^\dagger_{\rC}(\mathbf{k})\hat{\phi}_{\rC}(\mathbf{k})\rangle.
\end{equation} 
The use of Fourier transforms to convert
from the spatial to momentum representations of the \CF\  makes
evaluating such observables quite efficient.

In \fref{fig:posmtmden}(d)-(f) the momentum density in the $k_x=0$ plane is shown for the cases corresponding to the position space densities in \fref{fig:posmtmden}(a)-(c). Noting that the momentum density axis is logarithmic, we observe a strong peak forming  as the condensate fraction increases, providing an unambiguous signature of condensation. In \fref{fig:posmtmden}(g) the momentum density for the case in \fref{fig:posmtmden}(f) is shown in the $k_z=0$ plane, where the anisotropy  of the condensate mode (due to the anisotropy of the confinement potential in the $xy$-plane) is clearly apparent.

\begin{figure}[htbp]  
 {  \centering
   \includegraphics[width=5in]{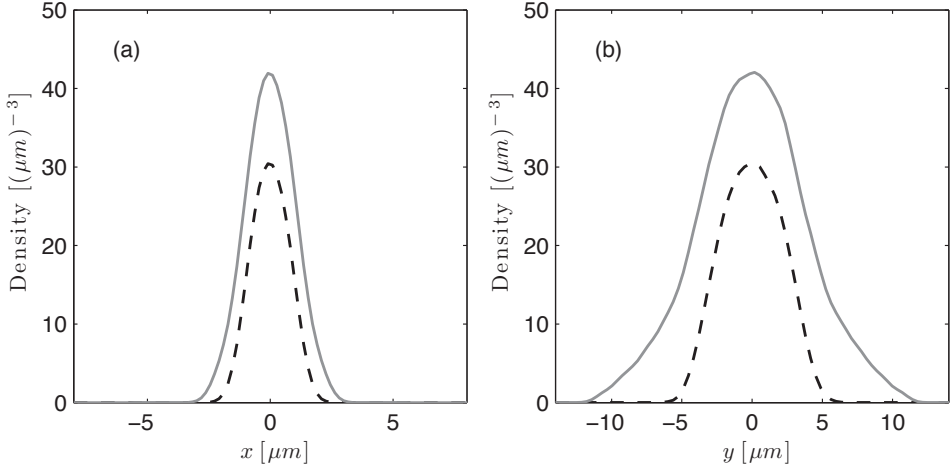} }
   \caption{Equilibrium density for the \CF\ region description of a Bose gas. Condensate density (dashed black line) and \CF\ region density (solid grey line) densities along (a) $x$-axis (more tightly confined direction)$x$ and (b) $y$-axis.
   Simulation for $E_{\rC}=20N_{\rC}\hbar\omega_z$ with other parameters given in \sref{SEC:PGPEparams}.
    }
   \label{fig:density}
\end{figure} 

 \paragraph{Condensate} 
It is important to quantify the amount and nature of condensate in the system. Unlike the uniform system, where the condensate mode is always the zero momentum mode, particle interactions in the trapped system have a strong effect on the shape of the condensate mode and cause it to be significantly different from the ideal case (see \eeref{EQ:tiGPE} for the regime $T\ll T_c$).  
 
According to the criterion provided by Penrose and Onsager \cite{Penrose1956a}, the condensate number $N_0$ is identified as the largest eigenvalue of the one-body density matrix, defined in terms of the field as 
\begin{equation}
G^{\rm 1B}({\mathbf{x}},{\mathbf{x}}^\prime)=\langle  \psi^*_{\rC}(\mathbf{x}) \psi_{\rC}({\mathbf{x}}^\prime)\rangle.
\end{equation}
The corresponding eigenvector is associated with the condensate mode of the system, $\xi_0(\mathbf{x})$, i.e.
\begin{equation}
\int d^3\mathbf{x}'\,G^{\rm 1B}({\mathbf{x}},{\mathbf{x}}^\prime)\xi_0(\mathbf{x}')=N_0\xi_0(\mathbf{x}).
\end{equation}
For the uniform system the one-body density matrix exhibits  the property of \emph{off diagonal long range order} (ODLRO) when a condensate is present  \cite{Yang1962}, i.e.,
\begin{equation}
G^{\rm 1B}({\mathbf{x}},{\mathbf{x}}^\prime)\to\frac{N_0}{V}\quad\rm{as}\quad|\mathbf{x}-\mathbf{x}'|\to\infty.
\end{equation}
In this case the condensate orbital is $\xi_0(\mathbf{x})\propto 1/\sqrt{V}$, where $V$ is the system volume and the thermodynamic limit is assumed.

Another widely used definition of the condensate ``order parameter" is given by
\begin{equation}
\xi_0(\mathbf{x})=\langle \hat{\psi}(\mathbf{x})\rangle,\label{Broksymm}
\end{equation}
based on the idea of spontaneously broken gauge symmetry. The time averaged value of $\langle \cf\rangle$ in the \CF\ approaches is typically zero, and so this definition is of limited use for quantifying condensate in this review. For a comprehensive discussion on the various definitions of condensate we refer the reader to the review article of Leggett \cite{Leggett2001}, who shows preference to the Penrose-Onsager definition, and states (on the topic of the broken symmetry definition) `` \ldots while possibly streamlining some calculations when judiciously used, is liable to generate pseudoproblems and is best
avoided".

 In our formalism $G^{\rm 1B}({\mathbf{x}},{\mathbf{x}}^\prime)$ is equivalently and much more efficiently computed in the mode basis as
$G_{mn}^{\rm 1B}=\langle \alpha^*_m\alpha_n\rangle$, which is quite feasible to compute for the typical classical region size ($\lesssim10^4$ modes in $\rC$)
  using time-averaging. In the spectral basis the condensate mode is specified by a vector $\alpha^0_n$ such that $\sum_nG_{mn}^{\rm 1B}\alpha^0_n=N_0\alpha_m^0$.
 
   In \fref{fig:density} we show the  time-averaged position density  along two coordinate axes obtained from a \CF\ simulation. In addition to the total \CF\ density $n_{\rC}$, we also show the condensate density $|\xi_0\xa|^2$. For reference, the condensate number, $N_0$, for simulations over a wide range of energies are given in \tref{tab:PGPEtable}. In the remainder of this section we develop techniques for extracting other thermodynamic quantities to attribute to these calculations: temperature and chemical potential in \sref{Sec:PGPEtemperature}, and incoherent region atoms in \sref{SEC:S3incohregion}.

 We note that the correlation functions discussed so far only apply to the \CF\ region. We return to this issue in \sref{SEC:S3incohregion}, when we consider including contributions from the incoherent region. We also mention that higher order correlation functions, including second order (e.g.~density fluctuations) correlations functions have been calculated using the PGPE approach, see references \cite{Blakie2005a,Bezett2008a,Simula2008a} (also see \cite{Kadio2005a}). A procedure for calculating two-point correlation functions is discussed in \sref{SEC:PGPEApp2ptcorrelns}.
   
\subsubsection{Thermodynamic quantities: temperature and chemical potential}
\label{Sec:PGPEtemperature} 
It is desirable to find a means to attribute a temperature to the thermalized state of a \CF\ simulation.  
Previous attempts to do this have been based on fitting the occupation
of high energy modes to perturbative calculations for the spectrum
based on Hartree-Fock-Bogoliubov (HFB) theory \cite{Davis2001a,Davis2002a} (see \sref{Sec:TandNmodes}).
For harmonically trapped gases, calculation of the HFB modes is much
more difficult, and limits temperature calculations to perturbative
regimes. 
However,  the temperature can be crudely estimated by fitting the
high momentum components of the system to a noninteracting distribution  \cite{Goral2002a}.

An alternative approach of general applicability is found by extending Rugh's
dynamical definition of temperature for classical Hamiltonian systems \cite{Rugh1997a} to the PGPE.
This scheme has the advantage that it is non-perturbative, and is
quite accurate.

Rugh's approach was formulated for a classical mechanical system, and it is convenient to write the \CF\ Hamiltonian as $H_{\rC}=H_{\rC}(\mathbf{\Gamma})$, where  $\mathbf{\Gamma}=\{Q_j,P_j\}$ is the vector of the canonical position and momentum coordinates introduced in \sref{truncatedWig} (equations (\ref{Qdef}), (\ref{Pdef})).
 We also need to explicitly account for the \CF\ normalization functional \eref{PN}, $\mathcal{N}_{\rC}=\sum_j|\alpha_j|^2$, which is another constant of motion and can also be written as a function of canonical coordinates, i.e.~$\mathcal{N}_{\rC}=\mathcal{N}_{\rC}(\mathbf{\Gamma})$.
The usual expression for the temperature of a system in the microcanonical ensemble is given by
\begin{equation}
\frac{1}{T} 
= \left(\frac{\partial S}{\partial E_{\rC}}\right)_{N_{\rC}},\label{EQforT}
\end{equation} 
where the entropy is defined by
\begin{equation}
S =
 k_B\ln\left\{\int d\mathbf{\Gamma} \;\delta[E_{\rC} - H_{\rC}(\mathbf{\Gamma})]\delta[N_{\rC} -\mathcal{N}_{\rC}(\mathbf{\Gamma})]\right\},
\end{equation}
with the delta functions ensuring our microcanonical description is one of fixed \CF\ energy and normalization.

There is several issues with using \eeref{EQforT} to determine the temperature in the PGPE approach. First, it is practically impossible to determine the entropy $S$ when a large number of modes are in the \CF\ region. Second,  \eeref{EQforT} cannot be evaluated using a single micro-canonical ensemble average, since $T$ depends on the derivative with respect to energy. In 1997 Rugh made a fundamental contribution to statistical mechanics by proving that a micro-canonical average could be used to calculate the temperature. This approach is now extensively used in the molecular dynamics community since the micro-canonical average can be replaced by a time-average, as we shall do here.

Rugh's result, proven using  differential geometry methods, showed that the temperature expression \eref{EQforT} could be equivalently written as \begin{equation}
\frac{1}{k_B T} = 
\bigg\langle \mathbf{\mathcal{D}} \cdot \mathbf{X}_T(\mathbf{\Gamma}) \bigg\rangle,
\label{eqn:temp_eqn}
\end{equation}
rigorously shown to work for Hamiltonian systems at energies where the 
energy surface is regular
\cite{Rugh1997a,Rugh1998a,Rugh2001a}. The components of the vector operator
  $\mathbf{\mathcal{D}} $ are
\begin{equation}
\mathbf{\mathcal{D}}_i = e_i \frac{\partial}{\partial \Gamma_i},
\end{equation}
where $e_i$ can be chosen to be any scalar value, including zero, and the vector field $\mathbf{X}_T$ can also be chosen freely within the constraints
\begin{eqnarray}
\mathbf{\mathcal{D}} H_{\rC} \cdot \mathbf{X}_T =1,\quad \mathbf{\mathcal{D}} \mathcal{N}_{\rC} \cdot \mathbf{X}_T =0.
\label{eqn:X_conditions}
\end{eqnarray}
Geometrically this means that the vector field $\mathbf{X}_T$ has a non-zero
component transverse to the $H_{\rC}(\bm{\Gamma})= E_{\rC}$ energy surface, and is parallel to the   $\mathcal{N}_{\rC}(\bm{\Gamma})=N_\rC$ surface. 
A vector field that satisfies these constraints is
\begin{equation}
\mathbf{X}_T=\frac{\mathbf{\mathcal{D}}H_{\rC}-\lambda_N\mathbf{\mathcal{D}}\mathcal{N}_{\rC} }{|\mathbf{\mathcal{D}}H_{\rC}|^2-\lambda_N\left(\mathbf{\mathcal{D}}\mathcal{N}_{\rC}\cdot\mathbf{\mathcal{D}}H_{\rC}\right) },\label{XT}
\end{equation}
where we have introduced the parameter $\lambda_N=\mathbf{\mathcal{D}}\mathcal{N}_{\rC}\cdot\mathbf{\mathcal{D}}H_{\rC}/|\mathbf{\mathcal{D}}\mathcal{N}_{\rC}|^2$.
The expectation value in  
Eq.~(\ref{eqn:temp_eqn}) is over all possible states in the microcanonical ensemble and  can be evaluated as a time-average for our ergodic \CF\ system.  In the interests of brevity we do not discuss the additional technical details of how the matrix elements in \eeref{XT} are evaluated, however point out that a procedure for doing this exactly and efficiently  is given in reference \cite{Davis2005a}.  

It is worth giving an example to illustrate the formalism. Consider a simple system of $M$ degenerate oscillators described by $H_{\rC}=\sum_j\Gamma_j^2$, with no normalization constraint. Taking $\mathcal{D}_i=\partial/\partial\Gamma_i$ we have that $(X_T)_i=\mathcal{D}_iH_{\rC}/|\mathbf{\mathcal{D}}H_{\rC}|^2=\Gamma_i/H_{\rC}$, where we have used that $|\mathbf{\mathcal{D}}H_{\rC}|^2=\sum_j\Gamma_j^2=H_{\rC}$. Finally, we have that $1/k_BT=\langle\mathbf{\mathcal{D}}\cdot\mathbf{X}_T\rangle=(M-1)/E_{\rC}$, which is the standard micro-canonical result.

  Similar to the discussion above, the chemical potential can be evaluated according to
  \begin{equation}
  \frac{\mu}{k_BT}=-\left(\frac{\partial S}{\partial \mathcal{N}_{\rC}}\right)_{E_{\rC}}=\langle \mathbf{\mathcal{D}}\cdot \mathbf{X}_{\mu}(\mathbf{\Gamma})\rangle,
  \end{equation}
where the conditions on the vector field $\mathbf{X}_{\mu}$ are
\begin{eqnarray}
\mathbf{\mathcal{D}} H_{\rC} \cdot \mathbf{X}_{\mu} =0,\quad \mathbf{\mathcal{D}} \mathcal{N}_{\rC} \cdot \mathbf{X}_{\mu} =1.
\label{eqn:Xmu_conditions}
\end{eqnarray}
The appropriate vector field is of the same form as the right hand side of \eeref{XT} but with $H_{\rC}$ and $\mathcal{N}_{\rC}$ interchanged.

 In \fref{fig:tempmu} we show instantaeous values of the  Rugh observables for temperature (i.e. $[k_B\mathbf{\mathcal{D}} \cdot \mathbf{X}_T]^{-1}$)  and chemical potential (i.e. $k_BT\mathbf{\mathcal{D}} \cdot \mathbf{X}_{\mu}$) evaluated from a PGPE evolution. The time-averaged results for these parameters over a broad range of initial energies are given in \tref{tab:PGPEtable}.

\begin{figure}[htbp]  
 {  \centering
   \includegraphics[width=0.95\textwidth]{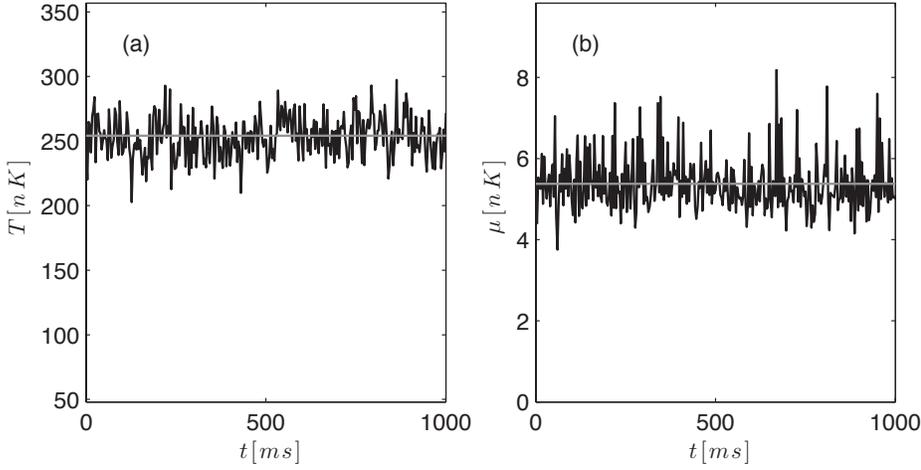} }
   \caption{Extracting dynamical thermal quantities. Instantaneous value of Rugh observable for (a) temperature and (b) chemical potential shown as black lines evaluated over one second of evolution. Average values of $T$ and $\mu$ shown as grey horizontal lines in (a) and (b) respectively.
   Simulation for $E_{\rC}=20N_{\rC}\hbar\omega_z$, with other parameters given in \sref{SEC:PGPEparams}. 
    }
   \label{fig:tempmu}
\end{figure} 

\subsubsection{Including the incoherent region atoms}\label{SEC:S3incohregion}
  To relate the PGPE results back to an experimental system we need to account for the sparsely occupied modes of the incoherent region, which we have so far ignored. To do this we take the classical region and the  incoherent region to be weakly-coupled systems in thermal and diffusive equilibrium (see \fref{Fig:weaklyintersystems}), i.e.~with the 
same temperature and chemical potential. The thermal cloud 
exists in the potential of the trap plus time-averaged \CF\ 
 density $n_{\rC}(\mathbf{x})$  determined from the PGPE
simulations. To model the incoherent region modes we use 
a  Hartree-Fock approximation. 
As discussed in \sref{sec:validity}, the mean-field approach provides a good description of the system away from the critical region (e.g. see \cite{Gerbier2004b}), and of modes well-above the energy scale $un_{\rC}$ (also see the discussion in \sref{s3validity}).  

\begin{figure}[t]
\begin{center}
\includegraphics[height=4.25cm]{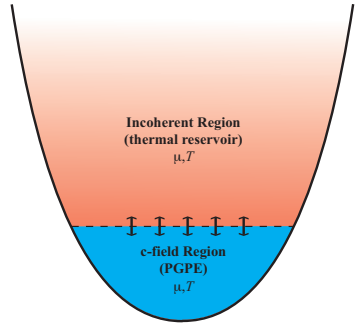}
\end{center}
\caption{\label{Fig:weaklyintersystems}%
Schematic view of the coupling between the \CF\  region,described by the PGPE, and the incoherent region. The systems are assumed to be weakly interacting and in thermal and diffusive equilibrium.}
\end{figure}

The average properties of the incoherent region can be calculated from
the one-particle Wigner distribution\begin{equation}
F_{\rI}(\mathbf{x},\mathbf{k})= \frac{1}{\exp(\beta[\epsilon_{\rm{HF}}(\mathbf{x},\mathbf{k})-\mu])-1},\label{eq:Wig3D}\end{equation} 
where 
\begin{eqnarray}
\epsilon_{\rm{HF}}(\mathbf{x},\mathbf{k}) & = & \frac{\hbar^2k^2}{2m}+ V_0(\mathbf{x})+2u(n_{\rC}(\mathbf{x})+n_{\rI}(\mathbf{x})),\label{eq:EHF3D}
\end{eqnarray}
is the Hartree-Fock energy, and $\mu$ is the chemical potential. The one-particle Wigner distribution is related to the one-body density matrix for the incoherent region (see \eeref{G1wig}), and should not be confused with the multi-mode (many-body) Wigner function discussed in \sref{WignerSec}.

In this semiclassical description $\mathbf{x}$ and $\mathbf{k}$ are treated as continuous (commuting) variables. However, care needs to be taken to ensure that Eq.~(\ref{eq:Wig3D}) is only applied to the appropriate region of phase space spanned by the incoherent region,  i.e.~single-particle modes of energy exceeding $\epsilon_{cut}$. Interpreted in phase space coordinates, this region is  
\begin{equation}\label{WIdef}
\Omega_{\rI}=\left\{ \mathbf{x},\mathbf{k}:\frac{\hbar^2k^2}{2m}+V_0(\mathbf{x})\ge\epsilon_{cut}\right\} .
\end{equation}

A quantity of particular interest for us to calculate is the incoherent region density
\begin{eqnarray}
n_{\rI}(\mathbf{x})&=& \int_{\Omega_\rI}\frac{d^3\mathbf{k}}{(2\pi)^3}\, F_{\rI}(\mathbf{x},k)\\
&=&\int_{K_{\rm{ cut}}(\mathbf{x})}^{\infty}\frac{dk}{2\pi^2}\,  k^{2}\, F_{\rI}(\mathbf{x},k),\label{eq:na3D}
\end{eqnarray}
where we have made use of the isotropic nature of the kinetic energy
term and have implemented the phase space restriction, $\Omega_{\rI}$, as a spatially dependent lower cutoff on the integral 
\begin{equation}
K_{\rm cut}(\mathbf{x})= \frac{\sqrt{2m\left[\epsilon_{cut}-V_0(\mathbf{x})\right]}}{\hbar}\,\theta\left(\epsilon_{cut}-V_0(\mathbf{x})\right),\label{eq:pmin3D}
\end{equation}
where $\theta(x)$ is the unit step function.
The incoherent region atoms interact with those in the \CF\ region, which can be accounted for by adding an effective potential $\delta V=2un_\rI\xa$ to the \CF\ description. To lowest order this shifts the system chemical potential by
$\Delta\mu\approx 2un_{\rI}(\mathbf{0})$. To ensure complete self-consistency, the \CF\ properties would need to be re-simulated including the effective potential, however this is often unnecessary as the incoherent region density is often quite small and approximately uniform in the spatial region of overlap with the \CF\ atoms.
 
We can also calculate the momentum density of the system as 
\begin{equation}
n_{\rI}(\mathbf{k})=\int \frac{d^{3}\mathbf{x}}{(2\pi)^3}\, F_{\rI}(\mathbf{x},\mathbf{k})\,\theta\left(\frac{\hbar^2k^2}{2m}+V_0(\mathbf{x})-\epsilon_{cut}\right).\label{eq:nap3D}\end{equation}

\begin{figure}[t]
\begin{center}
\includegraphics[width=0.95\textwidth]{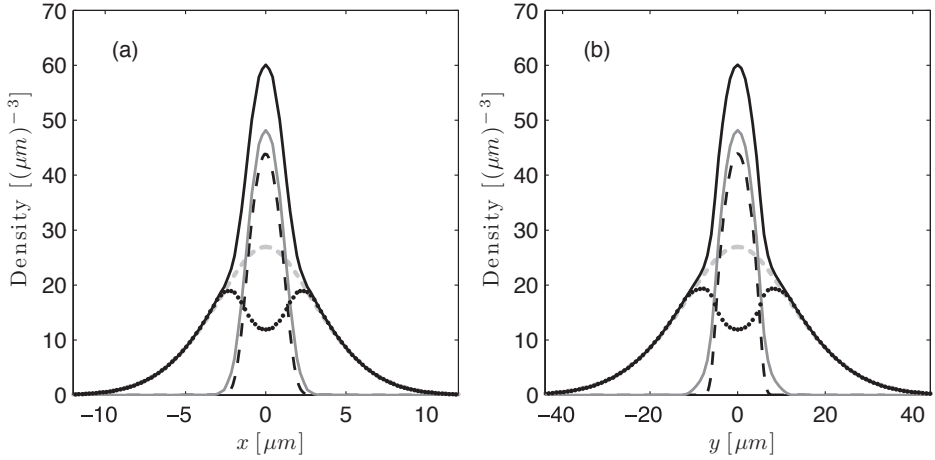}
\end{center}
\caption{\label{Fig:totaldensity} Total density profiles including the incoherent region. (a) Density along $x$-axis and (b) density along $y$-axis. (black dashed line) condensate density, (gray solid line) \CF\ region density, (black dots) incoherent region density, (black solid line) total system density, and the pure Hartree-Fock result for the total density (grey dashed line). Simulation for $E_{\rC}=15\hbar\omega_z$, with other parameters given in \sref{SEC:PGPEparams}.
}
\end{figure}

 Using the Hartree-Fock analysis we can now include the incoherent region atoms into the PGPE simulation results presented in the previous sections. 
In \fref{Fig:totaldensity} we show the typical profiles comparing the \CF\ and incoherent region density profiles, including the total density
\begin{equation}
n\xa=n_{\rC}\xa+n_{\rI}\xa.
 \end{equation}
 These results also allow us to ascribe the total number of atoms, $N=N_{\rC}+N_{\rI}$, where $N_{\rI}=\int d^3\mathbf{x}\,n_{\rI}\xa$, to the simulated systems. Using this analysis of the incoherent region in \tref{tab:PGPEtable} we can attribute total particle number to our PGPE simulations. For comparison, in \fref{Fig:totaldensity}  we have also shown the result of a pure Hartree-Fock analysis (as described above, but taking $\ecut\to0$ so that all modes are treated using mean-field theory). The pure Hartree-Fock result is for the same temperature and total number as the \CF\ calculation, yet predicts no condensate, since the temperature is a few nK above the mean-field critical temperature. Outside such critical regimes the difference between mean-field and   \CF\ calculations is generally much smaller.

The results in \tref{tab:PGPEtable} show that for fixed \CF\ region (i.e.~fixed cutoff $\ecut$ and $N_{\rC}$), the temperature and total number of particles grow rapidly as the \CF\ energy increases. In general this means to simulate a fixed total number of particles for various temperatures,  we must appropriately manipulate the macroscopic parameters defining our micro-canonical system, i.e.~$\ecut$, $N_{\rC}$ and $E_{\rC}$. 
We will see in \sref{sec:sgpe} that use of the stochastic PGPE formalism greatly eases the effort required to calculate systems at definite temperature by making use of a grand-canonical description.


 \begin{table}[htbp]
    \centering 
    \begin{tabular}{  cccccc  } 
     \hline
       $E_{\rC}  [N_{\rC}\hbar\omega_z]$  & $T [nK]$ & $\mu [\hbar\omega_z]$ & $N_0 [\times10^3]$ & $N [\times10^3]$ & $n_{\min}$  \\ 
       \hline     \hline
       
      $ 14.0$ &  $117$ &    $7.81$ &   $ 6.41$   & $180$ & $1.34$\\
     $15.0$ &  $141$   & $ 7.60$   & $5.59$ & $303$ & $1.66$\\
     $16.0$  & $165$ &  $7.34$ &  $4.79$ &    $477$ & $1.97$ \\
    $17.0$ & $189$ &   $7.07$ &   $4.02$ &   $712$ & $2.27$ \\
$18.0$ & $214$ &  $6.81$ &   $3.33$ &   $1,019$ & $2.58$ \\
$19.0$  & $238$ &  $6.58$ &    $2.59$ &   $1,400$ & $2.88$ \\
$ 20.0$ & $265$  &  $6.25$ &    $1.91$ &   $1,890$ & $ 3.20$ \\
$21.0$  & $289$ & $6.07$ &    $1.18$ &   $2,450$ & $3.50$\\
$22.0$  & $315$  &  $5.73$ &    $0.569$ &   $3,170$ & $3.83$\\
$23.0$  & $350$  &  $4.85$ &    $0.176$ &   $4,280$ & $4.18$\\
$ 24.0$  & $420$ &  $1.41$ &    $0.050$ &   $7,270$ & $4.63$\\
$25.0$  &  $602$ &  $-9.99$ &   $0.024$ &  $20,600$ & $5.32$\\ 
 
      \hline\hline
    \end{tabular}
    \vspace*{3mm}
    \caption{Summary of PGPE thermalization results for a \CF\  region with $N_{\rC}=10,000$ Rb-87 atoms.   Other parameters: $\{f_x,f_y,f_z\}=\{120,30,30\} \,\mathrm{  Hz}$ and $\ecut=33\hbar\omega_z$. For reference, the Thomas-Fermi ground state energy is $\  E_{\rm TF} \!=\!\frac{5}{7}N_{\rC}\mu_{\rm TF}\approx9.04N_{\rC}\hbar\omega_z$. Note: $T$ and $\mu$ are determined by the average of two different choices of Rugh temperature (see \cite{Davis2003a}), one of which is shown in \fref{fig:tempmu}.
}
    \label{tab:PGPEtable}
 \end{table}

\subsubsection{Validity conditions}\label{s3validity}
The high mode occupancy of \CF\ region described by the PGPE makes the validity requirements of this approach somewhat different from those listed for the truncated Wigner approach in \sref{sec:validity}. In particular, the dominance of classical fluctuations means that the thermalization of quantum noise is not a concern.  
Thus, the  conditions for the PGPE method to provide an accurate description of the \CF\ region are:
 \begin{enumerate}
 \item {\sl Good basis}: The cutoff has to be sufficiently large that the single
particle modes provide a good basis for describing the interacting
\CF\ region modes. This condition can be expressed in terms of the peak (central) \CF\  density as
$\epsilon_{cut}-\epsilon_0\gtrsim un_{\rC}(\mathbf{0})$, where $\epsilon_0$ is the ground single particle energy. This condition also ensures the validity of the separation into $\rC$ and $\rI$ regions in the critical regime (see discussion in \sref{sec:validity}).
\item {\sl High mode occupation}: the mean occupation of the highest energy single particle state is greater than unity. In general we refer to this quantity, extracted from simulations, as $n_{\min}$ which is also listed in \tref{tab:PGPEtable} to demonstrate the validity of those results.

\end{enumerate}

\subsection{Applications to the uniform Bose gas}\label{sAppsuniformBG}

\subsubsection{Temperature and quasi-particle modes of the uniform system}\label{Sec:TandNmodes}
\begin{figure}\centering 
\includegraphics[width=3.5in]{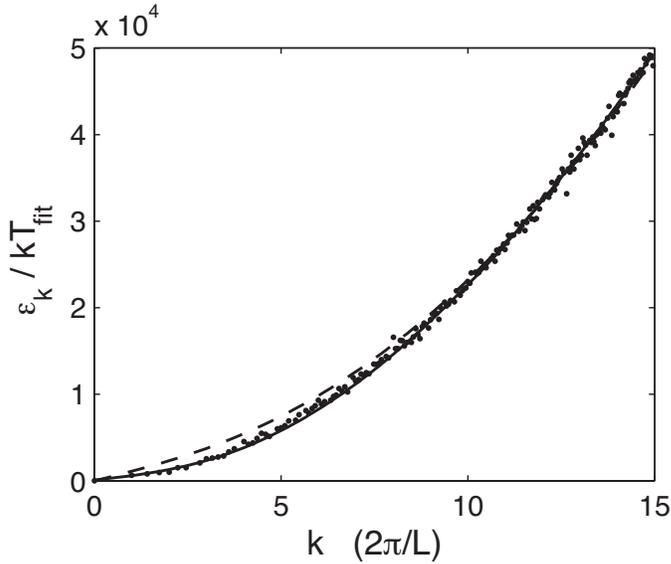}
\caption{Fits of the simulation quasiparticle  population data to dispersion
relations.  The dots are a plot of $(1/N_k- 1/ N_0
)$,  the solid curve is for dispersion relation predicted by second
order  theory, and the dashed curve is the dispersion relation predicted by
Bogoliubov theory. 
Simulation parameters: $u = 2000L^3\epsilon_L/N_{\rC}$, ${E}_{\rC} = 4000N_{\rC}\epsilon_L$,
and $ N_0 /N_{\rC} = 0.279$, where the unit of energy is $\epsilon_L=\hbar^2/2mL^2$, and the unit of temperature is $T_0=N_{\rC}\epsilon_L/k_B$.
Reproduced from \cite{Davis2002a} \APSCR{2002}.   
\label{fig:full_theory}}
\end{figure}
In a sufficiently weakly interacting Bose gas, the Hamiltonian for the system can be approximately diagonalized by a transformation to the Bogoliubov quasi-particle basis.  For the uniform gas, the interaction term only mixes modes of opposite momentum, and  the transformation from single-particle modes to Bogoliubov quasi-particles of a well-defined momentum $\mathbf{k}$ depends only on the product of the interaction strength $u$ and the condensate number $N_0$.  This is a quantity that can be determined from the PGPE calculations, and so the individual classical fields can be projected onto the Bogoliubov quasiparticle basis, and the time-averaged quasiparticle occupations $ N_{\mathbf{k}}$ can be accurately determined.

When the Bogoliubov quasiparticles form a good basis, we expecte that at thermal equilibrium the \CF\ method will result in the mean quasi-particle occupations being given by the equipartition relation \eref{nEQ}. 
If we define the condensate eigenvalue as $\varepsilon_0 \equiv \lambda$, and require that the condensate occupation also be given by the equipartition relation
\begin{equation}
N_0 = \frac{k_B T}{\lambda - \mu},
\end{equation}
then we can solve for the thermodynamic chemical potential $\mu = \lambda - k_B T/N_0$.\footnote{We note that the familiar result of $\lambda = \mu$  only holds in the thermodynamic limit, and for finite size systems it is often important to ensure that the chemical potential $\mu$ and condensate eigenvalue $\lambda$ are distinct.}
By substituting this result into \eref{nEQ} and rearranging we find
\begin{equation} 
\frac{{\varepsilon}_{k} - {\lambda}}{k_B{T}} = \left(\frac{1}{ N_{ k} } -
\frac{1}{ N_0 }\right), 
\label{eqn:fit} 
\end{equation} 
where the numerator of the left hand side is the quasi-particle energy \emph{relative} to the condensate.
This suggests a prescription for determining the temperature of a simulation.
The right hand side can be accurately measured by ergodic averaging in \CF\ simulations, and the left hand side can be evaluated using theoretical predictions of the spectrum ($\varepsilon_k-\lambda$) with the temperature forming a single fit parameter.
  This procedure was developed by Davis \etal \cite{Davis2001a,Davis2002a} before the application of the method of Rugh for determining temperature as described above in \sref{Sec:PGPEtemperature}.  We also note the non-projected classical field study of Brewczyk \etal \cite{Brewczyk2004a}.

\paragraph{Bogoliubov spectrum} 
In the limit of large condensate fraction $ N_0 /N_{\rC} \sim 1$, we
expect the Bogoliubov transformation to provide an accurate description of the system, with the dispersion relation
\begin{equation}
\varepsilon_k - \lambda = \left[\left(\frac{\hbar^2 k^2}{2 m}\right)^2
+ (c \hbar k)^2\right]^{1/2},
\label{eqn:bog_disperse}
\end{equation}
where $c = \sqrt{N_0 u / mL^3}$ is the speed of sound.

\paragraph{Second order spectrum} 
For sufficiently large
interaction strengths and temperatures, the cubic and quartic terms of the many-body Hamiltonian
that are neglected in the Bogoliubov transformation become important.  In
Ref.~\cite{Morgan2000a} Morgan develops a consistent extension of the Bogoliubov theory
to second order that leads to a gapless excitation spectrum.  This theory
treats the cubic and quartic terms of the Hamiltonian using perturbation theory
in the Bogoliubov quasiparticle basis, and results in energy-shifts of the
excitations away from the Bogoliubov predictions of Eq.~(\ref{eqn:bog_disperse}).

 The results in \fref{fig:full_theory} clearly show that second order theory provides a better description of mode occupations than Bogoliubov theory.  Other results in \cite{Davis2002a} show that as the interaction strength increases, initially better agreement with second order theory is observed, until the validity conditions of that theory are eventually surpassed. 
 In reference \cite{Davis2003a} the temperature, as determined from the spectral fitting procedure \cite{Davis2002a}, was shown to be in good agreement the Rugh (dynamical) temperature (discussed in \sref{Sec:PGPEtemperature}) in the regimes where spectral fitting was valid.

\subsubsection{Shift of $T_c$ for the uniform Bose gas}\label{shiftTcuniform}

\begin{figure}
\centering{\includegraphics{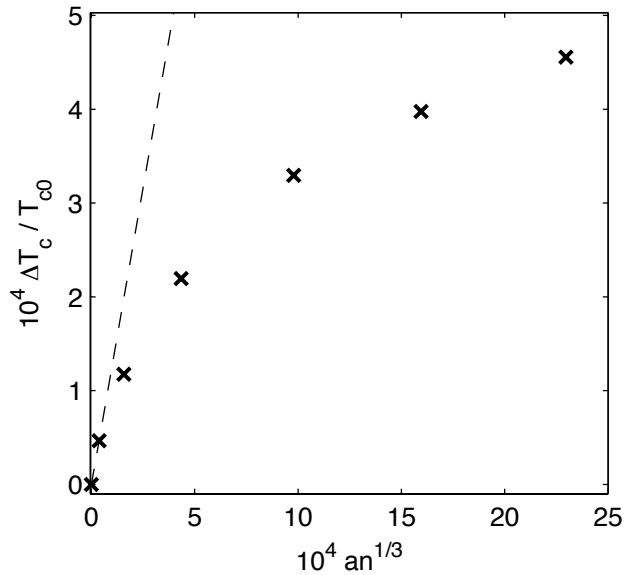}}
\caption{Shift in the critical temperature of a uniform Bose gas with interaction strength determined from
  PGPE simulations with  $N_{\rC} = 10^{10}$ for zero scattering length.  The dashed line
is a linear fit to the first two data
points and this has a slope of $ 1.3 \pm 0.4$. Reproduced from \cite{Davis2003a} \APSCR{2003}.}
\label{fig:uniTcshift}
\end{figure}
The shift in critical temperature $T_c$ with interaction
strength for the homogeneous Bose gas has been the subject of numerous studies and debate for almost fifty years since the first calculations of Lee and Yang
\cite{Lee1957a,Lee1958a}.  While there is a finite shift to the
chemical potential in mean-field (MF) theory, the shift of the critical temperature
is zero \cite{Baym2001a}. The leading order effect is due to long-wavelength  critical
fluctuations and is inherently non-perturbative. Using effective field
theory  it was determined that the shift is 
\begin{equation}
\Delta T_c/T_{c0} = c a n^{1/3},
\end{equation}
where $n$ is the particle number density, $a$ is the s-wave scattering
length, and $c$ is a constant of order unity \cite{Baym1999a}.  Until recently
results for the value of $c$ disagreed by an order of magnitude and even sign,
as summarised in Fig.~1 of \cite{Arnold2001c}.  However, two calculations
performed using lattice Monte Carlo 
have settled the matter, and
confirm that the shift is in the positive direction with combined estimate of $c
\approx 1.31 \pm 0.02$ \cite{Kashurnikov2001a,Arnold2001c}.  A number of recent
improved results broadly agree, and useful discussions are provided
by Andersen \cite{Andersen2004a} and Holzmann \emph{et
al.}~\cite{Holzmann2004a}.

Here, we briefly describe the procedure used by Davis \etal \cite{Davis2003a} 
  to calculate a value for $c$ using the (uniform gas) PGPE. 
 
\begin{enumerate}
\item For a given nonlinearity (i.e. scattering length) a randomized initial state of definite energy $E_{\rC}$ is evolved with the PGPE, and the temperature is determined by using the methods described earlier in \sref{Sec:PGPEtemperature}.  
\item  As the initial state energy is varied, the critical point is identified as where the Binder cumulant, $C_b=\langle N_0^2\rangle/\langle N_0\rangle^2$, with $N_0$ the population of the zero-momentum condensate mode. This Binder cumulant characterizes condensate  number fluctuations, and takes the universal value of $C_b^{\rm{crit}}=1.2430$ at the transition.  
\item The shift in the critical temperature is calculated as a function of interaction strength, parameterized by the s-wave scattering length.
\end{enumerate}


By fitting a straight line to the first two points as illustrated in \fref{fig:uniTcshift},
we get an estimate for the coefficient
\begin{equation}
c = 1.3 \pm 0.4,
\end{equation}
where the error specified is due to the uncertainty in the value of $T_c$ for the
data point. This agrees with the value determined in
Refs.~\cite{Arnold2001c,Kashurnikov2001a}.

\subsection{Applications to the trapped Bose gas}\label{sAppstrappedBG}

\subsubsection{Shift in $T_c$ for a trapped Bose gas: Comparison with experiment}\label{Tctrapped}

The behaviour of $T_c$ for the
harmonically confined Bose gas is drastically different from the uniform gas.   There
is a shift in $T_c$ due to finite size effects arising from the fact that the system is not in the thermodynamic limit \cite{Grossmann1995a}, and a first-order interaction shift  due
to mean-field effects  \cite{Giorgini1996a}.  

For a typical BEC experiment, the critical temperature deviates from the ideal
gas result only by a few percent.  Thermometry of Bose gases at this level
of accuracy is challenging: However, in 2004
Gerbier \etal~reported precise measurements of the critical temperature for a range
of atom numbers  \cite{Gerbier2004b}.
\begin{figure}
\centering{\includegraphics[width=4in]{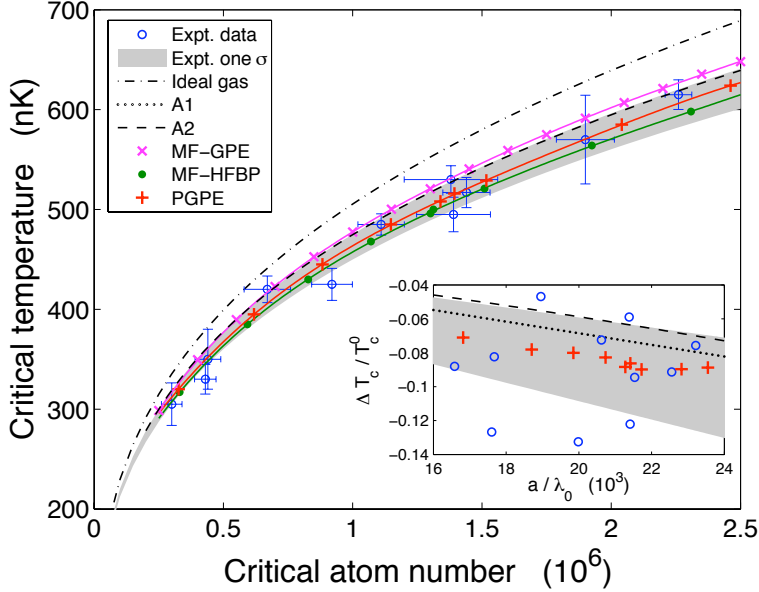}}
\caption{
Comparison of theoretical calculations
with experiment. The main figure plots $T_c$ vs $N_0$, whereas the
inset plots the shift of $T_c$ against the relevant small parameter
$a/\lambda_0$. Experimental results: data (open circles), one fit (gray
area). Theoretical results for $T_c$: ideal gas (dot-dashed line),
A1 (dotted line), A2 (dashed line), MF-GPE (crosses), MF-HFBP
(dots), PGPE (pluses). Solid lines through the data points
are polynomial fits. A1 is not shown in the main figure for clarity. The total number of atoms at the critical point is $N = 4.0 \times 10^6$ and $\lambda_0=h/\sqrt{2\pi mk_BT}$.
fits to the data. Reproduced from \cite{Davis2006a} \APSCR{2006}.
\label{fig:measureN}}
\end{figure}

Davis \etal  \cite{Davis2006a} used those measurements to make the first quantitative comparison of the PGPE formalism with experiment and other theories, which are summarized in \fref{fig:measureN}. The various other theories appearing in \fref{fig:measureN} are:

 {\bf  A1:} 
This is the meanfield analytic estimate as calculated by Giorgini 
\emph{et al.}~\cite{Giorgini1996a}, and was compared with the experimental data
in \cite{Gerbier2004a}. 

 {\bf  A2: }
This is the  analytic estimate as calculated by Arnold 
\emph{et al.}~\cite{Arnold2001a}, which includes next order fluctuation results, however is only strictly valid in \emph{broad} traps. 

 {\bf  MF-GPE:}
The GPE is solved
numerically using a variational Gaussian ansatz, and the thermal cloud
calculated using a semi-classical approximation \cite{Giorgini1996a}.  At each temperature the condensate and non-condensate are determined self-consistently with a fixed
number of particles, and the critical temperature is taken to be where the condensate fraction decreases to zero. This approach differs from theory A1 because it avoids using perturbation theory around the  ideal (saturated) gas density profile to estimate interaction effects.

 {\bf  MF-HFBP:} Here the condensate fraction is fixed, and the
temperature determined that gives an appropriate self-consistent condensate mode and
thermal density (The full Bogoliubov modes are used and the semiclassical approximation is avoided).  We have verified the results are unchanged for
equipartition or Bose-Einstein statistics.

The PGPE calculations appear to provide the best theoretical description of experiment, however, error bars in the experimental results are not yet small enough to definitively discriminate between results.

We also note that the Hartree-Fock-Bogoliubov Popov calculations (MF-HFBP) 
use the same procedure as in PGPE calculation to determine the
critical point, the above cutoff density and the total atom number, so that the difference between this \emph{best} meanfield calculation and the PGPE results is due to beyond meanfield fluctuation effects. This suggests if experimental accuracy in thermometry could improve by an order of magnitude, then effects of fluctuations on the critical temperature in this system could be directly investigated.

\subsubsection{Quasi-2D Bose gas}\label{sec:q2D}
The phenomena of superconductivity and superfluidity are striking manifestations of the role played by quantum statistics at low temperatures. Altering the temperature or effective dimensionality may radically change the physical properties of quantum degenerate systems. A well known consequence is that in contrast to 3D, there is no BEC for a homogeneous 2D ideal-gas in the thermodynamic limit at any finite temperature \cite{Mermin1966a,Hohenberg1967a}. Nevertheless, the Berezinskii-Kosterlitz-Thouless (BKT) vortex binding-unbinding phase transition allows the emergence of superfluidity in 2D systems \cite{Berezinskii1971a,Kosterlitz1973a}.   Although weak particle interactions alone are not sufficient to change the situation, an external confinement modifies the density of states in such a manner that the critical point of BEC is elevated to a finite temperature \cite{Bagnato1991a}. Therefore it is not certain \emph{a priori} whether the transformation from normal to superfluid in such systems is a BEC or BKT-type transition.

\begin{figure}[!th]
\centering{
\includegraphics[width=3in]{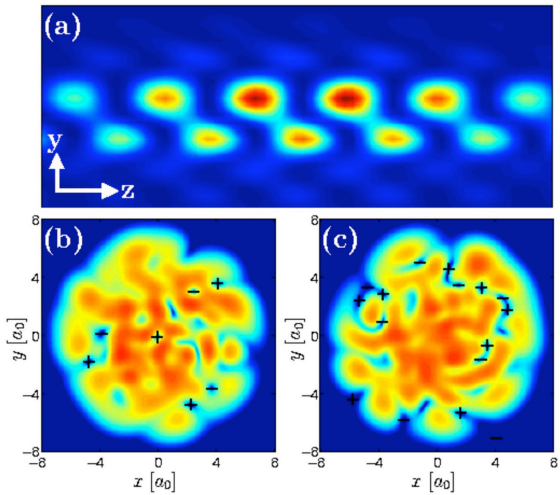}}
\caption{Interference pattern (a) produced by two independent \CF s (b) and (c) at temperature $T=0.86\;T_0$. The relevant particle numbers are $N_{\rm cl}=3.0\times 10^3$ and $N=4.0\times 10^4$. The ``zipper" structure in (a) is the telltale signature of the phase singularity associated with the central vortex in (b). The locations of vortices and antivortices are marked by $+$ and $-$ signs, respectively. Reproduced from \cite{Simula2006a} \APSCR{2006}.
\label{FigBKT}}
\end{figure}

This properties of the finite temperature trapped 2D system have proven difficult to analyze. Strong fluctuations mean that meanfield approaches are inapplicable, however since these fluctuations are classical in nature the PGPE approach is appropriate. Indeed, early studies of the uniform 2D Bose gas were performed by a \CF\  method, but sampled using Monte Carlo techniques \cite{Prokofev2001a,Prokofev2002a}. More recently Giorgetti \etal\ \cite{Giorgetti2007a} have developed an accurate semi-classical approach for simulating this system.

Simula \etal \cite{Simula2006a}  used PGPE   simulations of quasi-2D Bose-fields   to characterize the low temperature phases for such systems over a wide parameter range. These simulations show the emergence of thermally activated vortices (see \fref{FigBKT}(b)-(c)), their influence on interference patterns (see \fref{FigBKT}(a)), and provide strong evidence supporting the view that the BKT-type phase was observed in the recent experiment by Stock \emph{et al.} \cite{Stock2005a}. More recent work with  PGPE simulations have characterized correlation and collective mode properties of the system to quantify the BKT transition point \cite{Simula2008a,Bisset2008a}, and find good qualitative agreement with recent quantum Monte Carlo  simulations \cite{Holzmann2008a}.

\subsubsection{Two point correlation functions}\label{SEC:PGPEApp2ptcorrelns}
Recent experimental developments in ultra-cold gases \cite{Schellekens2005a,Ottl2005a,Jeltes2006a,Folling2005a,Greiner2005a,Rom2006a} have allowed atomic correlation measurements that are analogous to the photon correlations observed in the landmark experiments of Hanbury-Brown and Twiss \cite{hbt}. Such correlations are of particular interest in systems where many-body interactions are important \cite{Altman2004a,Toth2008a}, and  in the region of the phase transition, where critical exponents can be measured \cite{Donner2007a}.

The PGPE description is valid in this regime and can be used to calculate these correlations, and assess beyond meanfield effects (c.f \cite{Dodd1997b,Naraschewski1999a}). We now summarize an approach for calculating these correlations within the PGPE formalism that has been developed by Bezett \etal \cite{Bezett 2008a}.

The quantities of interest are the normally ordered first-order correlation function, $G^{(1)}(\x,\xd) \equiv\langle \hat{\psi}^\dagger\xa\hat{\psi}\xad\rangle$ (also known as the one-body density matrix), and second order correlation function, $G^{(2)}(\x,\xd) \equiv\langle \hat{\psi}^\dagger\xa\hat{\psi}^\dagger\xad\hat{\psi}\xad\hat{\psi}\xa\rangle$. From these functions  first- and second-order observables can be directly obtained, such as  the density-density correlation function.
 
Breaking the full quantum field into \CF\ and incoherent parts, the correlation functions can be written as
\begin{eqnarray}
G^{(1)}(\x,\xd) &=& G^{(1)}_{\rC}(\x,\xd) + G^{(1)}_{\rI}(\x,\xd),\label{G1full2}\\
G^{(2)}(\x,\xd) &=& G^{(2)}_{\rC}(\x,\xd) + G^{(2)}_{\rI}(\x,\xd) + 2G^{(1)}_{\rI} (\x,\xd)G^{(1)}_{\rC}(\x,\xd)\nonumber\\
&& + n_{\rI}(\x) n_{\rC}(\xd) + n_{\rI}(\xd) n_{\rC}(\x),\label{G2full2}
\end{eqnarray}
where $G^{(1)}_j (\x,\xd)= \langle\hat{\psi}_j^{\dagger}(\xd)\hat{\psi}_j(\x)\rangle$ and $G^{(2)}_j (\x,\xd)= \langle\hat{\psi}_j^{\dagger}(\xd)\hat{\psi}_j^{\dagger}(\x)\hat{\psi}_j(\x)\hat{\psi}_j(\xd)\rangle$
with $j=\{{\rI},{\rC}\}$ for the incoherent and classical regions respectively, and we have neglected any correlations between the \CF\ and incoherent regions.

 \begin{figure}
\centering{\includegraphics[width=5.2in, keepaspectratio]{\suffix{Fig17}}}
 \caption{\label{fig:2ptcorrelns} (color online) Two point position space correlation functions  of a harmonically trapped Bose gas of $N=3\times10^5$ $^{87}$Rb atoms at $T=159nK$. Other parameters:  $N_{0}=3540$,   $\ecut= 36\hbar\omega_x$, with $\{\omega_x,\omega_y,\omega_z\}=2\pi\,\{1,1,\sqrt{8}\}\times40\,s^{-1}$. Reproduced from \cite{Bezett 2008a} \APSCR{2008}.
 }
\end{figure}

While the \CF\  correlations can be evaluated using the approach detailed in \sref{SEC:PGPEdensitycondfrac}, for the incoherent region we can make use of the  one-particle Wigner function given in \eeref{eq:Wig3D}. Appropriately transforming the Wigner function we obtain the first order correlation function, i.e.
\begin{equation}
G^{(1)}_{\rI} (\x,\xd) = \int_{\Omega_{\rI}} {d^3\mathbf{k}} \,e^{-i\mathbf{k}\cdot(\x -\xd)}\;F_{\rI}\left(\frac{\x+\xd}{2},\mathbf{k}\right).\label{G1wig}
\end{equation}
As the $F_{\rI}$ Wigner description of the incoherent region is Gaussian, we can easily obtain the second order correlation function $G^{(2)}_{\rI}(\x,\xd) = n_{\rI}(\x)n_{\rI}(\xd) + |G^{(1)}_{\rI}(\x,\xd)|^2.$

Figure \ref{fig:2ptcorrelns} illustrates correlation functions for a trapped Bose gas at $T\approx T_c$. The results shown are for the case of two points along the $x$-axis of the system, and in \fref{fig:2ptcorrelns}(b) and (c) the normalized correlation functions, defined as $g^{(1)}(x,x')=G^{(1)}(x,x')/\sqrt{n(x)n(x')}$ and $g^{(2)}(x,x')=G^{(2)}(x,x')/{n(x)n(x')}$, are shown. The broad feature apparent in \fref{fig:2ptcorrelns}(a) and (b)  is the off diagonal long range order, arising from the emerging condensate in this system. The diagonal ridge is due to short range thermal correlations. We also note that  Holzmann \etal\ \cite{Holzmann1999a} have used a
quantum Monte Carlo method to obtain the pair distribution function for a trapped Bose gas.

\subsection{Applications of non-projected classical fields at finite temperature}
\label{sec:other_cfields}
As well as the work of the current authors on quantitative projected c-field techniques, there have been a number of other studies of finite temperature properties of degenerate Bose gases.  For completeness, in the final part of this section we briefly describe the results that have been obtained.

\subsubsection{Homogenous gas}\label{otherhomoBG}

In one of the first papers on classical fields, G\'{oral} \etal~\cite{Goral2001b} demonstrated the thermalization of a homogeneous multimode Bose gas in a similar manner to Davis \etal~\cite{Davis2001b,Davis2002a}.  They expanded the equation of motion for the c-field in terms of mode coefficients, and calculated the nonlinear terms by performing the appropriate summations, and implicitly correctly applied a projection operation.  
Brewczyk~\etal~\cite{Brewczyk2004a}  performed a Bogoliubov analysis of homogeneous Bose gas using a GPE classical field description. Zawitkowski~\etal~\cite{Zawitkowski2004a} attempt to describe not just the classical modes but the \rI\ region modes using only the GPE by fixing a grid cutoff such that the condensate fraction and temperature agree with that for the ideal Bose gas.  Doing this eliminates any possibility of describing e.g. describing the effects of interactions on the transition temperature, and includes a large number of modes in the problem that should not be described classically.  It should be clear from this review that the current authors strongly disagree with this approach.

Leadbeater \etal~\cite{Leadbeater2003a} studied the effect of condensate depletion on the critical velocity when an object is dragged through a superfluid. On a related note, Zawitkowski~\etal~\cite{Zawitkowski2006a} performed an interesting study of placing a homogenous moving condensate in a static thermal cloud, and investigated the decay of the superflow as a function of velocity and temperature.  Unfortunately it seems that the lack of projection caused some numerical issues in this work, such as the violation of momentum conservation.  

Witkowska~\etal~\cite{Witkowska2007a} related the dynamics of a nonlinear string to the weakly interacting Bose gas.  Nunnenkamp~\etal~\cite{Nunnenkamp2007a} made a comparison of three versions of a classical field theory for a one-dimenional Bose gas on a ring.  They found that an exact solution in the high temperature limit of a transfer integral method agreed well with both a molecular dynamics approach, and classical field simulations of the GPE.  Recently Sinatra~\etal~\cite{Sinatra2007a} found nondiffusive phase spreading of a 3D homogenous {B}ose-{E}instein condensate at finite temperature.

Connaughton and co-workers have studied condensate formation in the homogeneous gas using a GPE model \cite{Connaughton2005a,Josserand2006a}.  In an interesting application related to classical fields, Picozzi and co-workers have investigated the dynamics of equilibration in incoherent nonlinear optics both theoretically and experimentally.  See, for example, references \cite{Picozzi2007a,Pitois2006a,Lagrange2007a,Picozzi2005a}.  

\subsubsection{Trapped gas}\label{othertrappedBG}

G\'{oral} \etal~\cite{Goral2002a} were  the first to apply the condensation criterion of Penrose and Onsager \cite{Penrose1956a} to a classical field.  They solved a non-projected GPE for the trapped Bose gas at finite temperature, and developed some estimates of thermodynamics properties of the system.  Schmidt \etal~\cite{Schmidt2003a} applied the same simulation technique to investigate the decay of an off-centre vortex in a harmonic trap at finite temperature.
Recently Gawryluk~\etal~\cite{Gawryluk2007a} seeded a trapped $F=1$ spinor condensate with thermal fluctuations and studied the resulting spin dynamics.

The effect of thermal fluctuations in Bose gases is more significant in low dimensions.  Kadio~\etal~\cite{Kadio2005a} studied the coherence properties in a quasi-1D trapped Bose gas, and analysed the effects of phase fluctuations in a 3D elongated trap.  
Mebrahtu~\etal~\cite{Mebrahtu2006a} have analyzed coherence effects in the spatial splitting of a quasi-1D Bose-Einstein condensate and its subsequent merging at finite temperature.

\subsubsection{Superfluid turbulence}\label{otherSFturbulence}

Finally, we mention work in the area of superfluid turbulence, which attempts to describe the formation of tangles of vortices in the homogeneous superfluid transition and the subsequent relaxation to global phase coherence.  Many of the simulations of these systems make use of the GPE to describe finite temperature non-equilibrium dynamics, and hence are directly related to classical field techniques.  

Of particular interest to this review is
the work of Berloff and Svistunov~\cite{Berloff2002a}, who studied condensation from a strongly nonequilibrium state in a 3D homogeneous system using the GPE.  Their main interest was in the decay of superfluid turbulence, and the establishment of phase coherence, validating the scenario of superfluid growth as earlier described by Svistunov and Kagan  \cite{Svistunov1991,Kagan1992,Kagan1994,Kagan1997}.  Berloff subsequently studied the interactions of vortices and solitary waves and their role in the decay of superfluid turbulence~\cite{Berloff2004a}, as well as turbulence in a two-component system~\cite{Berloff2006a}.  Recently Berloff and Youd studied the decay of vortex rings in a homogeneous superfluid at finite temperature~\cite{Berloff2007a}.
In reference \cite{Kobayashi2005a}
 Kobayashi and Tsubota simulate a GPE with a dissipation at short wavelengths and obtain an energy spectrum consistent with the Kolmogorov law.

\section{Applications of the truncated Wigner PGPE to quantum matter wave dynamics}\label{sec:TWPGPEapps}

As more experimental investigations have begun probing beyond mean field quantum dynamics in BECs, theoretical applications have begun to explore the role of
thermal and quantum fluctuations using the truncated Wigner method. 
Here we give a brief survey of the background and recent developments of this method.

\paragraph{Background} The truncated Wigner approximation was introduced by Robert Graham in 1973~\cite{Graham1973} and has met with wide application in the field of laser physics. A precursor of work on trapped Bose gases was carried out by Carter \etal~\cite{Carter1987a} who applied phase space methods to the simulation of the quantum optical nonlinear Schr\"{o}dinger equation. The theoretical formulation and applications to Bose gas dynamics began with the work of Steel \emph{et al.} \cite{Steel1998a} who developed phase space techniques for atomic Bose fields and applied them to simulating the time evolution of a one dimensional homogeneous Bose gas. The truncted Wigner method was compared with the functional positive-P phase space method~\cite{Drummond1980a} in calculations of the first order coherence function $g^{(1)}(t)\equiv\langle
\hat{a}_0^\dag(t)\hat{a}_0(0)\rangle/\langle
\hat{a}_0^\dag(0)\hat{a}_0(0)\rangle$ for the condensate operator
$\hat{a}_0$. Different initial states of the condensate were sampled including the coherent state and the Bogoliubov state. A general conclusion of this work, which provides a reliable guide, is that the positive-P method, while exact, is unstable except for very short simulation times, whereas the truncated Wigner method, while approximate, is stable. Many subsequent works have considered aspects of the validity of the truncated Wigner method and its applications to dynamical Bose gases using both full phase space approaches, and the classical field method based on analysis of single trajectories.

A distinction between the formulation of \cite{Steel1998a} and the \CF\ description presented in \sref{SEC:Formalism}, is the emphasis placed on projection into a low energy subspace, both formally and numerically. In the TWPGPE formulation the projection operator imposes a formal UV-cutoff which allows a measure of control over the sometimes spurious effects of vacuum noise arising in the truncated Wigner method.



\begin{figure}[t]
\begin{center}
\includegraphics[height=7cm]{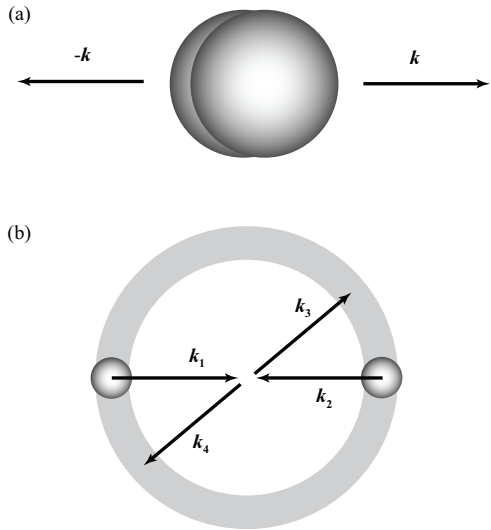}
\end{center}
\caption{ \label{Fig:BECcollisionSchematic} Schematic of condensate
collision scenario of \cite{Norrie2005a}. (a) Position space
densities of initially overlapping, counter-propagating condensate
wavepackets. (b) Momentum space representation of possible energy and
momentum conserving collisions between atoms in the two condensates
($\mathbf{k_1}, \mathbf{k_2}$) onto the allowed spherical scattering
halo ($\mathbf{k_3}, \mathbf{k_4}$), indicated by the gray annular
region in the collision plane.}
\end{figure}
\begin{figure}[htb]
\begin{center}
\includegraphics[height=12cm]{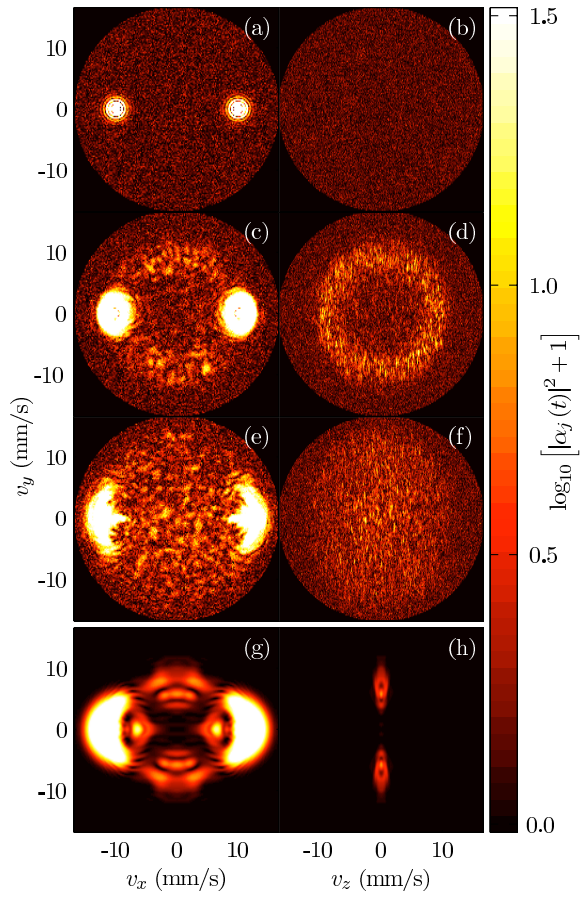}
\end{center}
\caption{\label{Fig:BECcollisions} (Color online) (a)-(f) Velocity
mode populations on the planes $v_z=0$ (left) and $v_x=0$ (right) for
the condensate collision described in the text at $t=0$ (top),
$t=0.5\;{\rm ms}$, and $t=2.0\;{\rm ms}$ (bottom). The spherical
momentum cutoff is clearly visible in the upper plots due to the
presence of quantum flucuations. (g)-(h) Mode populations at
$t=2.0\;{\rm ms}$ for an identical collision excluding vacuum noise.
Reproduced from \cite{Norrie2005a} \APSCR{2005}.
}
\end{figure}
\subsection{Condensate collisions in free space}\label{sec:BECcollisions}
The Bragg scattering of a condensate into a  superposition of states of momentum $\mathbf{0}$ and $2\hbar\mathbf{k}$ creates a well characterized non-equilibirum initial condition that is easily produced in experiments \cite{Kozuma99a,Stenger1999b,Blakie2002a,Steinhauer2003a}. This scenario is shown schematically in the centre-of-mass (COM) frame (in position space) in \fref{Fig:BECcollisionSchematic}a, where the original and scattered wavepackets move away from each other. In the subsequent dynamics, but while the two wave packets still overlap in position space,  pairs of atoms are scattered onto a spherical shell in momentum space (see \fref{Fig:BECcollisionSchematic}b). 
This scattering, often referred to as an S-wave halo, is clearly seen in experiments \cite{Chikkatur2000a,Perrin2007a},  but is absent in a GPE description. 

The truncated Wigner method was first used to model this process by Norrie \etal~\cite{Norrie2005a,Norrie2006a}.
Beginning with a condensate with mode function $\xi_0\xa$, Bragg scattering was assumed to scatter half the condensate, resulting in the  superposition of two wavepackets with momenta $\pm\hbar\mathbf{k}$ (in the COM frame) \cite{Blakie2000a}, i.e.,
\begin{equation}
\psi_0\xa=\frac{\xi_0(\x)}{\sqrt{2}}\left[e^{i\mbf{k}\cdot \x}+e^{-i\mbf{k}\cdot \x}\right].
\end{equation}  
The full initial condition (see \fref{Fig:BECcollisions}a-b)  was sampled by adding vacuum noise to modes orthogonal to $\psi_0$ (see \eeref{psivacformsample}).
In the truncated Wigner simulation modes on a spherical shell of radius $v\approx10$ mm/s in velocity space are seen to grow, while the initial wavepackets are situated at the poles of this sphere (see \fref{Fig:BECcollisions}c-d). The importance of vacuum fluctuations is clear: they seed the growth of the halo modes, thus \emph{mimicking} spontaneous processes. It was found that after the halo first develops (see \fref{Fig:BECcollisions}e-f), the stimulated evolution of these scattered modes leads to turbulent dynamics.
In contrast, no such spherical shell is seen to develop in the GPE simulations (see \fref{Fig:BECcollisions}g-h). 
 \begin{figure}[t]
\begin{center}
\includegraphics[width=7cm]{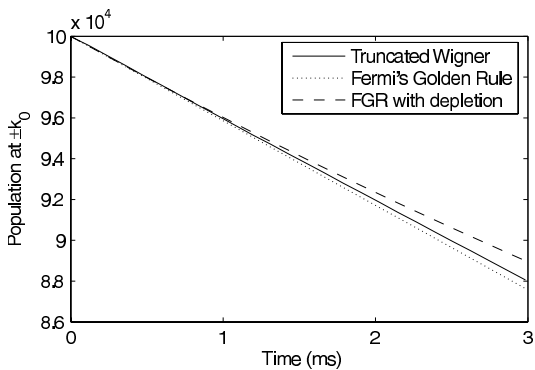}
\end{center}
\caption{\label{Fig:FGRTWA} Reduction in condensate population during a condensate collision.  Wigner simulations (solid), a linear fit to the
rate given by FermiÕs golden rule (dotted) and the solution to the
differential equation \eeref{FGR} that includes condensate depletion (dashed).
Reproduced from \cite{Ferris2008a} \APSCR{2008}.
}
\end{figure}

 We also note the work of Deuar \etal \cite{Deuar2007a} which used a positive-$P$ method (see \cite{Corney1999a,Corney2004a,Deuar2002a}) to simulate BEC collisions. The positive-$P$ method is a phase space approach, like the Wigner method, but does not require approximation (i.e. the \emph{truncation} of 3rd order derivates) to arrive at a set of stochastic equations for any Hamiltonian which is at most quartic in operators. The application to condensate collisions represents a significant success of the method for modeling real time dynamics of atomic Bose fields. While providing an exact mapping and being used widely for treating quantum optical systems (where dissipation is usually significant and interactions weaker), for pure Hamiltonian evolution the method suffers from stability problems limiting its use to short simulations and low density systems. Comparing with truncated Wigner simulations, Deuar \etal\ found discrepancies between the two approaches for the early time dynamics of initially unoccupied modes into which atoms were scattered. While directly probing such a discrepancy in an experiment would be difficult, this result emphasizes the need for care in interpreting TWPGPE results for lowly occupied modes.

\paragraph{Condensate depletion} \note{homogeneous gas}
A comparison between the truncated Wigner method and Fermi's second
golden rule was made by Ferris  \etal~\cite{Ferris2008a} for the case of colliding condensates in the uniform system. That study compared the early time depletion of the colliding condensates due to
spontaneous scattering with the Fermi golden rule prediction
\EQ{
\frac{dN_0}{dt}=\frac{u^2m|\mathbf{k}|}{2\pi\hbar^3V}N^2_0,\label{FGR}
}
where $N_0$ is the number of remaining (unscattered) condensate atoms, and $V$ is the system volume.
The condensate population from the truncated Wigner simulation is shown in \fref{Fig:FGRTWA}, where the good agreement with the Fermi golden rule estimates is evident at short times. The growing discrepancy at long times arises from  the depletion of the condensate and the stimulated dynamics of the scattered modes.
 
\subsection{Truncated Wigner treatment of three-body loss}\label{s3Bloss}
The three-body loss process is an inherently non-diffusive process in phase space (the generalized Fokker-Planck equation contains derivatives beyond second order in the phase space variables), and thus does not admit an exact formulation in terms of stochastic differential equations for any pseudo-probability distribution. 

An important consideration must be borne in mind at this point: \emph{practical} application of stochastic methods is not feasible for problems containing higher than second order derivatives for phase space variables. Thus, a more general statement of the TWA is that it should include all terms up to second order in phase-space variable derivatives, subject to the usual validity conditions of weak interactions and high mode occupation. This is the approach taken in the treatment of three-body loss.

The basic technical extension beyond the standard TWA presented in \sref{truncatedWig} is an additional stochastic term which models the diffusive effects of inelastic loss. Thus, while elastic two body collisions do not generate a stochastic equation of motion within TWA (second order terms are identically zero), three body inelastic collisions introduce a stochastic element to the evolution. 

The three-body loss master equation for the system density operator $\hat{\rho}$, which has been rigorously derived by Jack~\cite{Jack2002a,Jack2003a}, takes the form
\begin{equation}
\frac{\partial \hat{\rho}}{\partial t}\Big|_{3}=\frac{K_3}{6}\myint{\mbf{x}}\left\{2\hat{\psi}(\x)^3\hat{\rho}\hat{\psi}^\dag(\x)^3-\hat{\psi}^\dag(\x)^3\hat{\psi}(\x)^3\hat{\rho}-\hat{\rho}\hat{\psi}^\dag(\x)^3\hat{\psi}(\x)^3\right\},\label{3beom}
\end{equation}
which generates the time evolution for the total atom number 
\begin{equation}
\frac{d N}{dt}=-K_3\myint{\x}g_3(\x)n(\x)^3,\label{EQ3blossrate}
\end{equation}
where 
\begin{equation}
g_3(\x)=\frac{\langle \hat{\psi}^\dag(\x)^3\hat{\psi}(\x)^3\rangle}{\langle \hat{\psi}^\dag(\x) \hat{\psi}(\x)\rangle^3}.
\end{equation}
Within the truncated Wigner approximation \eeref{3beom} leads to a  stochastic differential equation for the \CF\ $\cf\xa$
\begin{equation}
d\cf(\x)=\PC\left\{-\frac{K_3}{2}|\cf(\x)|^4\cf(\x)dt+\sqrt{\frac{3K_3}{2}}|\cf(\x)|^2dW_3(\x,t)\right\},
\end{equation} 
 \cite{Norrie2006b}  (in addition to the terms already in the TWPGPE \eref{TWAPGPE}) where the noise term is given by\footnote{Note that while this SDE gives an exact correspondence with the TWA Fokker-Planck equation derived from \eeref{3beom}, the noise must be computed on a dense quadrature grid to ensure that the SDE preserves the correspondence (see \aref{sec:stochMapping} for a discussion of quadrature methods and SDEs).}
\begin{equation}
dW_3(\x,t)=\sum_{ n\in\rC} d\xi_{n}\phi_n\xa,
\end{equation}
with $d\xi_n(t)$ a complex Gaussian noise satisfying
\begin{eqnarray}
\overline{d\xi_{n}(t)d\xi_{n'}(t)}&=&0,\\
\overline{d\xi^*_{n}(t)d\xi_{n'}(t)}&=&\delta_{nn'}dt.
\end{eqnarray} 

\paragraph{Application to condensate collapse}
The three-body loss formalism was originally derived and applied to quantify the atom loses in condensate collisions \cite{Norrie2005a,Norrie2006a}, where it was shown to be a small effect. 
A regime where three-body corrections are more important is in the  description of the Bose-nova experiment performed by Donley \etal~\cite{Donley01a}. In that experiment a Feshbach resonance was used to suddenly change the scattering length from a value of $a\approx0$ (ideal stable BEC) to a negative value (i.e., attractive interactions), causing the system to collapse. During this process the condensate density increases significantly, and from \eeref{EQ3blossrate}, it is clear the three-body loss will become more important. This problem was studied with the TWPGPE approach by W\"{u}ster \etal \cite{Wuster2007a}, who assessed the effects of quantum and thermal fluctuations on the collapse process.  Where comparison was possible, the TWPGPE simulations of the collapse process agreed quantitatively with the results of Hartree-Fock-Bogoliubov theory, and both theories predicted slower collapse than observed in the experiment.

\subsection{Quantum reflection of a Bose-Einstein condensate}\label{sec:reflection}
Scott \etal \cite{Scott2006a,Scott2007a} used the GPE and the truncated Wigner approximation to model the collision of a BEC with an abrupt potential barrier, as studied experimentally by Pasquini \etal~\cite{Pasquini04a,Pasquini06a}. The system consists of a BEC held in a magnetic trap which is then accelerated at normal incidence toward a steep potential drop. Two regimes of behaviour for these reflections were characterized:
(i) For low approach velocities the BEC was observed to suffer disruption due to the interference of incident and reflected components. Most aspects of these slow collisions were adequately explained by the GPE, however for dense initial condensates the inclusion of vacuum fluctuations  was observed to have an appreciable effect on the dynamics through the formation of a scattering halo (see \fref{Fig:reflection}).
(ii) At higher velocities there is negligible disruption due to interference, so that the GPE results are relatively smooth. Studying this regime with the truncated Wigner approach, the inclusion of vacuum fluctuations cause a large scattering halo to develop.

In both regimes the experiments and the truncated Wigner results were found to be in quantitative agreement.

\begin{figure}[t]
\begin{center}
\includegraphics[width=8cm]{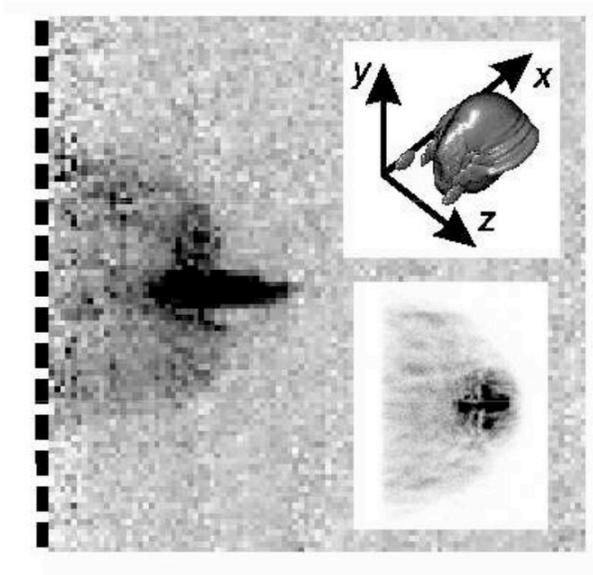}
\end{center}
\caption{Experimental absorption image of BEC for an impact velocity of $v_{x}=3.0$ mm s$^{-1}$ at $t=120$ ms, having reflected from the Casimir-Polder potential of a pillared silicon surface. The field of view is 500 $\mu$m, the vertical dashed line indicates the position of the barrier. Lower inset: corresponding simulated absorption image in the $y-x$ plane including quantum fluctuations for reflection from a barrier of height $V=1.67 \times 10^{-31}$ J. Upper inset: equivalent constant density surface excluding quantum fluctuations, axes are shown in the figure. The BEC in this simulation has a peak density of $5.2\times10^{12}$cm$^{-3}$ with its long axis perpendicular to barrier. Reproduced from \cite{Scott2006a} \APSCR{2006}. \label{Fig:reflection}}
\end{figure}

\subsection{Applications to optical lattices}\label{soptlatt}
There has been several studies of atom dynamics in 1D optical lattices using the truncated Wigner approach.

\paragraph{Dynamical instability of a BEC at the band edge of an optical lattice}\label{sec:bandEdge}
Ferris \etal~\cite{Ferris2008a} presented experimental results and truncated Wigner simulations of the dynamically unstable evolution of a BEC prepared in a band-edge state in a 1D optical lattice.  

The theoretical description was based on a full 3D simulation of the experimental system, and included degrees of freedom transverse to the lattice, and excited band states along the lattice direction. The large number of modes needed to accurately model the actual experimental system (i.e., in the combined lattice and weak harmonic potential), would violate the validity condition $N_\rC\gg M/2$  (see \sref{sec:validity}). To avoid this,  the theoretical model was simplified to a translationally invariant case, greatly reducing the number of basis modes required.
The truncated Wigner simulations showed that vacuum fluctuations have
 an important role in seeding the growth of unstable modes, leading to rapid depletion and heating of the condensate. Furthermore, the drastic modifications of energy and momentum conservation in the lattice were observed to have a substantial effect on the initial dynamics in the system, particularly the modes into which atoms were spontaneously scattered.
%

\paragraph{Quantum fluctuation effects on dipolar oscillations}
In \cite{Polkovnikov2004a} Polkovkinov \etal~considered the dipolar motion of a condensate displaced relative to the centre of the harmonic trap in a quasi-1D lattice. This study, conducted within the tight-binding Bose Hubbard description \cite{Jaksch1998a,Blakie2004a},  examined the nature of the damping, and how it is influenced by the quantum fluctuations included within the truncated Wigner treatment. An adiabatic mapping procedure was used to sample the initial Wigner distribution which allowed them to sample the ideal Bose gas in the harmonic and lattice potential \cite{Blakie2007d} as discussed in \sref{altsampling}. Their study presented evidence that there is a smooth crossover between the classical localization transition (overdamped oscillations that occur beyond a critical displacement) and the superfluid-to-insulator quantum phase transition in the limit of zero trap displacement.
Using a similar tight binding approach, the evolution of phase coherence in a deep quasi-1D lattice with a large number of atoms per lattice site was examined and compared with experiments by Tuchman \etal~\cite{Tuchman2006a}.

\paragraph{Number squeezing in 1D lattices}
 Ruostekoski and coworkers have also considered quasi-1D optical  lattice systems using the truncated Wigner approach, but used a  beyond tight binding description \cite{Isella2005a,Ruostekoski2005a}, which included excited band states.
 
 In reference \cite{Isella2005a} they consider the effect that lattice loading has on a quasi-1D gas initially prepared in a harmonic trap. The initial state was sampled using the Bogoliubov procedure \cite{Rey2003a,Wild2006a} (see \sref{subs:initStates}), and then evolved through a simulated lattice loading procedure. Coherence and number fluctuations were evaluated, and observed to be in qualitative agreement with the experiments of  Orzel \etal \cite{Orzel2001a}.

In later work Ruostekoski \etal~considered the quantum dynamics in shallow lattices \cite{Ruostekoski2005a}, and modelled experiments by Fertig \etal\cite{Fertig2005a} of  dipolar motion of a BEC   in an optical lattice. In their simulations the initial state was sampled using the Bogoliubov procedure for the combined harmonic and lattice potential. Using the truncated Wigner approach they modeled the sudden trap displacement and subsequent dynamics and found qualitatively good agreement with the damping behaviour observed in experiments. These results, which are for the low atom number regime (i.e.~$N_{\rC}\sim10^2$), where the strict validity conditions (see \sref{sec:validity}) for the Wigner approach are not satisfied, provides an indication that the Wigner method has an extended range of applicability. 

\paragraph{Dephasing in 1D interferometers}
Bistritzer \etal \cite{Bistritzer2007} examined the effect of quantum-phase fluctuations on a condensate split into two parts. Their system consisted of a pair of quasi-1D Bose gases 
which realize a basic model for an atom interferometer.  These 1D gases were modeled with a Bose Hubbard Hamiltonian using the Wigner method (note there was no explicit optical lattice potential, but the gases were treated in a \emph{lattice approximation}). In detail they used the adiabatic mapping procedure to sample the initial Wigner distribution as discussed in \sref{altsampling}. In this application an initial non-equilibirum state was prepared by imposing a relative phase between the two  quasi-1D systems. The dephasing between the (uncoupled) systems was then calculated as a function of time and shown to decay exponentially with little sensitivity to temperatures below a characteristic temperature $T^*$.
 
\paragraph{Quantum phase transition in a 1D optical lattice}
Dziarmaga \etal~\cite{Dziarmaga08a} investigated the quantum phase transition from Mott insulator to superfluid transition in a periodic 1D optical lattice. The Kibble-Zurek mechanism (KZM) predicts that when the lattice is ramped down in a time of order $\tau_Q$ (ramping up the tunneling rate) the winding number will grow through a random walk of BEC phases. In the high occupation regime the TWA leads to a discrete nonlinear Schr\"{o}dinger equation of motion, and the initial conditions were sampled to approximate a Mott insulator state $|n,n,n,\dots,n\rangle$, where $n$ is the (integer) number of particles per lattice site. The KZM scaling of the winding number with $\tau_Q$ was confirmed using truncated Wigner simulations of the transition dynamics.
\subsection{Dynamical instabilities and quasiparticle dynamics}\label{sec:deLaval}
\begin{figure}[t]
\begin{center}
\includegraphics[width=9cm]{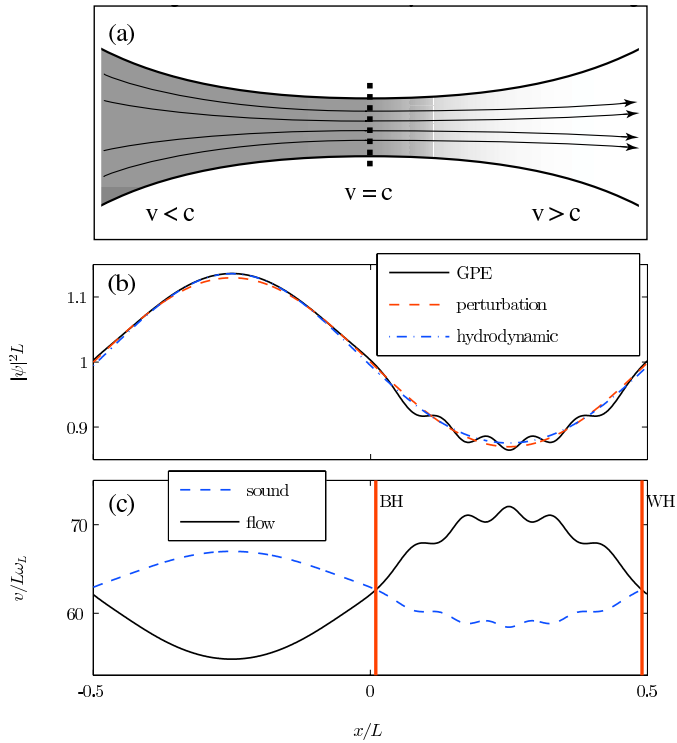}
\end{center}
\caption{Stationary flow in a doubly constricted toroid potential: (a) Schematic of a hydrodynamic de Laval nozzle. A flow that attains the speed of sound ($v=c$) at the narrowest point of the nozzle becomes supersonic beyond the nozzle. (b) Stationary GP solution in a quasi-one dimensional toroidal geometry perturbed by the potential \eref{VJain}, with winding number $w_0=10$  (other parameters are given in \cite{Jain2007a}). The full solution is compared with the results of hydrodynamic and perturbation theory approaches showing the importance of beyond hydrodynamic corrections in the supersonic region. (c) Flow velocity and sound velocity corresponding to the GP solution of (b), showing the position of the black hole (BH) and white hole (WH) horizons for sound waves. The energy unit of the toroid is $\hbar\omega_L=\hbar^2/mL^2$. Reproduced from  \cite{Jain2007a} \APSCR{2007}.\label{fig:deLavalScenario} }
\end{figure}
\paragraph{Quantum de Laval nozzle}
The quantum dynamics of a dynamically unstable supersonic current have been investigated by Jain \etal \cite{Jain2007a}. 
They considered the stationary flow of a condensate in a 1D toroidal trap that was  modified to form a double \emph{de Laval nozzle} geometry by the inclusion of a spatially varying potential (around a torus of length $L$) of the form
\begin{equation}\label{VJain}
V(x)=-V_0\sin^2{\left(\frac{2\pi x}{L}\right)}
\end{equation}
which has periodicity 2 over the region $-L/2\leq x\leq L/2$.
There are persistent current solutions for this system which have distinct spatial regions of subsonic and supersonic flow, with two acoustic horizons for sound waves (phonons): one where the flow goes supersonic (black hole, see \fref{fig:deLavalScenario}a) and the other where it returns subsonic (white hole). A typical stationary solution of the GP equation which has this property is shown in \fref{fig:deLavalScenario}b and the flow scenario is shown in \fref{fig:deLavalScenario}c. Supersonic flows are known to be energetically unstable and will decay in the presence of dissipation unless the decay is topologically prohibited such as in a toroidal trapping configuration. It was found that the system can also be dynamically unstable in certain scenarios, and the quantum dynamics of the instability were investigated with the truncated Wigner approach and compared with the predictions of Bogoliugov theory.  

For a system with a dynamical instability the usual Bogoliubov expansion \eref{psibogform} is inadequate \cite{Leonhardt2003a,Jain2007a}, as modes arise with complex eigenvalues. For these unstable modes  the Bogoliubov description acquires an irreducible off-diagonal component, which takes the form (equation (47) of \cite{Jain2007a})
\begin{eqnarray}\label{eq:bdgH2amplifier}
\hat{H}_{2}&=&\sum_j \hbar\omega_j \Big[ (\hat{b}^{\dagger}_{j+}
\hat{b}_{j+}\!-\!\hat{b}^{\dagger}_{j-} \hat{b}_{j-})-\int d x \, (|V_j^{+}|^2\!-\!|V_j^{-}|^2) \Big] \nonumber \\ 
&&+\!\sum_j i \,\hbar\gamma_j \Big[(\hat{b}_{j+}
\hat{b}_{j-}\!-\!\hat{b}^{\dagger}_{j+}
\hat{b}^{\dagger}_{j-})+\int d x \, (U_j^{+} V_j^{-} - U_j^{+*} V_j^{-*})\Big]
\end{eqnarray}
where $\{U_j^+,V_j^+\}$ and $\{U_j^-,V_j^-\}$ are the positive and negative energy Bogoliubov modes respectively, $\rm{e}_j^\pm=(\pm\omega_j-i\gamma_j)\hbar$ are the respective (complex) eigenvalues for these modes with annihilation operators $\hat{b}_{j\pm}$ \cite{Leonhardt2003a}. The second line is analogous to the interaction Hamiltonian for non-degenerate down-conversion of light by a nonlinear crystal. 
\begin{figure}
\begin{center}
ÊÊÊÊÊÊÊ\includegraphics[width=0.6\textwidth]{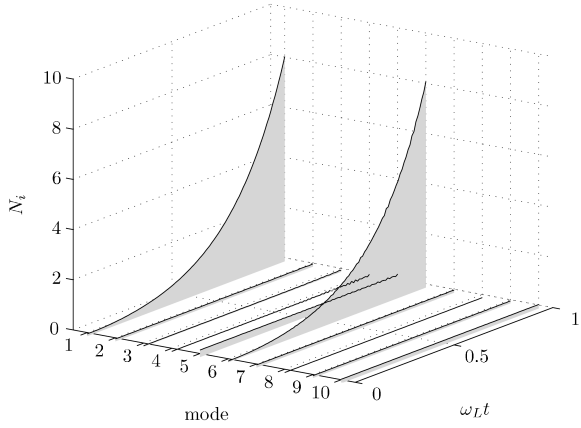}
ÊÊÊÊÊÊÊ\end{center}
ÊÊÊÊÊÊÊ\caption{Mode populations from averaging 40 trajectories of the truncated Wigner evolutions for the Quantum de Laval nozzle. Modes 1 and 6 are dynamically unstable, corresponding to the  negative and positive energy modes respectively. Reproduced from  \cite{Jain2007a} \APSCR{2007}. \label{Fig:unstablegrowth}}
\end{figure}
In regimes where only a pair of modes are coupled then $\hat{H}_2$   describes the formation of a two-mode squeezed state \cite{Walls1994a}. In this case, tracing over the  negative energy mode, the density matrix for the positive energy state  is of the form of that for a thermal state    with mean occupation $\langle n\rangle_+ = \sinh^2{\gamma t}$. 
This is analogous to the Hawking effect in that pairs of quasi-particles are produced at no energy cost: one enters the negative energy state (in the supersonic regime) and the other is promoted to positive energy and emerges at the horizon.

While the Bogoliubov analysis is useful in predicting the regions of instability, it cannot describe the  process dynamics, as the unstable modes grow exponentially (at least initially) and rapidly invalidate the linearized Bogoliubov analysis. Simulations with the truncated Wigner approach (see \fref{Fig:unstablegrowth}) avoid such limitations as they include the nonlinear interactions between excitations and their back-reaction. These features of the Wigner approach have seen several recent applications to the study of cosmological analogue models in BEC systems, such as
particle production in an expanding universe  \cite{Jain2007b}, and studies of Hawking radiation \cite{Carusotto08}.

In relation to the treatment of instabilities, also note the work of Polkovnikov \cite{Polkovnikov2003a} on the evolution of the macroscopically entangled states in optical lattices, where the truncated Wigner approach was used to deal with an unstable system where the usual Bogoliubov treatment breaks down. In that work the author presents arguments that this formalism should be able to adequately describe collapse and revival dynamics of the condensate.

Another application to quasiparticle dynamics was performed by Modugno \etal~\cite{Modugno2006a} who investigated the possibility of driving a parametric resonance in a toroidally trapped BEC. It was shown that specific quasiparticle modes could be resonantly excited, and then individually observed via expansion imaging. Starting from a zero temperature BEC, the quasiparticle excitation was initiated from vacuum fluctuations present in the initial state of the Wigner representation and driven by modulating the trapping potential.

\begin{figure}[t]
\begin{center}
	\includegraphics[width=0.95\textwidth]{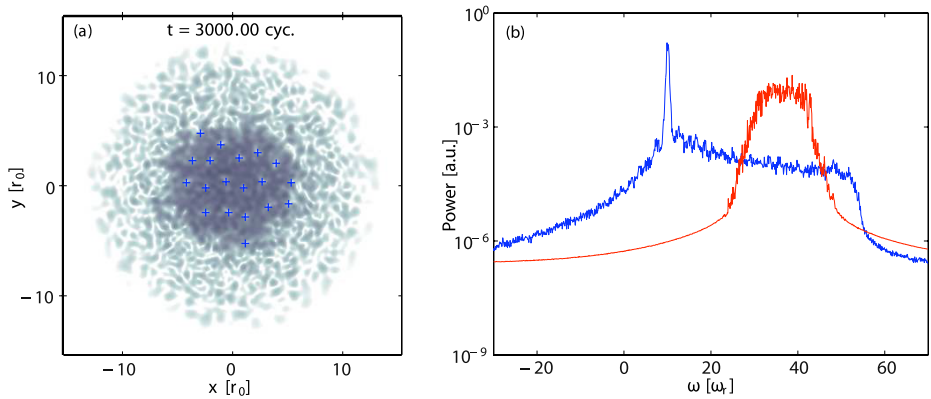}
\end{center}
	\caption{\label{Tod_density_plot} (Color online)(a) Classical field
density of Êequilibrium state of stirred condensate. ÊVortices are
indicated by $+$ symbols, and are indicated only where the surrounding
density of the fluid exceeds some threshold value. Ê(b) Power spectral
density traces at particular radii. The black (blue) and dark grey
(red) lines
correspond to radii $ r = 3.1894r_0$ and $r = 11.9575r_0$
respectively. Data corresponds to the period t = 9900 - 9910 trap
cycles. The plots in (a) and (b) are from the same simulation, in which the
condensate was initially in a trap of frequency $\omega_r$, at
temperature $T=0$, Êchemical potential $\mu_i = 14\hbar\omega_r$. The
elliptical perturbation rotates continuously at $\Omega = 0.75\omega_r
$. The spatial scale is the oscillator length $r_0 = \sqrt{\hbar/m
\omega_r}$.
Reproduced from \cite{Wright08} \APSCR{2008}.}
\end{figure}

\subsection{Vortex formation in a stirred BEC}\label{stirredBECs}
A number of experiments have shown that rotationally stirring a low
temperature ÊBEC can lead to formation of a vortex lattice e.g.
\cite{Madison00,Abo-Shaeer01,Abo-Shaeer02,Hodby02}. In the most
typical scenario, the stirring excites
a dynamical instability of the condensate \cite{Sinha01}, which is transformed into a
highly turbulent state, and then Êafter a long period Êevolves into a
rotating state containing a regular vortex lattice. This system
provides a challenging test for dynamical theories of cold bosonic
gases, Êbecause while the process (ideally) involves only Êan initial
pure ground state subject to a ÊÊconserving (Hamiltonian) process,
dissipation is required for the system to
evolve Êinto a state in equilibrium with the stirrer, e.g. see
\cite{Tsubota02a,Kasamatsu03}. A number of approaches have been given
based either on the pure GPE, ÊÊe.g. Ê\cite{Lundh03,Parker05a} or the
GPE supplemented with phenomenological damping terms
\cite{Tsubota02a}, but the description of the turbulent state is
clearly beyond the validity of the GPE. ÊAn {\em a priori} description
of the thermalisation which is central to the process, and provides
the necessary dissipation mechanism, was first given Êby ÊLobo \etal
\cite{Lobo2004a} using a classical field method. ÊThey
simulated a three dimensional condensate stirred Êby an elliptically
perturbed Êrotating harmonic trap, and considered the case of Êan
initial Ê$T=0$ condensate, (which they modelled as the ground state of
the GP equation), and also the case of an initial finite temperature
condensate.
Their simulations for an initially Ê$T=0$ condensate showed evolution similar to that seen in earlier approaches (e.g. \cite{Kasamatsu03,Lundh03,Parker05a}), with
vortices eventually entering the high density region of the field Êa
few hundred trap cycles after the creation of the turbulent state.
The vortices Êsettle into an ordered lattice after another period of a
few hundred trap cycles, and then the lattice slowly damps over a
further period of about 1000 trap cycles. Lobo \etal~made an
approximate estimate of the total energy Êtransferred
irreversibly out of the condensate, and by assuming equipartition over
the available modes, obtained a temperature of the thermal cloud which
they assumed was responsible for the dissipation. More recently,
Wright \etal \cite{Wright08}
have treated this stirring problem using a ÊTWPGPE approach. The initial vacuum noise gives an irreducible mechanism for
seeding the
dynamical instability, and their method 
ensures particle number and rotating frame energy are conserved to very high accuracy over the length of the simulation. Furthermore the Êbasis choice and numerical
method they use is free from grid method and boundary artifacts such
as aliasing and spurious damping at high momenta, so that any thermalisation and damping observed can be unambiguously
attributed to the intrinsic field theory, rather than numerical
artifact. The authors considered systems in
`pancake' traps, which are effectively two-dimensional.
A typical final state, after the system has been subjected to a constantly rotating elliptical perturbation for 3000 trap cycles, is shown in
\fref{Tod_density_plot}(a).

This treatment allows a detailed and quantitative description of the
thermalisation of the
condensate. The thermal cloud Êcreated is initially located primarily
in an outer annulus, and quickly obtains a classical moment of
inertia, while the Êcentral region of the field is irrotational until
penetrated by vortices. The temperature and chemical potential of the
thermal cloud Êare obtained by a self-consistent fit, and good
agreement is obtained to an analytic estimate. The Penrose-Onsager
criterion (see \sref{SEC:PGPEdensitycondfrac}) for identifying the
condensate component fails in this system, due to the complex phase
and amplitude structure associated with the vortex array. As an
alternative method of characterising the coherence properties of the
system, Êthe authors examine the local behaviour of  the temporal
power spectra of the classical field about time $t_0$, namely 
\begin{equation}
	H(\mathbf{x},\omega;t_0) = \left| \frac{1}{T}\int_{t_0-T/2}^{t_0+T/2} \cf(\mathbf{x},t) e^{-i\omega t}dt\right|^2.	
\end{equation} 
The sampling period $T$ is chosen to be a number of trap cycles, 
and is long compared to the timescales characterising the phase evolution of the
  condensate ($ \hbar/\mu$), but  short compared to the relaxation time of  the field. Spectra such as shown in \fref{Tod_density_plot}
(b) are obtained. The spectrum from the central region of the field has a prominent
and narrow peak Êat $\omega \approx 10\omega_r$, which is interpreted as the
condensate eigenvalue. At larger radii ($r\gtrsim 9r_0$) Êthe spectrum is broadened and is approximately that of the non-interacting gas.
The local correlation times obtained from these data (by Fourier transform of $H$) allow an
unambiguous distinction to be made between superfluid turbulence  and
thermal gas (which has a much shorter correlation time). We note that a feature of the 2D system is that the
final state is not a regular Abrikosov lattice, but instead is a
spatially disordered vortex liquid state. This can be interpreted as a
thermally excited vortex lattice, and indeed both the simulations and an analytic prediction given in \cite{Wright08} show that a considerable amount of thermal energy is generated
in the stirring process, and that 
therefore  the final condensate state must have considerable thermal excitation. The final state of the simulations is consistent with the condensate being in thermal and rotational Êequilibrium with the
thermal cloud. 

\begin{figure}[t]{
\begin{center} 
\includegraphics[width=0.5\textwidth]{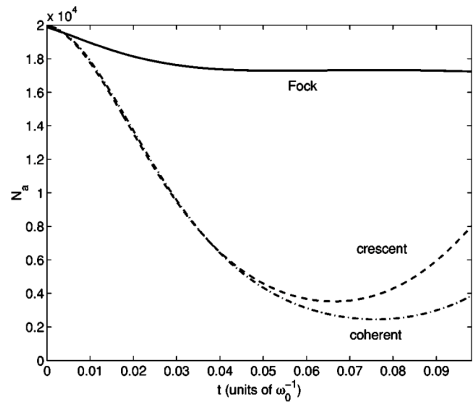}
\caption{\label{fig:photoAssoc} Comparison of atom number in photoassociation dynamics for Fock (solid line), coherent (dash-dot) and crescent quantum states of the initial atomic BEC. Reproduced from \cite{Olsen2004} \APSCR{2004}.}
\end{center}}
\end{figure}

\subsection{Quantum statistical effects in Superchemistry}\label{sec:superChem}
The field of \emph{superchemistry} was defined by Heinzen \etal~\cite{Heinzen2000a} as ``the coherent stimulation of chemical reactions via macroscopic occupation of a quantum state by a bosonic chemical species".  
Truncated Wigner simulations were used by Olsen \etal~\cite{Olsen2004,Olsen2004b,Olsen2004c,Olsen2004d,Olsen2003} to investigate superchemistry based on photo-association of trapped BECs into molecular dimers. The atom-molecule coupling occurs through a Raman two-photon transition for which the interaction Hamiltonian can be written as~\cite{Olsen2004c}
\begin{eqnarray}\label{HintMol}
\hat{H}_{\rm int}&=&\frac{i}{2}\intV{\x}\chi(\x)\left(\hat{\psi}_a^{\dag2}(\x)\hat{\psi}_{m^*}(\x)-\hat{\psi}_a^2(\x)\hat{\psi}_{m^*}^\dag(\x)\right)\nonumber\\
&&+i\intV{\x}\Omega(\x)\left(\hat{\psi}_{m^*}^\dag(\x)\hat{\psi}_{m}(\x)-\hat{\psi}_{m^*}(\x)\hat{\psi}_{m}^\dag(\x)\right)
\end{eqnarray}
where $\hat{\psi}_a(\x)$ is the atomic field, $\hat{\psi}_{m^*}(\x)$ the excited molecular field, and $\hat{\psi}_{m}(\x)$ is the molecular ground state field. $\chi(\x)$ is the Rabi frequency of the transition from atoms to excited molecules and $\Omega(\x)$ is the Rabi frequency for the transition from excited to ground molecular states. 

A notable feature of this work was the departure from the mean field predictions (see also~\cite{Hope01a}). The quadratic dependence of \eref{HintMol} on atomic fields combined with the relatively short timescale of the atom-molecule transition combine to make the process highly sensitive to quantum statistical effects such as squeezing (e.g.~see \fref{Fig:wignerfun} for comparison of some different quantum states for the condensate). Results of those studies demonstrated a regime where the quantum statistics of the atomic condensate play a crucial role in superchemistry dynamics and confirmed that photo-association of a trapped BEC into molecules provides a signature of the quantum state of the BEC. This can be seen in \fref{fig:photoAssoc} where photoassociation dynamics are compared for different quantum states of the initial BEC.

\subsection{The quantum linewidth of an atom laser}\label{satomlaser}
Johnsson \etal~\cite{Johnsson2007a} used the TWA to model the process of weak outcoupling from a trapped Bose-Einstein condensate and to determine the linewidth of the outcoupled beam. Semi-classical analysis predicts the linewidth will be essentially Fourier limited, scaling as the inverse of the outcoupling time~\cite{Johnsson07b}. Determining the quantum linewidth requires a multimode quantum theory of the atom laser outcoupler which the was implemented using the TWA. 
For Raman transition based outcouplers the primary source of linewidth broadening in the weak outcoupling regime comes from phase fluctuations of the condensate.
%
The main source of phase fluctuations arises from the nonlinear interactions in the condensate which convert number fluctuations into phase fluctuations \cite{Wiseman2001a}. A simple estimate for the quantum linewidth, $\Delta E$, of the output coupled atom laser beam from condensate containing  $N_0$ atoms in a harmonic trap 
is given by
\begin{equation}
\Delta  E = \frac{\partial \mu_{\rm TF}}{\partial N_0}\Delta N_0,\label{poslw}
\end{equation}
where ${\mu_{\rm TF}}$ is the Thomas-Fermi chemical potential (see \eeref{muTF}).
If the quantum state of the BEC is approximately a coherent state then the fluctuations in condensate number are  Poissonian, i.e.~$\Delta N_0=\sqrt{N_0}$, and an analytic expression for $\Delta E$ can be calculated.

In \cite{Johnsson2007a} one and two dimensional truncated Wigner simulations were used to obtain the atom laser linewidth as a function of output coupling time, and these results were compared against GPE simulations. For short times the linewidth was found to be inversely proportional to the output coupling time, a feature adequately described by the GPE. However, on longer timescales the linewidth predictions of the two theories differed:  the truncated Wigner simulations plateaued towards the Poissonian-limit \eref{poslw}, whereas the GPE continued to narrow. 
\section{The stochastic projected Gross-Pitaevskii equation}\label{sec:sgpe}
The stochastic projected Gross-Pitaevskii equation (SPGPE) is a truncated Wigner theory of Bose gases which takes into account the interactions between the atoms in the \CF\ region and the \rI\ region. The theory is 
valid for sufficiently large systems for temperatures from about $0.5 T_c$ 
to just above the BEC transition at $T=T_c$ when the \rI\ region contains many weakly populated thermal modes. For a trapped system with largest trap frequency $\omega$ the condition $\hbar\omega\ll k_BT$ must be satisfied; in this sense it is a high temperature theory, extending the PGPE theory treated in \sref{SEC:PGPE} and complimenting the low temperature TWPGPE approach described in \sref{sec:TWPGPEapps}. An important point of difference is that the SPGPE theory is a grand canonical theory, in contrast to the microcanonical approaches described in previous sections. In its simplest implementation the theory is parameterized by the (in general time dependent) temperature $T$ and chemical potential $\mu$ of the thermal reservoir comprised of the thermally occupied modes contained in the \rI\  region. Thus, in contrast to microcanonical approaches for which temperature must be determined a posteriori, the SPGPE formalism allows direct control of the temperature of the interacting system. In general the \rI\ and \rC\ regions may be out of equilibrium (such as during quench cooling towards $T_c$). The dynamics of the condensation process are particularly well suited to treatment with this theory. 
\subsection{Formalism}\label{s5formalism}
The treatment of thermal processes using stochastic methods has a long history, beginning with the theory of Brownian motion~\cite{Einstein1905a,Langevin1908a}. The theory of open \emph{quantum} systems couples the modes of interest in the quantum system to a reservoir~\cite{Gardiner1998a}. The precise details of the derivation can be found in~\cite{Gardiner2002a,Gardiner2003a,Bradley2006a,Bradley2008a}. As a full treatment is rather lengthy, here we will briefly outline the development of the theory.

The essential conditions for a treatment of the degenerate Bose gas with minimal complexity are i) the system and the reservoir are uncorrelated, ii) the reservoir is in \emph{local equilibrium}, described by the semi-classical Bose-Einstein distribution. In many systems of interest these conditions can be readily satisfied by appropriately choosing the cutoff energy $\epsilon_{\rm cut}$. In essence, the stochastic PGPE theory extends the PGPE theory by including the coupling of \rC\ region atoms with atoms in \rI\ which form a grand-canonical reservoir. The coupling generates additional damping and stochastic terms in the PGPE, which necessarily takes the form of a stochastic differential equation.  

The assumption of local equilibrium for the \rI\ region is convenient, but does not pose a fundamental limitation of the formalism.  In principle it is possible to derive a quantum Boltzmann-like kinetic equation for the particles in \rI\ coupled to a stochastic c-field equation for the particle in \rC.  However, this would result in further computational complexity which is not necessary for a broad range of applications.  Such a description remains a goal for the future, and would result in a near complete model of Bose gas dynamics at high temperature.

\subsubsection{Background}\label{s5backgrnd}
The theory of finite temperature BEC dynamics of Zaremba, Nikuni, and Griffin~\cite{Zaremba1999a}, which was developed along the lines of the two fluid theory of superfluid helium, provided the foundations of generalized GPE theory from a hydrodynamic point of view. The essential assumption of the theory is that atoms enter and leave the condensate so as to enforce \emph{local} energy and momentum conservation. The resulting description takes the form of a finite temperature GPE for the condensate, coupled to a Boltzmann equation for the non-condensate. The theory has the advantage that the condensate and non-condensate are described on an equal footing, making the description of coupled dynamics of thermal cloud and condensate tractable. While being intuitively appealing and a providing a powerful approach for extending zero temperature mean-field theory, enforcing local conservation of energy and momentum also has disadvantages which are removed by the SPGPE approach. Firstly, as a mean field theory it cannot be valid near the BEC phase transition where the condensate is relatively small or non-existent. Furthermore, a locally conserving theory neglects the non-locality of quantum mechanics, which plays a fundamental r\^{o}le in determining the dynamics of atoms stimulated into a highly occupied field.

The SGPE formulation of Stoof is closely related to the SPGPE theory presented here: the formulation leads to a stochastic differential equation for the condensate which is driven by a noise term associated with condensate growth. However, there are two important differences. Firstly, in \cite{Stoof1999a} the reservoir is chosen to contain all modes with energy in excess of the chemical potential $\mu$. The stochastic GP-equation so obtained involves self-energy functions, necessitating a many-body $T$-matrix treatment of interactions. As well as being difficult to implement numerically, the \CF\ of the theory only describes the condensate and few very low energy excitations. As discussed in \sref{secCR}, in the vicinity of the transition there are typically $10^3-10^4$ degenerate modes warranting a \CF\ treatment. Secondly, the approach neglects scattering terms which conserve population but transfer energy from the reservoir to the \CF\ region. The inclusion of these terms in the SPGPE stems from the explicit use of a high energy projector, which renders them finite and tractable.  

\paragraph{Stochastic PGPE} The Quantum Kinetic theory of Bose-Einstein condensation~\cite{QKIII,QKV,QKVII} has been shown to provide a good description of the process of condensate formation~\cite{QKPRLI,QKPRLII,QKPRLIII} at the level of condensate population dynamics provided condensation occurs into the absolute ground state of the system.
Recent work by Gardiner \etal~\cite{Gardiner2002a,Gardiner2003a,Bradley2005b,Bradley2006a,Bradley2008a} developed the SPGPE theory to explicitly include a high energy cutoff, thus unifying the PGPE theory with the reservoir theory of high temperature BECs. 

Since its inception, the theory of finite temperature BECs has presented many challenges and a rich body of work has accumulated~\cite{Proukakis2008a}. Several key problems are: the consistent treatment of two body interactions~\cite{Morgan2000a,Hutchinson2000a}, the description of thermal cloud coupling and dynamics~\cite{Zaremba1999a,Stoof1999a}, and the unification of GPE approaches with dissipative finite temperature phenomenology~\cite{Choi98a,Tsubota02a}.
In these respects, the PSGPE theory has some notable computational and physical advantages which we briefly summarize:
\begin{enumerate}
\item \emph{Consistent UV-cutoff:} The imposition of a high energy cutoff using the methods of \sref{SEC:PGPE} imposes a consistent cutoff, even for trapped systems. As noted in \sref{sec:Pimport}, if the cutoff is only imposed in momentum space the precise wavelength for the cutoff is position dependent. The PGPE formalism addresses this problem by imposing a cutoff in the single particle basis which approximately diagonalises the many-body Hamiltonian at high energies.
\item \emph{Two body T-matrix:} By imposing the cutoff at high energies the need to use the many body $T-$matrix to describe scattering is eliminated, requiring only the two body $T-$matrix description of S-wave scattering: $T(0)\to 4\pi\hbar^2 a/m$.
\item \emph{Nonlocal description of condensate growth---} Fundamentally the theory is nonlocal: atoms which leave the high energy cloud enter the \CF\ region so as to minimize the difference between the chemical potential of the high energy region and the GP operator acting on the \CF. Beyond hydrodynamic effects are thus explicitly included.
\item \emph{Scattering terms:} Imposing a cutoff at high energies allows the so-called \emph{scattering} terms --- reservoir interactions that do not directly change the populations of the reservoir or \CF\ --- to be consistently included in the theory. Analogous terms proved to play an important role in the Quantum Kinetic theory description of condensate growth~\cite{QKVI}. In the SPGPE theory the scattering terms couple to dynamical excitations in the \CF\ region.
\item \emph{Valid at the BEC transition:} A notable feature of the SPGPE is that it is particularly well suited for dynamical studies in the regime $T\sim T_c$ since the conditions of validity are \emph{high temperature} ($\hbar \bar{\omega} \ll k_BT$) and moderate occupation of modes, both of which are readily satisfied near the BEC transition. 
\item \emph{Reservoir dynamics:} In principle the dynamics of the incoherent region can be treated with a Quantum Boltzmann equation with little additional formalism provided the region remains in approximate local equilibrium.
\item \emph{Consistent mean field treatment of dissipation} The mean field theory recovered by setting all noises to zero gives a GP equation of motion with extra dissipative terms. The dissipation evolves the \CF\ to a ground state with the chemical potential of the reservoir. In contrast to phenomenological approaches~\cite{Choi98a,Tsubota02a}, the dissipation rate arising in the equation of motion is the physical rate of the theory which can be derived from a Boltzmann integral\footnote{The difference between the effective dissipation rates in the two descriptions is usually small for weak dissipation, but the equations of motion are formally distinct. See reference \cite{Bradley2006a} for a more detailed comparison of  dissipative mean field approaches to vortex formation in BECs.}~\cite{Bradley2008a}.
\end{enumerate}
\subsubsection{The system and its separation}\label{s5syssep}
We again consider a dilute Bose gas held in a trapping potential defined by \eref{EQ:H0}, but extend the PGPE description of \sref{SEC:PGPE} to consider the coupling of the \CF\ region to the 
 \rI\ region.  To accommodate applications such as the formation of vortex lattices in BECs (discussed below in \sref{SPGPEapps}), we include the possibility that the \rI\ region may be rotating.
This requires either that the trapping potential is axially symmetric, or time-independent in a rotating frame of reference (corresponding to elliptical stirring at a constant angular frequency). For simplicity we restrict our attention to systems where either i) the \rI\ region is stationary in the laboratory frame, or ii) the \rI\ region is rotating in an axially symmetric trap. For the latter case the theory is conveniently formulated in the frame rotating at the frequency of the \rI\ region, which we denote by $\Omega$. Choosing the symmetry axis of the system to be the z-axis, the single particle Hamiltonian for the system transformed to the rotating frame is
\begin{equation}\label{HspPSGPE}
H_{\rm sp}=H_0=-\frac{\hbar^2\nabla^2}{2m}+V_0\xa-\Omega L_z,
\end{equation}
where $L_z=-\i\hbar(x\partial_y-y\partial_x)$ is the z-component of the angular momentum operator. It has been shown that when the incoherent region is in rotational equilibrium in a harmonic trap the theory is modified by transforming the \CF\ description to the rotating frame and making the replacement $\omega_r\to\omega_\perp$, where
\begin{equation}\label{Wperp}
\omega_\perp=\sqrt{\omega_r^2-\Omega^2},
\end{equation}
in the dissipation rates of the SPGPE~\cite{Bradley2008a}. We can thus treat the rotating case in the formalism by including the effects of rotating frame transformation in the \CF\ description \eref{HspPSGPE}, parameterized by $\Omega$. We return to rotating systems in more detail in \sref{RBEC}, but in what follows the formalism applies to systems that are either non-rotating and in general non-axisymmetric, or rotating and axi-symmetric ($\omega_x=\omega_y\equiv\omega_r$).

As described in \sref{SEC:ProjOps}, the field operator for the full system is decomposed into a \CF\ and an incoherent field.
The SGPE takes the form of an equation of motion for the \CF\ $\psi_\rC(\x)$ with terms arising from interaction with the incoherent region \rI. The two regions are treated using different approximations.

\subsubsection{Treatment of the incoherent region}\label{s5incohreg}
The local equilibrium assumption for the state of the incoherent region allows all higher order correlation functions arising in the theory to be factorized into products of second order correlation functions. At this level of approximation the essential reservoir interaction physics can be reduced to functions of the single particle Wigner function for the incoherent region
\begin{equation}
F_\rI(\x,\mbf{K})=\int d^3\xd\left\langle \hat{\psi}_\rI^\dag(\x+\xd/2)\hat{\psi}_\rI(\x-\xd/2)\right\rangle e^{i\mbf{K}\cdot\xd},
\end{equation}
previously introduced in \sref{SEC:S3incohregion}.
Introducing the variables $\mbf{u}\equiv(\mbf{x}+\mbf{x}^\prime)/2$, 
$\mbf{v}\equiv\mbf{x}^\prime-\mbf{x}$, we can write
\EQ{\label{rncexp1}
\av{\hat{\psi}_\rI^\dag(\mbf{x^\prime})\hat{\psi}_\rI(\mbf{x},\tau)}&=&\av{\hat{\psi}_\rI^\dag(\mbf{u}+\mbf{v}/2)\hat{\psi}_\rI(\mbf{u}-\mbf{v}/2,\tau)},\nonumber\\
&\approx&\frac{1}{(2\pi)^3}\int_{\Omega_\rI} d^3\mbf{K}\;F_\rI(\mbf{u},\mbf{K})e^{-i\mbf{K}\cdot\mbf{v}-i\omega(\mbf{u},\mbf{K})\tau},\\ \label{rncexp2}
\av{\hat{\psi}_\rI(\mbf{x^\prime})\hat{\psi}_\rI^\dag(\mbf{x},\tau)}
&\approx&\frac{1}{(2\pi)^3}\int_{\Omega_\rI} d^3\mbf{K}\;[1+F_\rI(\mbf{u},\mbf{K})]e^{i\mbf{K}\cdot\mbf{v}+i\omega(\mbf{u},\mbf{K})\tau},
}
where phase space integration over the incoherent region $\Omega_\rI$ constrains the coordinates to satisfy $\hbar\omega(\mbf{x},\mbf{K})>\ecut$ (see \ref{WIdef} and \ref{eq:pmin3D}) and the energy in the frame rotating with angular frequency vector $\mbf{\Omega}=\Omega\hat{z}$ has the semi-classical form
\begin{equation}\label{Wdef}
\hbar\omega(\mbf{x},\mbf{K})=\frac{\hbar^2\mbf{K}^2}{2m}-\hbar\mbf{\Omega}\cdot(\mbf{x}\times\mbf{K})+V_0(\mbf{x}).
\end{equation}

The dissipation rates of the theory are time-integrated products of such functions, as can be seen from \eref{Gdef} and \eref{Mdef}. 
\paragraph{Semi-classical Bose-Einstein distribution}
For many applications the \rI\ region may be described by a semi-classical Bose-Einstein distribution
\begin{equation}\label{SCBE}
F_\rI(\mbf{x},\mbf{K})=\frac{1}{\exp{[(\hbar\omega(\mbf{x},\mbf{K})-\mu)/k_BT]}-1},
\end{equation}
where we note that for high energies where this is assumed to apply, the gas is always rapidly thermalized and interactions with the \CF\ region are a small correction to the single particle energy \eref{Wdef}.
This form has been used to evaluate the dissipation rates of the SPGPE theory~\cite{Bradley2008a}.

The properties of the ideal gas description of \eref{SCBE} and \eref{Wdef}, including the effect of the cutoff, can be expressed in terms of the  incomplete Bose-Einstein function defined as~\cite{Bradley2008a}
\begin{equation}
\begin{split}
 g_\nu(z,y)&\equiv\frac{1}{\Gamma(\nu)}\int_y^\infty dx\;x^{\nu-1}\sum_{l=1}^\infty(ze^{-x})^l,\\
&=\sum_{l=1}^\infty\frac{z^l}{l^\nu}\frac{\Gamma(\nu,yl)}{\Gamma(\nu)},
\end{split}
\end{equation}
where $\Gamma(\nu,x)\equiv \int_x^\infty dy\;y^{\nu-1}e^{-y}$ is the incomplete Gamma function. In analogy with the reduction to an ordinary Gamma function $\Gamma(\nu,0)=\Gamma(\nu)$, we have $g_\nu(z,0)=g_\nu(z)\equiv\sum_{l=1}^\infty z^l/l^\nu$, reducing to the ordinary Bose-Einstein function. We then find for the \rI\ region density, for example
\begin{eqnarray}
\begin{split}
n_\rI({\bf x})=&\int_{\Omega_\rI} \frac{d^3{\bf k}}{(2\pi)^3}F_\rI({\bf x},{\bf k})\\
=&\lambda_{\rm dB}^{-3}g_{3/2}\left(e^{\beta\left[\mu-V_\perp({\bf x})\right]},\beta\hbar^2 K_{\rm cut}({\bf x})^2/2m\right),
\end{split}
\end{eqnarray}
where $\hbar^2 K_{\rm cut}({\bf x})^2/2m={\rm max}\left\{\ecut-V_\perp({\bf x}),0\right\}$ and we have introduced the effective potential 
\begin{equation}
V_{\perp}(\x)=\frac{m}{2}(\omega_\perp^2r^2+\omega_z^2z^2),
\end{equation}
which accounts for the transformation to the rotating frame \eref{Wperp}.
Setting $\ecut=0$, we recover the standard form for the semi-classical particle density of the ideal gas~\cite{Dalfovo1999a}. The total number in \rI\ region is given by 
\begin{equation}\label{NNC}
N_\rI=g_3\left(e^{\beta\mu},\beta \ecut\right)/(\beta\hbar\bar{\omega})^3,
\end{equation}
where $\bar{\omega}=(\omega_z\omega_\perp^2)^{1/3}$ is the geometric mean frequency in the rotating fame. In this way the usual semi-classical expressions can be generalized to include a cutoff in terms of the incomplete Bose-Einstein function.

\subsubsection{Treatment of the \CF\ region: deriving the equation of motion} \label{s5cregion}
The standard procedure of phase space methods for open systems involves deriving a master equation for the reduced system by eliminating the reservoir degrees of freedom. The master equation may then be mapped to a generalized Fokker-Planck equation (FPE) of motion for a quasi-probability distribution, such as the Wigner representation, by making use of operator correspondences (e.g. \eref{opcorr1}--\eref{opcorr4}). Provided the FPE contains derivatives which are at most second order (representing diffusion), an equivalent stochastic differential equation may be found which can be conveniently simulated numerically. The truncated Wigner approximation involves neglecting third order terms in the FPE (see \eref{HCthirdorder}), and in the context of the SPGPE theory the truncated Wigner approximation reduces the FPE to second order, allowing a formulation of the problem in terms of stochastic differential equations.
\paragraph{Validity of the SPGPE treatment of the \CF}
In addition to treating the \rI\ region semi-classically, the SPGPE formalism makes the truncated Wigner approximation which requires that the modes under consideration are highly occupied. An approximate master equation that can be mapped to a stochastic differential equation is then obtained by truncating the interaction between \rC\ and \rI\ at first order in powers of $\hbar\omega/k_B T$, where $\omega={\rm max}\{\omega_i\}$ is the largest oscillator frequency of the system.

A feature of the the high temperature theory is that the dissipation arising from the reservoir coupling acts to smooth out sharp phase space structure that would otherwise generate significant third order term corrections~\cite{Zurek1998a}. In this sense the high temperature Bose gas is particularly well suited to treatment using the truncated Wigner method. 

\subsubsection{Stochastic projected Gross-Pitaevskii equation}\label{s5SPGPE}
In the rotating frame the full non-local form of the SPGPE is given by the following stochastic differential equation in Stratonovich form
\begin{subequations}
\label{SPGPE}
\begin{eqnarray}
(S)d\psi_\rC(\mathbf{x},t)&=&\PC\Bigg\{ -\frac{i}{\hbar}L_\rC\psi_\rC(\mathbf{x})dt \label{SPGPEa}\\
&&+\frac{G(\mathbf{x})}{k_BT}(\mu-L_\rC)\psi_\rC(\mathbf{x})dt+dW_G(\mathbf{x},t) \label{SPGPEb}\\
&&+\intV{\mathbf{x}^\prime}M\left(\mbf{x}-\mbf{x}^\prime\right)\frac{ i\hbar\nabla\cdot \bm{j}_\rC(\mbf{x}^\prime)}{k_BT} \psi_\rC(\mbf{x})dt
+i\psi_\rC(\mbf{x})dW_{M}(\mbf{x},t)\Bigg\}.\;\;\;\;\;\;\;\;\;\; \label{SPGPEc}
\end{eqnarray}
\end{subequations}
The first line of the SPGPE \eref{SPGPEa} describes Hamiltonian evolution according to the PGPE introduced in section \ref{sec:Hcfield} and developed in \sref{SEC:PGPE}.
The projector $\PC$ appears as a natural consequence of formally imposing a high energy cutoff in the definition of the c-field region.
The operator $L_\rC$ is the Hamiltonian evolution operator for the \CF\ region defined via equations
 \eref{PH} and \eref{PHeom}. Its explicit form is
\EQ{\label{LCdef}
L_\rC\psi_\rC(\x)\equiv \left(H_{\rm sp}+u|\psi_\rC(\x,t)|^2\right)\psi_\rC(\x).
}
The second line of the SPGPE \eref{SPGPEb} is directly responsible for condensate growth from scattering between two \rI\ region atoms as illustrated in \fref{processschematic}(a).  The $\mu$ and $T$ that arise are respectively the chemical potential and temperature of the thermal reservoir of particles in the \rI\ region.  
The quantity $G(\mathbf{x})$ is a spatially dependent collision rate, specified by a quantum Boltzmann integral over the \rI\ region as discussed in more detail in section \sref{sec:growth} below.  The complex noise associated with growth is $dW_G(\mbf{x}^\prime,t)$ and satisfies
\begin{eqnarray}
\langle dW_G^*(\mbf{x},t)dW_G(\mbf{x}^\prime,t)\rangle&=&2G(\mbf{x})\delta_C(\mbf{x},\mbf{x}^\prime)dt,\\
\langle dW_G(\mbf{x},t)dW_G(\mbf{x}^\prime,t)\rangle&=&\langle dW_G^*(\mbf{x},t)dW_G^*(\mbf{x}^\prime,t)\rangle=0.
\end{eqnarray}
The third line of the SPGPE \eref{SPGPEc} represents number conserving scattering processes between atoms in \rC\ and \rI\ and provides a mechanism for energy transfer between the two regions
as illustrated in \fref{processschematic}(b).
This couples to the divergence of the \CF\ region current given by
\begin{equation}
\bm{j}_{\rC}(\x)\equiv \frac{i\hbar}{2m}\Big([\nabla\psi_\rC^*(\mbf{x})]\psi_\rC(\mbf{x})
-\psi_\rC^*(\mbf{x})\nabla\psi_\rC(\mbf{x})\Big)-(\mbf{\Omega}\times\mbf{x})\;|\psi_\rC(\mbf{x})|^2,\label{JCdef}
\end{equation}
where the last line is a rigid-body rotation term arising from the transformation to the rotating frame. The limit $\bm{j}_\rC(\x)=0$ gives the laboratory frame velocity field ${\bf v}=\mbf{\Omega}\times\mbf{x}$, corresponding to the irrotational system mimicking rigid body rotation. 
The rate function $M\left(\mbf{x}-\mbf{x}^\prime\right)$ is specified by a second quantum Boltzmann integral, and is discussed in detail below in section \sref{sec:scattering}.
\begin{figure}[t]{
\begin{center} 
\includegraphics[width=8cm]{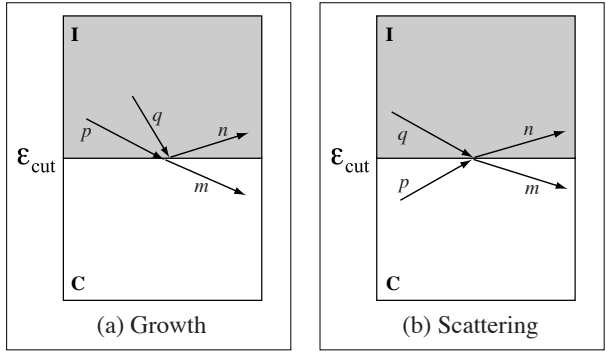}
\caption{Schematic of the processes arising from the interactions between the \CF\ and incoherent regions. In (a) two \CF\ region atoms collide, with a significant fraction of the collision energy transferred to one of the atoms, with the other atom passing into the \CF\ region.  The time-reversed process also occurs. In (b) a \CF\ region atom collides with a incoherent region atom with no change in \CF\ region population. 
\label{processschematic}}
\end{center}}
\end{figure}

The real noise $dW_{M}(\mbf{x},t)$ associated with scattering  is specified by
\begin{eqnarray}
\langle dW_{M}(\mbf{x},t)dW_{ M}(\mbf{x}^\prime,t)\rangle &=&2M\left(\mbf{x}-\mbf{x}^\prime\right) dt.
\end{eqnarray}

\paragraph{Grand canonical equilibrium}
As described in \sref{SEC:PGPEdensitycondfrac}, the PGPE provides a means to sample the microcanonical ensemble of equilibrium states. By including interactions with the incoherent region we have arrived at a grand canonical description, parameterized by the chemical potential and temperature of \rI. 
Irrespective of the form of $G(\x)$ and $M(\x)$, the SPGPE evolves the system to the grand canonical equilibrium distribution
\begin{equation}
W_s\propto \exp{\left(\frac{\mu N_\rC-H_\rC}{k_BT}\right)},
\end{equation}
corresponding to the density matrix
\begin{equation}
\hat{\rho}_s\propto \exp{\left(\frac{\mu \hat{N}_\rC-\hat{H}_\rC}{k_BT}\right)},
\end{equation}
in the truncated Wigner approximation. Once the \CF\ reaches equilibrium single trajectories may be used to sample the grand canonical ensemble. 

\subsection{Growth and scattering in the SPGPE}\label{s5growscatt}
We now discuss the properties of the dissipative terms in the SPGPE \eref{SPGPE}. 
We give the explicit form of the rate functions $G(\mbf{x})$ and $M(\mbf{x})$, the regimes under which they may be evaluated in closed form, and discuss details of their physical interpretation.
\subsubsection{Growth terms}
\label{sec:growth} 
\paragraph{Growth rate} The explicit form of the growth rate is~\cite{Gardiner2003a,Bradley2008a}
\begin{eqnarray}\label{Gdef}
G(\mbf{x})&\equiv&\frac{u^2}{(2\pi)^5\hbar^2}\iiint_{\Omega_\rI}d^3\mbf{K}_1d^3\mbf{K}_2d^3\mbf{K}_3 F(\mbf{x},\mbf{K}_1)F(\mbf{x},\mbf{K}_2)\nonumber
\\
&&\times[1+F(\mbf{x},\mbf{K}_3)]\Delta_{123}(0,0),
\end{eqnarray}
where $\Delta_{123}(\mbf{k},\epsilon)\equiv\delta(\mbf{K}_1+\mbf{K}_2-\mbf{K}_3-\mbf{k})\delta(\omega_1+\omega_2-\omega_3-\epsilon/\hbar)$ conserves energy and momentum during the collision. 
\begin{figure}[t]{
\begin{center} 
\includegraphics[width=8cm]{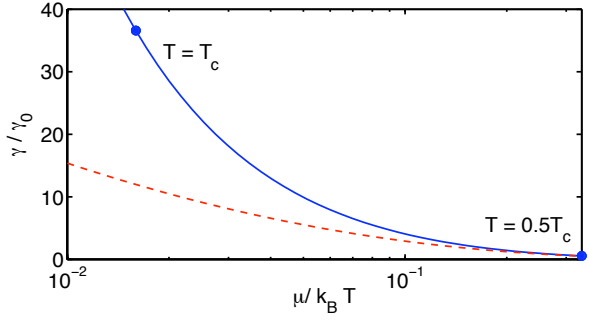}
\caption{Dependence of the growth rate $\gamma$ \eref{gamdef} on $\mu$ for the choice $\ecut=3\mu$. The full rate \eref{gamdef} (solid line) is compared with the logarithmic term (dashed line). The two points are calculated for $N=10^6$ $^{87}$Rb atoms in a trap with geometric mean frequency $\bar{\omega}=2\pi\times 25$Hz using the ideal gas relation for $T_c$. The chemical potentials are estimated from $\mu\approx3\hbar\bar{\omega}/2$ at $T=T_c$, and $\mu\approx\mu_{\textrm{TF}}(N_0)$ at $T=0.5T_c$.
\label{gammaDep}}
\end{center}}
\end{figure}
The rate $G(\mbf{x})$ can be calculated in the regime where the \rI\ region 
is quasi-static, and well-approximated by an ideal semi-classical Bose-Einstein distribution \eref{SCBE} with slowly varying $\mu(t)$ and $T(t)$.
For the inner spatial region of a harmonic trap satisfying $V(\mbf{x})\leq 2E_R/3$
we find that $G(\mbf{x})\equiv\gamma$ is independent of position with
\begin{eqnarray}\label{gamdef} \gamma&=&\gamma_0\Bigg\{\left[\ln{\left(1-e^{\beta(\mu-\ecut)}\right)}\right]^2+e^{2\beta(\mu-\ecut)}\sum_{r=1}^\infty\;e^{r\beta(\mu-2\ecut)}\left(\Phi[e^{\beta(\mu-\ecut)},1,r+1]\right)^2\Bigg\},\;\;\;\;\;\;\;\;\;
\end{eqnarray}
where $\gamma_0= 4m(ak_BT)^2/\pi\hbar^3$ and $\Phi[x,y,z]$ is the Lerch transcendent. 
Outside this region there is a weak spatial dependence which can be neglected for most purposes \cite{Bradley2008a}. The bare rate, $\gamma_0$, has been used as an estimate in the literature, often requiring a `fudge factor' (usually chosen as $\sim 3$) to obtain a rate that gives physically reasonable damping times. In \fref{gammaDep} we show $\gamma$ for a fixed choice of $\ecut=3\mu$. The full picture is more complicated than shown in \fref{gammaDep} because the choice $\ecut=3\mu$ would not be appropriate near $T_c$, but instead a much higher $\ecut$ would be necessary. In practice, the more accurate form typically increases the ratio $\gamma/\gamma_0$ by a factor which is in the range 1--10. 
\paragraph{Dissipative dynamics of condensate growth} 
By neglecting the scattering and noise terms in the SPGPE \eref{SPGPE} it is possible to show that 
\EQ{\label{HcNc}
\frac{\partial (H_\rC-\mu N_\rC)}{\partial t}=-\frac{2\gamma}{\hbar}\intV{\x}|(\mu-L_\rC)\psi_\rC(\x,t)|^2,
}
where we used the approximation $G(\mbf{x})\approx\gamma$ as given by \eeref{gamdef}. As the RHS of \eeref{HcNc} is a strictly non-positive term, we can see that the growth term acts to minimize the effective grand-canonical Hamiltonian $K_\rC\equiv H_\rC-\mu N_\rC$. The equilibrium solution is the ground state of the PGPE \eref{Pgpe} with chemical potential $\mu$. The growth terms describe the Bose-stimulated transfer of particles between the \rC\ and \rI\ regions during two body collisions.

\subsubsection{Scattering terms}
\label{sec:scattering}
\paragraph{Scattering rate}\label{sec:Mrate} The rate function for the scattering term of the SPGPE \eref{SPGPE} is most easily calculated by transforming to momentum space $M(\mbf{x})=\int d^3\mbf{k}\;e^{-i\mbf{k}\cdot\mbf{x}}\;\tilde{M}(\mbf{k})$, where we find
\begin{eqnarray}\label{Mdef}
\tilde{M}(\mbf{k})&=&\frac{2 u^2}{(2\pi)^5\hbar^2}\iint_{\Omega_\rI}  d^3\mbf{K}_1 d^3\mbf{K}_2\;\Delta_{12}(\mbf{k},0) F(\mbf{u},\mbf{K}_1)[1+F(\mbf{u},\mbf{K}_2)],
\end{eqnarray}
with $\Delta_{12}(\mbf{k},\epsilon)\equiv\delta(\omega_1+\omega_2-\epsilon/\hbar)\delta(\mbf{K}_1+\mbf{K}_2-\mbf{k})$. It has been shown that to a good approximation this expression is independent of $\mbf{u}$ \cite{Gardiner2003a,Bradley2008a}, and for this reason we suppress the argument in the present definitions.

Using the same approximation of a  quasi-static thermal cloud as used in the calculation of the rate $G(\mbf{x})$ in the previous section we find that
\begin{equation}
\tilde{M}(\mbf{k})=\frac{16\pi a^2 k_B T}{(2\pi)^3\hbar |\mbf{k}|}\frac{e^{\beta(E_R-\mu)}}{\left(e^{\beta(E_R-\mu)}-1\right)^2}\equiv \frac{\cal M}{(2\pi)^3|\mbf{k}|},
\end{equation}
and thus
\begin{equation}\label{Mfull}
M(\mbf{x})=\frac{{\cal M}}{(2\pi)^{3}}\int d^3\mbf{k}\frac{e^{-i\mbf{k}\cdot\mbf{x}}}{ |\mbf{k}|}.
\end{equation}
so that $M(\mbf{x})$ is a spatially dependent function over the whole \rC\ region.
At first glance this term appears somewhat pathological, but well defined results are obtained since $M(\x)$ is convolved with functions of condensate band fields. Such functions are both UV- and IR- cutoff, giving a finite result for the convolution in \eref{SPGPEc}.

\paragraph{Effect of scattering on hydrodynamic collective modes} To gain some physical insight into the nature of the scattering we note that the evolution according to the deterministic part of \eref{SPGPEc} can be written as a real effective potential
\begin{equation}
i\hbar\frac{\partial \psi_\rC(\x,t)}{\partial t}\Bigg|_{M}=\PC \left\{V_M(\x,t)\psi_\rC(\x,t)\right\},
\end{equation}
where
\begin{equation}
V_M(\x,t)=-\intV{\mathbf{x}^\prime}M\left(\mbf{x}-\mbf{x}^\prime\right)\frac{ \hbar^2}{k_BT} \nabla\cdot \bm{j}_\rC(\mbf{x}^\prime).
\end{equation}
To first approximation we can neglect all dissipation as relatively weak corrections to the PGPE evolution \eref{SPGPEa}, giving the continuity equation $\nabla \cdot \bm{j}_\rC(\x)\approx-\partial n_\rC(\x)/\partial t$ for the \CF\ region and
\begin{equation}\label{VMapprox}
V_M(\x,t)\approx\intV{\mathbf{x}^\prime}M\left(\mbf{x}-\mbf{x}^\prime\right)\frac{ \hbar^2}{k_BT} \frac{\partial n_\rC(\mbf{x}^\prime)}{\partial t}.
\end{equation}
Thus the scattering term generates an effective potential from dynamical density fluctuations in the \CF\ region.
\begin{figure}[t]
\centering
\includegraphics[width=10cm]{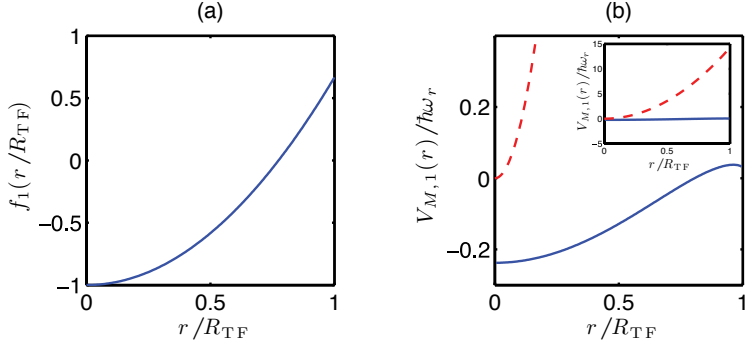}%
\caption{Scattering potential for the breathing mode $\delta n_1(r,t)$ of a Thomas-Fermi condensate. (a) The radial shape of the mode $f_1(r/R_{\rm TF})$ (see text). (b) The scattering potential for a breathing mode excitation with amplitude $A=0.1$ (solid line, shown at $t=0$) and the harmonic trapping potential (dashed line). We have used $\ecut=3\mu$ and $\mu\approx \mu_{\rm TF}(N_0)$ for $T=0.7 T_c$ and have estimated $N_0$ using the ideal gas relation for $10^6$ $^{87}$Rb atoms in a trap with radial frequency $\omega_r=2\pi\times 25$Hz. For these parameters $V_{M,1}(r)$ is small compared to the harmonic trap for significant radii (shown inset).
\label{scattV}}
\end{figure}
As a specific example we consider excitations of a system with spherical symmetry, since the momentum distribution will have the same symmetry as the scattering kernel \eref{Mfull}. We can then evaluate the effective potential for collective modes in the hydrodynamic and Thomas-Fermi approximations. For a spherically symmetric trap the condensate wavefunction in the Thomas-Fermi approximation is $n_{\rm TF}(\x)=(\mu_{\rm TF}/u)(1-(r/R_{\rm TF})^2)$, with Thomas-Fermi radius $R_{\rm TF}=\sqrt{2\mu_{\rm TF}/m\omega_r^2}$. The density profile for spherically symmetric modes~\cite{Stringari1996a}  with amplitude $A$ can be written as
\begin{equation}\label{breathe}
\delta n_n(\x,t)=\frac{A\mu_{\rm TF}}{u}\sin{(\omega_n t)}f_{n}(r/R_{\rm TF}),
\end{equation}
where the radial form is given by the Jacobi polynomials  $f_n(x)=\binom{n+1/2}{1/2}^{-1}P_{n}^{(0,1/2)}\left(2x^2-1\right)\theta(1-x)$ and $\theta(x)$ is the unit step function. Since $f_n(0)=1$ the peak density of the excitation is $A\mu_{\rm TF}/u$. The modes have frequencies $\omega_n=\omega_r\sqrt{2n^2+3n}$, 
 for example the breathing mode has frequency $\omega_1=\sqrt{5}\omega_r$. 
Evaluating equations \eref{VMapprox} and \eref{Mfull} leads to
\EQ{\label{VMbreathing}
V_{M,n}(\x,t)&=&A\hbar\omega_n\cos{(\omega_n t)}\left(\frac{4a R_{\rm TF}^3}{\pi a_r^4}\right)\frac{e^{\beta(\ecut-\mu)}}{(e^{\beta(\ecut-\mu)}-1)^2}F_n(r/R_{\rm TF}),
}
where $a_r=\sqrt{\hbar/m\omega_r}$ is the radial harmonic oscillator length, and 
\EQ{
F_n(x)=\int_0^{\infty}dk\;\frac{\sin{(kx)}}{kx}\int_0^{1} dy\;y\sin{(k y)}f_n(y).
}
In \fref{scattV} we show the effect of scattering on the breathing mode. At $t=0$ the density modulation \eref{breathe} vanishes, but has maximal rate of change, reflected by the phase of \eref{VMbreathing} relative to \eref{breathe}. The radial shape of the mode $f_1(r/R_{\rm TF})$ is shown in \fref{scattV}a, corresponding to outward flow at $t=0$. The potential, shown in \fref{scattV}b (at $t=0$), generates damping of the excitation by imposing an additional potential gradient that acts to oppose the outward flow. The same qualitative result holds for higher modes, where $V_{M,n}(r)$ is found to have the same overall shape as $f_n(r/r_{\rm TF})$ and a relative phase so as to oppose the excitations with an additional potential gradient. From \fref{scattV}b it is clear that $V_{M,1}(r)$ can be a significant correction to the bare trapping potential near the center of the trap, but is unimportant near $R_{\rm TF}$ (see inset).
\subsection{Simple growth SPGPE}\label{s5simplegrow}
The scattering term of the SPGPE \eref{SPGPEc} does not alter the population of the condensate band directly, nor does it affect the grand canonical equilibrium of the c-field.  It is generally expected to be less important to the c-field dynamics in comparison to the growth term \eref{SPGPEb}, which directly describe the dominant collision processes resulting in Bose-Einstein condensation.
Combined with the difficulty in its numerical implementation arising from the nonlocality of the deterministic part and the multiplicative nature of the associated noise, it seems reasonable to neglect it in the first instance.  This results in the simple growth SPGPE
\begin{equation}
\begin{split}
d\psi_\rC(\mbf{x},t)=&\PC\left\{ -\frac{i}{\hbar}L_\rC\psi_\rC(\mbf{x},t)dt+\frac{\gamma}{k_BT}(\mu-L_\rC)\psi_\rC(\mbf{x},t)dt+dW_\gamma(\mbf{x},t)\right\},
\label{SGPEsimp}
\end{split}
\end{equation}
where
\begin{equation}
\label{GamCor}
\langle dW_\gamma^*(\mbf{x},t)dW_\gamma(\mbf{x}^\prime,t)\rangle=2\gamma\delta_C(\mbf{x},\mbf{x}^\prime)dt.
\end{equation}
The numerical implementation of this simplified form is relatively straightforward, being only somewhat more complicated than the PGPE. As the only noise term is additive, the simple growth SPGPE can be integrated using high-order algorithms, such as a modified fourth-order Runge-Kutta algorithm \cite{Milstein04}.
\begin{figure}[t]
\centering
\includegraphics[width=12.5cm]{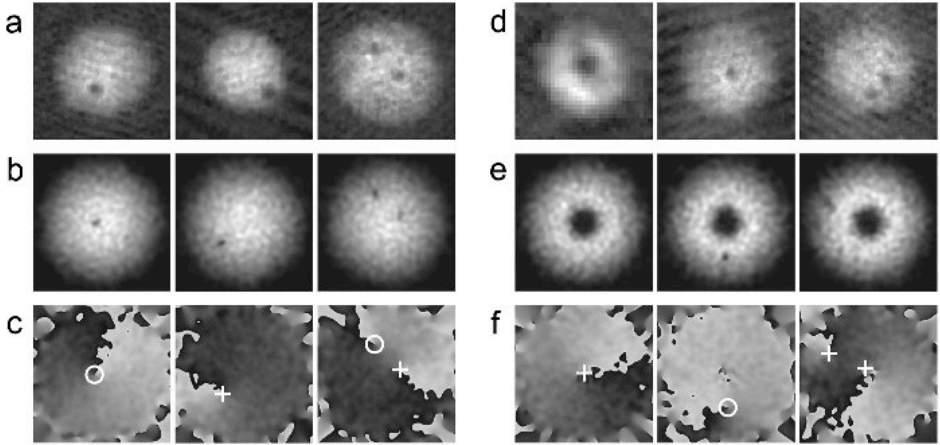}%
\caption{Spontaneous formation of vortices during Bose-Einstein condensation. \textbf{a}, 200-$\mu$m-square expansion images of BECs created in a harmonic trap, showing single vortices (left, centre) and two vortices (right).  \textbf{b, c}, Sample simulation results from evaporative cooling in a harmonic trap, showing in-trap integrated column densities along $z$ (in \textbf{b}) and associated phase profiles in the $z=0$ plane (in \textbf{c}), with vortices indicated by crosses and circles at $\pm 2\pi$ phase windings.  \textbf{d}, Left image: 70-$\mu$m-square phase-contrast experimental image of a BEC in a toroidal trap.  Remaining images: vortices in 200-$\mu$m-square expansion images of BECs created in the toroidal trap.  \textbf{e, f},   Simulations of BEC growth in the toroidal trap show vortices (as in \textbf{b},\textbf{c}) and persistent currents. Reproduced from \cite{Weiler08a}\label{fig:spontVort1}}
\end{figure}
\subsection{Applications to the dynamics of partially condensed Bose gases}\label{SPGPEapps}
\paragraph{Background} The first stochastic Gross-Pitaevskii treatment of Bose gases was developed by Stoof \cite{Stoof1999a} using a functional integral formulation of the Keldysh method. The first application of SGPE theory was carried out by Stoof and Bijlsma~\cite{Stoof2001a} who used the SGPE theory developed in \cite{Stoof1999a} to study finite temperature dynamics of a one dimensional Bose gas. Focussing on the scenario of growth into a one dimensional dimple trap the authors studied reversible condensate formation (see~\cite{Stamper-Kurn98a}), and the frequencies and damping rates of collective modes.
The SGPE theory has also been used in conjunction with variational techniques to study finite temperature collective excitations~\cite{Stoof2001a,Duine2001b}, and dissipative vortex dynamics~\cite{Duine2004a}. More recently, Proukakis \etal~\cite{Proukakis2003a} have used the method to investigate quasicondensate growth into a one dimensional dimple trap. For a deep dimple the dynamics were found to involve shock-wave propagation in addition to quasicondensate formation.

Applications of the SPGPE theory are still relatively few in number.  The current authors have only recently completed the implementation of the simple growth SPGPE for a harmonic trap in two dimensions with rotation, and in three dimensions without.
However, even the dissipative mean field equation obtained by neglecting all noise terms in the SPGPE has given significant insight into the dynamics of finite temperature BECs. Here we briefly review the applications that have appeared to date.

\subsubsection{Spontaneous vortex formation during Bose-Einstein condensation}\label{s5SpontVortex}
The formation of topological defects in symmetry-breaking phase transitions has been a topic of interest in both cosmological~\cite{Kibble1976a} and condensed matter~\cite{Zurek1985a} scenarios. 
Until recently quantitative models of the formation dynamics of trapped Bose-Einstein condensates have only been studied at the level of populations without allowing for the possibility of topological excitations~\cite{QKPRLI,QKPRLII,QKVII,Bijlsma2000a,QKPRLIII,QKVIII}.  The possibility of the spontaneous formation of vortices in the growth of a trapped Bose-Einstein condensate has been suggested previously by Anglin and Zurek~\cite{Anglin1999} and Svistunov~\cite{Svistunov2001a}, but until recently had not been confirmed by experiment and the possibility has been the subject of much speculation~\cite{Stoof2007a}.  Recent work by Weiler \etal~\cite{Weiler08a} has reported the observation of spontaneous vortex formation in the growth of a trapped BEC, and compared the statistics of formation with predictions of the simple growth SPGPE \eref{SGPEsimp}. 
\begin{figure}[t]
\centering
\includegraphics[width=12.5cm]{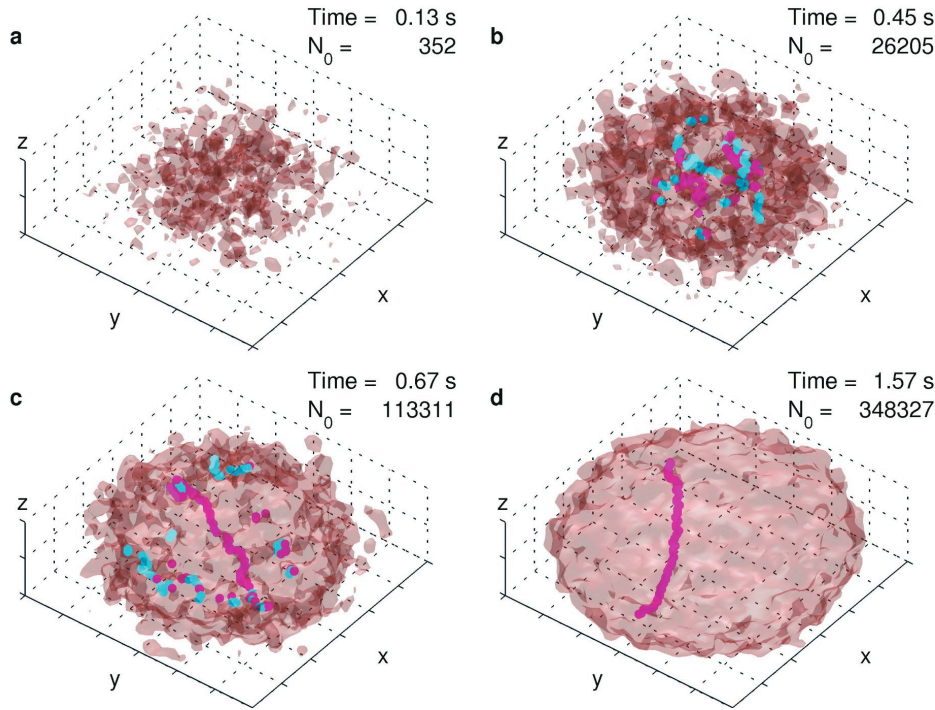}%
\caption{BEC growth dynamics. \textbf{a--d}, Four snapshots during the simulated growth of a BEC showing isodensity surfaces (in light red) in a three-dimensional rendering.  Vortex cores of opposite charges about the $z$ axis are indicated  as magenta and cyan lines.  The corresponding times are \textbf{a}, 0.13 s;  \textbf{b}, 0.45 s;  \textbf{c}, 0.67 s;  \textbf{d}, 1.57 s, where $t=$0 s is the time when the quench is initiated in the simulation. Reproduced from \cite{Weiler08a}\label{fig:spontVort2}}
\end{figure}
\paragraph{Experiment}
The experiments of Weiler \etal~\cite{Weiler08a} evaporatively cooled a dilute gas of $^{87}$Rb atoms from slightly above $T_c$ in both an oblate harmonic trap and a toroidal trap formed by the addition of a Gaussian barrier from a tightly-focussed blue-detuned laser beam along the symmetry axis.  The oblate nature of the trapping potential
resulted in an energy penalty for vortices not aligned with the symmetry axis, hence improving the fidelity of vortex detection.
 After condensate formation and sufficiently long time-of-flight expansion, vortex cores were observed with a probability ranging from 20--60\%. 
\paragraph{Theory}
The results of Weiler \etal~\cite{Weiler08a} were modelled using the simple growth SPGPE \eref{SGPEsimp} matched to the growth of the condensate number in the experiments by ramping the chemical potential and temperature of the thermal cloud.
For each experimental evaporative cooling ramp an ensemble of $\sim 300$ SPGPE trajectories was computed\footnote{See online supplementary information for \cite{Weiler08a} for a subset of trajectories showing possible outcomes for each scenario of the experiment.} and the vortex observation statistics were compared with the experimental measurements. A comparison between experimental absorption images and column densities of representative SPGPE trajectories is shown in \fref{fig:spontVort1}, with clear qualitative similarity. A comparison of the vortex observation statistics yielded  quantitative agreement between experiment and the SPGPE theory. In \fref{fig:spontVort2} a time series of density iso-surfaces is shown for a particular trajectory. The emergence of the final BEC is seen to be a turbulent process, in this case resulting in a metastable vortex.

\subsubsection{Rotating Bose-Einstein condensation}\label{RBEC}

Bradley {\em et al.} \cite{Bradley2008a} have used the simple growth equation \eref{SGPEsimp} to model the dynamics of the formation of a \emph{rotating} Bose-Einstein condensate~\cite{Haljan2001a} where it is possible for stable vortices to form \emph{during} condensation.
The following question  arises: does a vortex-free condensate form before it is penetrated by vortices, or does condensation proceed into a state with vortices already present?  

It is helpful to consider the single particle energy spectrum (in the rotating frame) of the cylindrically symmetric harmonic trap in three dimensions
\begin{equation}\label{singlePS}
\epsilon_{nlm}=\hbar\omega_r(2n+|l|+1)-\hbar\Omega l+\hbar\omega_z(m+1/2),
\end{equation}
where $n, l, m$ are the radial, angular and axial quantum numbers. 
In the absence of interactions BEC will form in the ground state mode with energy $\epsilon_{000}=\hbar\omega_r+\hbar\omega_z/2$. Vortices arise from occupation of states with nonzero angular momentum and the positive angular momentum part of the spectrum behaves as $\hbar(\omega_r-\Omega)l$ leading to near-degeneracy of positive angular momentum modes as $\Omega\to\omega_r$. States with angular momentum $\langle \hat{L}_z\rangle=\hbar l>0$ are increasingly easy to populate as the rate of rotation increases. 

For a rapidly rotating BEC transition, vortices can play a dominant role at all stages in the BEC growth process. It is then possible that atoms condense into a vortex-filled but spatially disordered state---a \emph{vortex liquid}. 
\paragraph{SPGPE treatment of rotation}
In our development of the SPGPE we have included the possibility that the thermal cloud occupying \rI\ may be in rotational equilibrium in a symmetric trap. 
The transformation has consequences for the choice of $\ecut$ and the basis of single particle states. In harmonic oscillator units the single particle modes corresponding to \eref{singlePS} are also eigenstates of $L_z$:
\begin{equation}
\phi_{nlm}(r,\theta,z)={\cal N}_{nlm}e^{i\theta}r^{|l|}e^{-r^2/2}L_n(r^2)e^{-z^2/2}H_m(z)
\end{equation}
which is the appropriate basis for introducing a projection operator to separate the \rC\ and \rI\ regions. Imposing the cutoff in the rotating frame leads to a bias (in the direction of the rotation) for the angular momentum for the \CF\ region modes. In the limit of very fast rotation ($\Omega\to \omega_r$), $z$-excitations are strongly suppressed and only the lowest Landau level (LLL) is contained in the \CF\ if $\ecut<\hbar\omega_z$.

\begin{figure}[t]
\centering
\includegraphics[width=12.5cm]{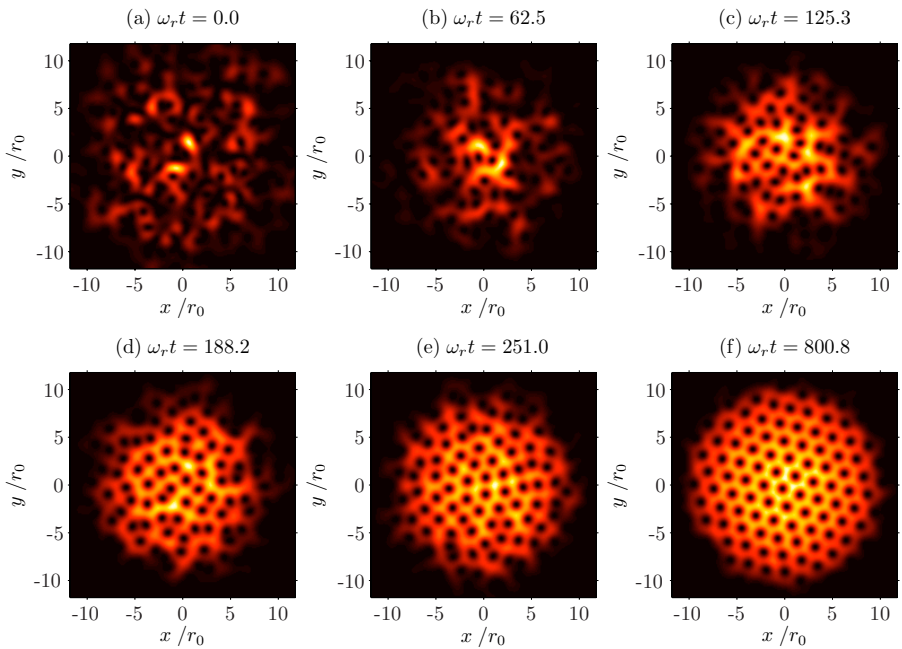}%
\caption{Rotating Bose-Einstein condensation: \CF\ region density for a single trajectory of the SPGPE (in the frame of the thermal cloud rotating at $\Omega_0=0.979\omega_r$). (a) Initial state for $1.3\times 10^6$ $^{87}{\rm Rb}$ atoms at $T_0=12 {\rm nK}$, with $\mu_0=0.5\epsilon_{000}$. At time $t=0$ the non-condensate band is quenched to $T=1 {\rm nK}$, $\mu=3.5\epsilon_{000}$, and $\Omega=\Omega_0$ preserving the rotation rate. (b)-(d) The \CF\ region undergoes rapid growth into a vortex liquid state. (e)-(f) At this low temperature the vortices then assemble into a regular Abrikosov lattice. Reproduced from \cite{Bradley2008a} \APSCR{2008}.\label{fig:RBEC}}
\end{figure}

In \fref{fig:RBEC} we show a representative SPGPE simulation of the BEC transition in a rapidly rotating system (not in the LLL regime). An initial thermal state with no condensate present is evolved subject to a sudden change in the thermal cloud temperature and chemical potential consistent with a quench below the critical point. When the final temperature after the quench is comparatively low ($T\ll T_c$ shown here) the dynamics show a phase of rapid growth into a vortex liquid, followed by a much slower ordering phase where the vortices assemble into a regular lattice. At higher temperatures ($T\sim T_c$), condensation into a vortex liquid still occurs but the ordering phase is frustrated by thermal fluctuations. 


\section{Conclusion}\label{sconclus}

In this review we have outlined a unified \CF\ theory for studying the dynamics and statistical mechanics of ultra-cold Bose gases. The various \CF\ approaches considered apply to situations as diverse as the zero-temperature quantum dynamics of colliding Bose-Eintein condensates, through to the effects of critical fluctuations at the condensation transition. 
Perhaps one of the most surprising aspect of these approaches is that underlying them is the well-known, and computationally obliging,  Gross-Pitaevskii equation. 
Two essential adaptions to the Gross-Pitaevskii theory bring this power to describe quantum and thermal effects: stochastisation of the field and evolution equation, and projection onto the \CF\ region.

Recently there has been increasing use of classical field techniques in the ultra-cold atom community, and particularly in the truncated Wigner approach to describe beyond-mean-field dynamics in BECs at zero temperature.  Most of the classical field calculations performed at finite temperature have not made controlled use of projectors, other than the \emph{accidental} projection intrinsic to the numerical representation. One of the consequences is that these results can often not be easily related to experiment.
 In this review we have extolled the virtues of a consistent energy projector used to define the \CF\ region, and have shown that this allows a theory that can be applied to quantitatively describe experiment.

The future development of \CF\ techniques is closely related to the direction in which many ultra-cold gas experiments are heading. We close this review by pointing to a few such areas under investigation:
\paragraph{Full implementation of SPGPE} To date the SPGPE scattering term has yet to be implemented numerically, primarily due to technical challenges. This term is expected to have an effect on strongly non-equilibrium scenarios such as condensate growth and unstable vortex dynamics, although it is currently not clear in what regimes it would give rise to measurable differences. Developing a formulation that can handle the dynamics of the incoherent region remains an important future challenge which would complete the theoretical description for most practical purposes. 
\paragraph{General interactions} Long-range interactions, such as dipole interactions, and strong s-wave interactions using Feshbach resonances are now routinely available in experiments. For these systems finite temperature and incoherent processes appear to become more important than in the weakly interacting s-wave case,  and should be well-suited to a \CF\ description.  
\paragraph{Fermionic systems} The quantum statistics of bosons allows for the modes of the atomic field to be highly occupied, providing the coherence central to the \CF\ description of the ultra-cold  Bose gas. In contrast, quantum statistics prohibits multiple fermions from occupying a single mode, and thus invalidating a \CF\ approach at face value. A fundamental question to address is whether there is some other pathway to providing a useful \CF\ description of Fermi gases.
 
\section*{Acknowledgments} \label{sAcknowledge}
The authors gratefully acknowledge discussions and interactions with:
D.~Baillie,
A.~H.~Bezett,
R.~N.~Bisset,
K.~Burnett,
C.~W.~Clark,
J.~F.~Corney,
B.~J.~D\c{a}browska-W\"uster,
P.~D.~Drummond,
A.~J.~Ferris,
C.~J.~Foster,
T.~I.~A.~Fudge,
D.~A.~W.~Hutchinson,
P.~Jain,
K.~V.~Kheruntsyan,
M.~D.~Lee,
W.~Moore,
S.~A.~Morgan,
A.~A.~Norrie,
M.~K.~Olsen,
B.~Schneider,
R.~G.~Scott,
T.~P.~Simula,
A.~G.~Sykes,
T.~M.~Wright,
S.~W\"{u}ster,
as well as many others who have played an important role in shaping the development of the theories and applications discussed in this review.

This work was financially supported by the University of Otago, the New Zealand Foundation for
Research Science and Technology under the contracts NERF-UOOX0703: Quantum Technologies, and UOOX0801, Marsden Contract No. UOO509, the
University of Queensland, and the Australian Research Council Centre of Excellence for Quantum-Atom
Optics (project number CE0348178).  

\pagebreak
\appendices
\section{Numerical technique for the harmonically trapped system}\label{SEC:Numerics}
In this appendix we overview a numerical method that allows an efficient  and accurate solution of the projected Gross-Pitaevskii equation in a harmonic potential.
In \aref{FTnumerics} we briefly discuss a method for the uniform gas and refer  to  references \cite{Blakie2005a,Blakie2008a,Bradley2008a,Wright08} for a more detailed discussion of implementation details and applications to rotating systems.

\subsection{Numerical requirements}\label{snumreq}
The modes of the system are of central importance in the
assumptions used to derive the various \CF\ methods presented in this review, and care must be taken in numerical implementations to ensure the modes are faithfully represented. Any useful
simulation technique must satisfy the following requirements.
\begin{itemize}
\item[(i)]The space spanned by the modes of the simulation 
should match the \CF\ region as closely as possible.  
\item[(ii)] All  modes in the \CF\ regime must be propagated accurately.
\end{itemize}

The case we  examine here is the PGPE equation for the harmonically trapped system, i.e. where
\begin{equation}
V_{0}(\mathbf{x})=\frac{1}{2}m\omega^2\left(x^2+y^2+z^2\right).
\end{equation}
To simplify the discussion we have
taken the harmonic trapping potential to be isotropic\footnote{This restriction simply allows us to avoid using cumbersome notation to account for different
spectral bases in each direction, and the fully anisotropic case is of no additional computational complexity.} and will not consider the perturbation potential $\delta V$. 
We take the \CF\ region to be defined by an energy cutoff in the single particle basis, i.e.~eigenstates of $H_0=p^2/2m+V_0(\mathbf{x})$ with energy less than $\ecut$. 

\subsection{Spectral representation of the PGPE}\label{sspecrep}
For convenience, we write the PGPE in dimensionless units to simplify the discussion, and explicitly indicate all dimensionless quantities in this section by use of tildes. We do this by introducing a unit of distance $x_0=\sqrt{\hbar/m\omega}$ and time $t_0=1/\omega$. These choices immediately imply computational units for  energy $E_0=\hbar\omega$ and momentum $p_0=\sqrt{\hbar m\omega}$. So, for example, our dimensionless distance variable is defined as $\tilde{x} = x/x_0$, dimensionless time, $\tilde{t}=t/t_0$, and \CF, $\tilde{\psi}_{\rC} = \psi_\rC x_0^{3/2}$. The coefficient of the nonlinear term in the Gross-Pitaevskii equation is given by the product $u$. In dimensionless units we define this as the nonlinearity constant $C_{\rm{NL}}\equiv ut_0/\hbar x_0^3$.

In dimensionless units, the PGPE takes the form
\begin{eqnarray}
i\frac{\partial\tilde{\psi}_{\rC}}{\partial \tilde t} & = & \tilde H_{0}\tilde{\psi}_{\rC}+ \PC\left\{C_{\rm{NL}} |\tilde{\psi}_{\rC}|^{2}\tilde{\psi}_{\rC}\right\},\label{eq:numGPE1}\end{eqnarray} 
where
\begin{equation}
\tilde{H}_0=-\frac{1}{2}\tilde{\nabla}^2+\frac{1}{2}(\tilde{x}^2+\tilde{y}^2+\tilde{z}^2).
\end{equation}

The \CF\ is expanded in a spectral basis as 
\begin{equation}
\tilde{\psi}_{\rC}(\tilde{\mathbf{x}},\tilde t)=\sum_{n\in\rC}c_{n}(\tilde t)\,\tilde\phi_{n}(\tilde{\mathbf{x}}),\label{eq:psibasis}\end{equation}
 where $\{\tilde\phi_{n}(\tilde{\mathbf{x}})\}$ are the harmonic oscillator eigenstates of  $\tilde H_0$ with respective eigenvalues $\tilde\epsilon_{n}$, and the $\{c_n\}$ are complex amplitudes. 
 The projection
is explicitly implemented by limiting the summation indices in \eref{eq:psibasis} to the set of values
\begin{equation}
\rC=\{n:\tilde\epsilon_{n}\leq \tecut\},\label{eq:Cset}\end{equation}  
i.e. the field $\tilde{\psi}_{\rC}$ only contains the modes of interest. 

\subsection{Mode evolution}\label{smodeevol}
Having used the modes of $\tilde{H}_0$ as the spectral basis and to realize the projector,  we follow the Galerkin approach  (i.e.~projecting   \eref{eq:numGPE1} on to our spectral basis) to obtain the amplitude evolution equation 
\begin{eqnarray}\label{eq:GPEshobasis}
\frac{\partial c_{n}}{\partial\tilde t} & = & -i\left[\tilde\epsilon_{n}c_n +C_{\rm{NL}}G_{n}\right],\end{eqnarray}
 where
\begin{eqnarray}\label{eq:GNL}
G_{n}&\equiv&\int d^{3}\tilde{\mathbf{x}}\:\tilde\phi_{n}^{*}(\tilde{\mathbf{x}})|\tilde{\psi}_{\rC}(\tilde{\mathbf{x}},\tilde t)|^{2}\tilde{\psi}_{\rC}(\tilde{\mathbf{x}},\tilde{t}),\end{eqnarray}
is the nonlinear matrix element. Once this matrix elements is evaluated, the evolution of the system can be calculated using  numerical algorithms for  systems of ordinary differential equations, e.g. the Runge-Kutta algorithm (e.g.~see \cite{Press92a}). Since this is a well-understood area of numerical mathematics we do not concern ourselves with the details of the propagation algorithm, but instead focus on evaluating \eref{eq:GNL}.  

We can point out the central issue for numerical implementation. Expanding the fields in expression \eref{eq:GNL} into the mode basis we obtain
\begin{equation}
G_{n}=\sum_{pqr}\left\{\int d^{3}\tilde{\mathbf{x}}\: \tilde\phi_{n}^{*}(\tilde{\mathbf{x}})\tilde\phi_{p}^{*}(\tilde{\mathbf{x}})\tilde\phi_{q}^{}(\tilde{\mathbf{x}})\tilde\phi_{r}^{}(\tilde{\mathbf{x}})\right\}\: c^*_pc_qc_r.
\end{equation}
While the matrix elements within the brackets can be exactly calculated in advance,  computing all $G_n$ values  using this expression requires $O(M^4)$ floating point operations, where $M$ is the number of \CF\ region modes. Such scaling would be prohibitive for performing realistic calculations. In what follows we show how to compute these matrix elements with a scheme that only requires $O(M^{4/3})$ operations. Such spectral representations have also been considered for the zero-temperature (non-projected) Gross-Pitaevskii equation in references \cite{Dion2003a,Bao2005a}.

\subsection{Separability} \label{sAsep}
 An important feature of the basis states (i.e.~eigenstates of $\tilde{H}_0$) is that they are separable into 1D eigenstates, i.e.
\begin{eqnarray}
\tilde{\phi}_n(\tilde{\mathbf{x}})&\leftrightarrow&\tilde\varphi_{\alpha}(\tilde{x})\tilde\varphi_{\beta}(\tilde{y})\tilde\varphi_{\gamma}(\tilde{z}),\label{EQ:PW1Dstates}\\
\tilde{\epsilon}_n&\leftrightarrow&\tilde{\varepsilon}_{\alpha}+\tilde{\varepsilon}_{\beta}+\tilde{\varepsilon}_{\gamma},\label{EQ:PW1Denergies}\\
c_n&\leftrightarrow&c_{\alpha\beta\gamma}, \label{EQ:PW1Damps}
\end{eqnarray} 
where $\{\tilde\varphi_{\alpha}(\tilde x)\}$ are eigenstates of the 1D harmonic
oscillator Hamiltonian, i.e.\begin{equation}
\left[-\frac{1}{2}\frac{d^{2}}{d\tilde x^{2}}+\frac{1}{2}\tilde x^{2}\right]\tilde\varphi_{\alpha}(\tilde x)=\tilde\varepsilon_{\alpha}\tilde\varphi_{\alpha}(\tilde x),\label{eq:sho1D}\end{equation}
 with eigenvalue $\tilde\varepsilon_{\alpha}=(\alpha+\frac{1}{2})$, for $\alpha$ a non-negative integer.

For clarity we use Greek subscripts to label the 1D eigenstates, so that the specification of the \CF\  region in \eref{eq:Cset} becomes \begin{equation}
\rC=\{\alpha,\beta,\gamma:\tilde\varepsilon_{\alpha}+\tilde\varepsilon_{\beta}+\tilde\varepsilon_{\gamma}\le\tecut\}.\label{sepCR}
\end{equation} 
Within the \CF\ region there exists $M_x$ ($\approx\tilde{\epsilon}_{\rm{cut}}$) distinct 1D eigenstates (i.e.~$\tilde{\varphi}_{\alpha}$) in each direction, and thus $M\approx \frac{1}{6}M_x^3$ 3D basis states ($\tilde{\phi}_n$) in the \CF\ region (see the left subplot of \fref{fig:Numtrans}).

\subsection{Evaluating the matrix elements} \label{sec:matrixEval}
 An important observation made in reference \cite{Dion2003a} was that the nonlinear matrix element given in \eref{eq:GNL} can be computed exactly with an appropriately chosen Gauss-Hermite quadrature. To show this we note that because the harmonic oscillator
states are of the form $\tilde\varphi_{\alpha}(\tilde x)=h_{\alpha} H_{\alpha}(\tilde x)\exp(-\tilde x^{2}/2),$
where $H_{\alpha}(\tilde x)$ is a Hermite polynomial of degree $\alpha$, the field (at any instant of time) can be written as\begin{equation}
\tilde \psi_\rC(\tilde{\mathbf{x}},\tilde t)=Q(\tilde x,\tilde y,\tilde z)e^{-(\tilde x^{2}+\tilde y^{2}+\tilde z^{2})/2},\label{eq:PsiQpoly}\end{equation}
 where 
 \begin{equation}Q(\tilde x,\tilde y,\tilde z)\equiv \sum_{\{\alpha\beta\gamma\}\,\in\,\rC}c_{\alpha\beta\gamma}(\tilde t)\,h_{\alpha}H_{\alpha}(\tilde x)h_{\beta}H_{\beta}(\tilde y)h_{\gamma}H_{\gamma}(\tilde z),
 \end{equation}is a polynomial that, as a result of the cutoff,
is of maximum degree $M_x-1$ in the independent variables.

Similarly, it follows that because the interaction term (\ref{eq:GNL}) is fourth
order in the field, it can be written in the form\begin{equation}
G_{\alpha\beta\gamma}=\int d^{3}\tilde{\mathbf{x}}\: e^{-2(\tilde x^{2}+\tilde y^{2}+\tilde z^{2})}P_{\alpha\beta\gamma}(\tilde x,\tilde y,\tilde z),\label{eq:FNLquad}\end{equation}
 where 
\begin{eqnarray}P_{\alpha\beta\gamma}(\tilde x,\tilde y,\tilde z)&\equiv& h_{\alpha}H_{\alpha}(\tilde x)h_{\beta}H_{\beta}(\tilde y)h_{\gamma}H_{\gamma}(\tilde z)|Q(\tilde x,\tilde y,\tilde z)|^2Q( \tilde x,\tilde y,\tilde z),
\end{eqnarray}
 is a polynomial of maximum degree $4\,(M_x-1)$
in the independent variables. To evaluate these integrals, we note the general form of the $N_Q$ point Gauss-Hermite quadrature
\begin{equation}
\int_{-\infty}^{+\infty}d\tilde{x}\,w(\tilde{x})f(\tilde{x})\approx\sum_{j=1}^{N_Q}w_jf(\tilde{x}_j),
\end{equation}
where $w(\tilde{x})$ is a Gaussian weight function, and the $N_Q$ values of $w_j$ and $x_j$ are the quadrature weights and roots, respectively (see \cite{abramowitz1964a}). This quadrature   is exact if $f(\tilde{x})$ a polynomial of maximum degree $2N_Q-1$.

 Identifying the exponential term in \eref{eq:FNLquad} as
the usual weight function for quadrature,  the integral
can be exactly evaluated using a three-dimensional spatial 
grid of $8\,(M_x-1)^3$ points (i.e. $2\,(M_x-1)$ points in each direction \footnote{Since a polynomial of degree $2N-1$ is integrated exactly using an $N$-point quadrature.}), i.e.
\begin{equation}
G_{\alpha\beta\gamma} =\sum_{ijk}w_iw_jw_kP_{\alpha\beta\gamma}(\tilde{x}_i,\tilde{x}_j,\tilde{x}_k),\label{Gabc}
\end{equation}
where $\tilde x_i$ and $w_i$ are the $2\,(M_x-1)$ roots and weights of the 1D Gauss-Hermite quadrature with weight function $w(\tilde x)=\exp(-2\tilde x^2)$ \cite{abramowitz1964a}. Note, the due to the isotropy of the trapping potential, the quadrature grids in all spatial directions are identical.

\subsection{Overview of numerical procedure}\label{SEC:HARMnummeth}
\begin{figure}[htbp]  
 \centering{\includegraphics[width=5in]{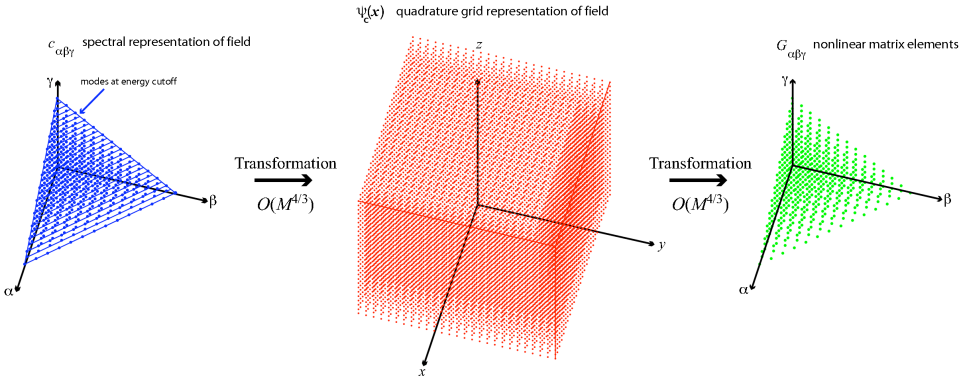} }
   \caption{Schematic of numerical procedure to evaluate the nonlinear matrix elements $G_{\alpha\beta\gamma}$.  \label{fig:Numtrans}
}
  
\end{figure} 
Here we briefly overview how the quadrature described  above can be efficiently implemented numerically. We require the transformation matrices, given by
 1D basis states evaluated on the quadrature grid, i.e.
\begin{equation}
U_{i\alpha}=\tilde\varphi_{\alpha}(\tilde{x}_i),\label{harmU}
\end{equation}
to be pre-calculated.
 Starting from the basis set representation of the field (i.e. $\{c_{\alpha\beta\gamma}\}$) at an instant of time $\tilde t$, the steps for calculating the matrix elements are as follows (also see \fref{fig:Numtrans}):
\begin{enumerate}
\item Transform from spectral to spatial representation:
\begin{equation}
\tilde{\psi}_{\rC}(\tilde{\mathbf{x}}_{ijk},\tilde t)=\sum_{\{\alpha\beta\gamma\}\,\in\,\rC}U_{i\alpha}U_{j\beta}U_{k\gamma}\,c_{\alpha\beta\gamma}(\tilde t),\label{Harm2pos}
\end{equation}
where $\tilde{\mathbf{x}}_{ijk}=(\tilde{x}_i,\tilde{x}_j,\tilde{x}_k)$.

\item The quadrature integrand of the nonlinear matrix element  \eref{eq:GNL}  is constructed by appropriately dividing by the weight function and pre-multiplying by the weights\footnote{Here we form $e^{2|\tilde{\mathbf{x}}_{ijk}|^2}|\tilde{\psi}_{\rC}|^{2}\tilde{\psi}_{\rC}$ as this corresponds to the polynomial  ($P$) required for the quadrature (see \eref{Gabc}).}, i.e.
\begin{eqnarray} 
g(\tilde{\mathbf{x}}_{ijk})&\equiv&w_iw_jw_ke^{2|\tilde{\mathbf{x}}_{ijk}|^2}|\tilde{\psi}_{\rC}(\tilde{\mathbf{x}}_{ijk},\tilde{t})|^{2}\tilde{\psi}_{\rC}(\tilde{\mathbf{x}}_{ijk},\tilde{t}).\label{Eqgtrans}
\end{eqnarray}

\item Inverse transforming these integrand functions yields the desired matrix elements:
\begin{eqnarray} 
G_{\alpha\beta\gamma} &=& \sum_{ijk}U_{i\alpha}^*U_{j\beta}^*U_{k\gamma}^*g(\tilde{\mathbf{x}}_{ijk}).
\end{eqnarray}
\end{enumerate}
  The slowest step in this procedure is carrying out the basis transformation, which requires $O(M_x^4)$, i.e.~ $O(M^{4/3})$ floating point operations when carried out as a series of matrix multiplications.  Typical simulations, where we evolve a \CF\ field with $M\approx 2,000$ modes for 100 trap periods, take about 2 hours.

\section{Numerical technique for the uniform system}\label{FTnumerics}
\subsection{Spectral representation}\label{FTspecrep}
The basic quadrature arguments presented for the harmonic oscillator case can be applied to the numerical description of the uniform Bose gas. In this section  we will briefly discuss the uniform case, referring to results from \aref{SEC:Numerics} where they are the same.
The system of interest is taken to be in a cuboid volume with linear dimensions $\{L,L,L\}$ and subject to periodic boundary conditions. 

The dimensionless PGPE takes the same form as in \eeref{eq:numGPE1}, but with a basis Hamiltonian of the form
\begin{equation}
\tilde{H}_0=-\frac{{\tilde\nabla}^{2}}{(2\pi)^2},\label{EQ:H0UniformNumerical}
\end{equation}
where we take periodic boundary conditions,
\begin{equation}
\tilde{\psi}_{\rC}(\tilde{x}+1,\tilde{y},\tilde{z})=\tilde{\psi}_{\rC}(\tilde{x},\tilde{y}+1,\tilde{z})=\tilde{\psi}_{\rC}(\tilde{x},\tilde{y},\tilde{z}+1)=\tilde{\psi}_{\rC}(\tilde{x},\tilde{y},\tilde{z}),\label{EQ:unifBDs}
\end{equation}
and have used $x_{0}=L$  and $t_{0}=mL^{2}/\pi h$
as the units of length and time.

As in the harmonic case (see equations \eref{EQ:PW1Dstates} - \eref{EQ:PW1Damps}) the basis states are separable into 1D eigenstates (i.e. $\tilde{\phi}_n(\tilde{\x})=\tilde{\varphi}_{\alpha}(\tilde{x})\tilde{\varphi}_{\beta}(\tilde{y})\tilde{\varphi}_{\gamma}(\tilde{z})$) of the form
\begin{equation} 
\tilde{\varphi}_{\alpha}(\tilde{x})=e^{i\tilde{k}_{\alpha}\cdot\tilde{x}},\label{EQ:pwstates1D}\end{equation}
with the wavevectors
$\tilde{k}_{\alpha}$ chosen as harmonics of the perodicity interval, i.e.,
$\tilde{k}_{\alpha}=2\pi \alpha,$ with $\alpha$ an integer, and 
respective eigenvalues $\tilde{\varepsilon}_{\alpha}=\alpha^{2}$. The values of the indices $\{\alpha,\beta,\gamma\}$ specifying the \CF\ region is given by \eeref{sepCR}, which defines  a sphere of radius $\sqrt{\tecut}$ in $\alpha\beta\gamma$-space for the uniform system.
For later convenience
we define $\alpha_{\max}$ as the maximum value of $\alpha$  that occurs in $\rC$, i.e. the highest order  basis state in each direction. For the planewave case we have $\alpha_{\max}\simeq\sqrt{\tecut}$, and thus in the \CF\ region we have a total of $M_x=2\alpha_{\max}+1$ distinct 1D basis states (i.e., $\tilde{\varphi}_\alpha$) in each direction, with $M\approx\frac{\pi}{6}M_x^3$ 3D basis states (i.e., $\tilde{\phi}_n$). 

\subsection{Evaluating the matrix elements}\label{FTmatelems}
In the planewave spectral representation the PGPE takes the form \eref{eq:GPEshobasis}, for which the main challenge is evaluating the nonlinear matrix element \eref{eq:GNL}. We now show how a quadrature approach can be used to evaluate this matrix element exactly. The  essence of this approach is to transform the field to a spatial representation where the nonlinear term is local.  

In each spatial dimension, the quadrature grid of interest (for the uniform case) consists of $N_{\rm Q}$ equally-spaced points  given by \begin{equation}
\tilde{x}_{j}=j\,\Delta\tilde{x},\qquad1\le j\le N_{\rm Q},\label{EQ:PWxgrid}\end{equation}
with spacing $\Delta\tilde{x}=1/N_{\rm Q}$, which spans the spatial region $(0,1]$. The quadrature expression for an integral of an arbitrary  function $f$ is
\begin{equation}
\int_0^1d\tilde{x}\,w(\tilde x)\,f(\tilde x)\approx\sum_{j=1}^{N_{\rm Q}} w_j f(\tilde x_j),
\end{equation} 
where $w(\tilde x)=1$ is the weight function, and $w_j=\Delta\tilde{x}$. That is, for the planewave approach, the quadrature rule is the well-known \emph{rectangular rule} from elementary numerical analysis.
  
The requirement that our quadrature will exactly calculate the nonlinear matrix elements is equivalent to the requirement that the 1D integrals between are all products of four $\tilde{\varphi}_j(\tilde{x})$ are evaluated exactly, i.e.
\begin{eqnarray}
I_{\alpha\beta\gamma\delta}&=&\sum_{j=1}^{N_{\rm Q}}\Delta\tilde{x}\,\tilde{\varphi}_{\alpha}^{*}(\tilde{x}_{j})\tilde{\varphi}_{\beta}^{*}(\tilde{x}_{j})\tilde{\varphi}_{\gamma}(\tilde{x}_{j})\tilde{\varphi}_{\delta}(\tilde{x}_{j}), \quad -\alpha_{\max}\le\alpha,\beta,\gamma,\delta\le\alpha_{\max},\label{NLMEPW}\\
&=&\delta_{\alpha+\beta,\gamma+\delta},\end{eqnarray}
which holds for the quadrature described above if we take $N_{\rm Q}\ge 2M_x$. Thus the most efficient and accurate representation is when we choose $2M_x$ grid points in each spatial dimension.

\subsubsection{Fourier interpretation}\label{FTinterp}
The quadrature grid requirement ($N_{\rm Q}=2M_x$) can be interpreted in terms of Fourier properties of the spatial grid. To represent a maximum wavevector of $\tilde{k}_{\rm{cut}}=2\pi\alpha_{\max}\approx\pi M_x$, the Nyquist requirement for the spatial grid is that the distance between points should be $\Delta\tilde{x}=1/M_x$ (or smaller), which requires atleast $M_x$ points over the interval $(0,1]$. However, our quadrature argument above was that to evaluate the nonlinear matrix elements correctly we need twice as many grid points, i.e $N_{\rm Q}=2M_x$. Such a grid is sufficient to satisfy the Nyquist condition for wavevectors of magnitude up to $2\tilde{k}_{\rm{cut}}$. To understand why we need so many points consider the \emph{worst case} for the matrix element in \eref{NLMEPW}: the case $-\alpha=-\beta=\gamma=\delta=\alpha_{\max}$, i.e. where all modes occur with the maximum magnitude wavevector  $\tilde{k}_{\rm cut}$. The integrand of \eref{NLMEPW}, i.e. the product of these four modes, is itself a planewave with wavevector $4\tilde{k}_{\rm cut}$. On the spatial grid with $2M_x$ points this cannot be represented unambiguously (i.e. it exceeds the Nyquist limit of $2\tilde{k}_{\rm{cut}}$), and is aliased. However, for the choice of $2M_x$-points, this aliasing does not map the wavevector into the region $[-\tilde{k}_{\rm{cut}},\tilde{k}_{\rm{cut}}]$,  and hence does not effect the matrix elements evaluated for the \CF\ region. For any fewer points the aliased wavevector maps into $[-\tilde{k}_{\rm{cut}},\tilde{k}_{\rm{cut}}]$, and gives rise to spurious dynamics. 

\subsection{Overview of numerical procedure}\label{SEC:PWnummeth}
We could apply an identical procedure to that discussed in \sref{SEC:HARMnummeth} to evaluate the matrix elements with a computational cost per evaluation of $O(M^4)$. However, for the planewave case the basis transformation between spectral (momentum space) and position space quadrature grids (i.e. steps (i) and (iii) in \sref{SEC:HARMnummeth}) is equivalent to a fast Fourier transformation, which has a computational cost of $O\left(M^3\log(M)\right)$.  
For more details on the planewave procedure we refer the reader to reference \cite{Blakie2008a}.

\section{Mapping to stochastic equations}\label{sec:stochMapping}
The utility of phase space methods requires that the equation of motion for the quasi-probability distribution (here \eref{HCthirdorder}) can be mapped to an equivalent stochastic differential equation, which is comparatively much easier to solve. A projected functional Fokker-Planck equation of the form
\begin{eqnarray}\label{FPEgen}
\frac{\partial P}{\partial t}&=&\intV{\mathbf{x}}\Bigg\{-\DDP{\psi_\rC(\mathbf{x})}A\left(\psi_\rC(\mathbf{x}),\psi_\rC^*(\mathbf{x}),t\right)+{\rm h.c.}\nonumber\\
&&+\DPDP{\psi_\rC(\mathbf{x})}{\psi_\rC(\mathbf{x})}D_{11}\left(\psi_\rC(\mathbf{x}),\psi_\rC^*(\mathbf{x}),t\right)+{\rm h.c.}\nonumber\\
&&+\DPDP{\psi_\rC(\mathbf{x})}{\psi_\rC^*(\mathbf{x})}D_{12}\left(\psi_\rC(\mathbf{x}),\psi_\rC^*(\mathbf{x}),t\right)+{\rm h.c.}\Bigg\}P,
\end{eqnarray}
with drift vector $\bm{A}=[A,A^*]$ and diffusion matrix $\bm{D}\equiv [D_{11},D_{12};D_{12}^*,D_{11}^*]$ has an equivalent stochastic equation if the diffusion matrix is positive semi-definite. A factorization of the diffusion matrix in the form $\bm{D}=\bm{B}\bm{B}^T$ can then be found, and the stochastic differential equation is given by
\begin{equation}\label{SGPEgen}
\bm{d\psi}_\rC(\x,t)=\bm{\PC}\left\{\bm{A}(\x,t)dt+\bm{B}(\x,t)\bm{dW}(\x,t)\right\}
\end{equation}
where $\bm{ d\psi}_\rC=[d\psi_\rC,d\psi_\rC^*]^T$, $\bm{\PC}=[\PC, \PC^*]^T$ and $\bm{dW}(\x,t)$ is a vector of  noises. In general, mapping to ordinary stochastic differential equations is only possible if the equation of motion for the quasi-probability is strictly a Fokker-Planck equation (derivatives up to second order). 

There is an important technical point regarding the equivalence of \eref{FPEgen} and \eref{SGPEgen}. The strict equivalence holds only if the projector $\PC$ is implemented with sufficient care. In the language of \sref{sec:matrixEval}, the quadrature chosen to compute \eref{SGPEgen} must be sufficient to generate a \CF\ delta function of the appropriate numerical order upon stochastic averaging. The standard proofs of equivalence \cite{Gardiner1985a} can be adapted to show that for a diffusion term of polynomial degree $2D_x$, giving $\bm{B}$ of degree $D_x$, the delta function must to be a true delta function for terms up to order $D_x+M_x$, requiring expansion of the noise up to states of degree $D_x+M_x$ and implementation of \eeref{SGPEgen} using a numerical quadrature rule sufficient to integrate terms of order $2D_x+2M_x$ or a rule of order $D_x+M_x-1$ to generate the appropriate equivalence. \note{careful here, check result!}

\pagebreak

\bibliographystyle{plain}
\bibliography{cfields}
\end{document}